\documentclass[preprint,authoryear,11pt,letterpaper]{elsarticle}
\usepackage[top=0.75in, bottom=0.75in, left=0.75in, right=0.75in]{geometry}

\usepackage{chngcntr}
\usepackage{amsmath,amsthm}
\usepackage{amssymb}
\usepackage{bbm}
\usepackage{float}
\usepackage{amsfonts}
\usepackage{graphicx}
\usepackage{subfigure}
\graphicspath{{graphics/}}
\usepackage[ruled,vlined,algo2e]{algorithm2e}

\usepackage{algorithm}
\usepackage{algorithmic}
\usepackage{tabularx,booktabs}

\usepackage{graphicx,color}
\usepackage[T1]{fontenc}

\usepackage{xcolor}
\definecolor{darkblue}{rgb}{0.0,0.5,0.5}
\definecolor{blue}{rgb}{0.0,0.0,1}
\usepackage[colorlinks]{hyperref}
\hypersetup{colorlinks,breaklinks,linkcolor=blue,urlcolor=blue,anchorcolor=blue,citecolor=blue}
\usepackage{booktabs,caption}
\usepackage{multirow}
\usepackage{lscape}
\usepackage{epstopdf}
\usepackage{lineno}

\usepackage{microtype}
\usepackage{lineno}
\newcommand{\given}{\;\middle|\;}

\journal{arxiv}

\begin{document}
\begin{frontmatter}

\title{Bayesian inference for link travel time correlation of a bus route}

\author{Xiaoxu Chen}
\ead{xiaoxu.chen@mail.mcgill.ca}
\author{Zhanhong Cheng}
\ead{zhanhong.cheng@mail.mcgill.ca}%
\author{Lijun Sun\corref{cor1}}
\ead{lijun.sun@mcgill.ca}

\cortext[cor1]{Corresponding author.}
\address{Department of Civil Engineering, McGill University, Montreal, QC H3A 0C3, Canada}

\begin{abstract}
Estimation of link travel time correlation of a bus route is essential to many bus operation applications, such as timetable scheduling, travel time forecasting and transit service assessment/improvement. Most previous studies rely on either independent assumptions or simplified local spatial correlation structures. In the real world, however, link travel time on a bus route could exhibit complex correlation structures, such as long-range correlations (e.g., a delayed bus is more likely to be further delayed due to bus bunching), negative correlations (e.g., a bus that goes ahead of schedule may intentionally slow down to follow a pre-defined timetable), and time-varying correlations (e.g., different correlation patterns for peak and off-peak hours). Therefore, before introducing strong  assumptions, it is essential to empirically quantify and examine the correlation structure of link travel time from real-world bus operation data. To this end, this paper develops a Bayesian Gaussian model to estimate the link travel time correlation matrix of a bus route using smart-card-like data. Our method overcomes the small-sample-size problem in correlation matrix estimation by borrowing/integrating those incomplete observations (i.e., with missing/ragged values and overlapped link segments) from other bus routes. Next, we propose an efficient Gibbs sampling framework to marginalize over the missing and ragged values and obtain the posterior distribution of the correlation matrix. Three numerical experiments are conducted to evaluate model performance. We first conduct a synthetic experiment and our results show that the proposed method produces an accurate estimation for travel time correlations with credible intervals. Next, we perform experiments on a real-world bus route with smart card data; our results show that both local and long-range correlations exist on this bus route. Finally, we demonstrate an application of using the estimated covariance matrix to make probabilistic forecasting of link and trip travel time.

\end{abstract}


\end{frontmatter}

\date{} 					

\section{Introduction}

Understanding travel time characteristics of buses is not only vital in providing better services for passengers (e.g., better travel time estimation), but also essential for transit agencies to design efficient and economical operation strategies (e.g., better route and timetable optimization) \citep{liao2020disparities}. A bus route can be viewed as a directed chain network, where each node represents a bus stop and each link represents the road section between two adjacent bus stops. Link travel time correlation of a bus route is essential to understanding the characteristics of the bus route and improving bus travel time estimation \citep{dai2019bus}.

However, most existing studies on estimating link travel time and the corresponding correlation structure mainly centers on passenger car traffic, while such analysis is often inappropriate for bus systems due to the unique operational properties of bus services. A major limitation that prevents us from using existing link travel time analysis models is that an accurate estimation of link travel time correlation matrices requires a large number of complete observations, while buses are essentially sparse in general road traffic and incomplete trip observations are pretty normal from bus operations, particularly for a long route with many links. As a result, the scale of many studies is confined to only a few links \citep[e.g.,][]{gajewski2005estimating}. To better utilize the limited data, another approach is to use simplified/parsimonious correlation structure to model link travel time. For instance, many statistical models assume only the travel times of adjacent/near links are correlated \citep[e.g., ][]{chen2012reliable,jenelius2013travel, srinivasan2014finding}. Although this is an intuitively reasonable assumption, it is inappropriate for bus travel time since link travel time correlations of a bus route have much more complex spatiotemporal characteristics. First, the link travel time of a bus route may have long-range correlations due to factors such as bus bunching (e.g., a delayed bus tends to be further delayed). Second, the correlation might be negative; for example, a bus that goes ahead of schedule may intentionally slow down to follow a pre-defined timetable. Moreover, link travel time correlations vary in different periods; the correlations of peak hours and the off-peak period could be completely different due to the time-varying service frequency and road traffic. This further limits the available sample size to conduct link travel time analysis for a bus route over a pre-defined time window. In summary, it is difficult to adapt existing link travel time analysis for car traffic to bus operation due to (1) small sample size: the limited number of complete observations are usually insufficient to estimate the link travel time correlation accurately; (2) oversimplified assumptions on the correlation structure: assuming only local spatial correlation is insufficient to capture the complex characteristics of actual link travel time for bus operation. 

If having access to a large amount of bus operation data (e.g., automatic vehicle location data, smart card data), we can infer the arrival time of a bus at a bus stop. Then, we can build a vector of link travel time using such in-out-stop records for each service run (from the first stop to the last stop), and then estimate the mean and covariance from samples for multiple service runs in a similar way as for car traffic. However, such data in practice is often not readily available due to the following issues. First, the arrival time at a stop becomes inaccessible when there is no boarding/alighting passenger or when the stop is skipped by a bus, bringing many unknown values in arrival time (and thus link travel time). A unique property in bus operation is that, for the stop-skipping case we still can obtain the sum of travel time of several adjacent links from the arrival time of upstream and downstream stops; we refer to this special type of missing values as \textit{ragged values}. Ragged values are quite common in bus systems and contain valuable information for enhancing link travel time correlation estimation. Secondly, buses are essentially sparse in traffic, and the number of operational buses per bus route per day is usually very small. For instance, a high-frequency bus route with a 10-min headway will only generate a sample size of 6/hour (if fully observed), which is much smaller than that of general car traffic. The small sample size and the ragged pattern prevent us from having a robust estimation of link travel time correlation, especially when quantifying link travel time correlations of a specific period (such as morning peak).

To address the above issues, in this paper we develop a Bayesian probabilistic model to estimate the link travel time correlations in a bus route. In particular, we aim to address the missing/ragged value problem and limited sample size problem for a target bus route by incorporating data from other bus routes that have overlapped links/stops. We assume the travel time of links in a bus route follows a multivariate Gaussian distribution. The task is to estimate the covariance matrix and the Bayesian credible interval of each entry in the matrix (correlation matrix can be obtained from the covariance matrix). In particular, our method makes use of incomplete observations with missing, ragged values and route segments from multiple bus routes. We point out that the conditional distribution of missing and ragged values can be viewed as a multivariate Gaussian distribution truncated on the intersection with a hyperplane. Next, we develop an efficient Markov chain Monte Carlo (MCMC) sampling algorithm to marginalize over the missing and ragged values and obtain the posterior distribution of the covariance matrix. In a test with synthetic data, we found our method produces accurate estimation for link travel time covariance. The MCMC scheme also allows us to exploit the posterior distribution of each entry in the covariance/correlation matrix. In addition, the incorporation of incomplete data substantially improves the estimation. Moreover, we use our model to empirically quantify the link travel time correlations of a twenty-link bus route in Guangzhou, China; results reveal strong local and long-range correlation patterns in link travel time of the bus route. Finally, we demonstrate an example of probabilistic forecasting of link/trip travel time in a bus route using the estimated covariance matrix; our forecasting method is more accurate than the historical average.

The contribution of this paper is twofold. First, we propose a Bayesian model that can use incomplete/corrupted vectors of link travel time from multiple bus routes to estimate the link travel time correlations of a target bus route. This model overcomes the small sample size of a single bus route by integrating incomplete data that are unusable in other models. A Gibbs sampling algorithm is developed to obtain the posterior distribution of the covariance/correlation matrix. Second, we verify the robustness and applicability of the proposed model by a synthetic example and a real-world case study. Results show our model can accurately estimate the link travel time correlations with incomplete observations, and the model applies to problems at a practical scale. The estimated correlations are beneficial to system understanding/evaluation and bus travel time estimation/forecasting.

The remainder of this paper is organized as follows. In Section~\ref{relatedwork}, we review previous studies on link travel time correlation. In Section~\ref{sec:problem}, we describe the problem of the link travel time correlation estimation in a bus route and introduce notations. In Section~\ref{methodology}, we present the Bayesian probabilistic model and the inference method based on MCMC. Next, in Section~\ref{casestudy}, we demonstrate the capability of our model through three experiments. Finally, we conclude our study, summarize our main findings, and discuss future research directions in Section~\ref{conclusion}.

\section{Related Work}
\label{relatedwork}
Most previous studies have concentrated on link travel time correlation or covariance matrix estimation for car trips. During the 1990s, Advanced Traveler Information System (ATIS) was deployed rapidly, which aims to provide information to assist surface transportation travelers in moving from a starting location (origin) to their desired destination \citep{schofer1993behavioral}. This system collects data from probe automobiles, prompting the emerging research in travel time estimation and forecasting. \citet{sen1999variances} pointed out the covariance of link travel times which are close together, may not be zero, and they proposed estimating the correlation matrix of link travel time as an open problem for future research. A straightforward solution is to infer the correlation matrix using asymptotic theory (i.e., correlation formula), which is the traditional estimation of correlation. \citet{bernard2006correlation} used the straight method to estimate link travel speed correlations, which are similar to link travel time correlations. Nevertheless, \citet{gajewski2005estimating} figured out that the classical estimation method lacks interpretability and is complicated due to the nonparametric nature of the estimator and the covariance between links. Then they adopted a Bayesian approach to estimate link travel time correlation, which had benefits in terms of interpretation and ease of use. The authors only experimented on three links because they could collect many full observations of the three links. However, for one road network with many links, the number of full observations dramatically decreases, and this simple Bayesian model can not estimate the correlation matrix accurately.

Link travel time correlation is essential to the stochastic routing problem, as it helps to consider the reliability of travel time. Many early studies \citep{cheung1998iterative,miller2001adaptive,seshadri2010algorithm} ignored link travel time correlation because of the low computational efficiency for large networks. Some studies use origin-destination trip data to estimate link travel times and they usually make the independent link travel time assumptions \citep[e.g., ][]{hunter2009path,zhan2013urban,zhan2016bayesian,sun2015integrated}. Many studies focus on estimating the distribution of route travel time, but their methods do not model link travel time correlations \citep[e.g.,][]{rakha2006estimating,jenelius2017urban,woodard2017predicting, huang2021bus}. 
Next, correlations between link travel times are explored in \citet{waller2002online} and \citet{fan2006optimal}. Both of them consider local spatial correlations between adjacent links. Many studies followed this assumption that only the adjacent link travel times are correlated. Although the local spatial correlation assumption is strong, it seems reasonable because we would expect the impact of a link on another decreases with the increase of distance, and it becomes a popular choice in the literature \citep{chen2012reliable,srinivasan2014finding} partially due to the model simplicity and the lack of empirical evidence. \citet{rachtan2013spatiotemporal} adopted three regression models to estimate the correlation by considering various combinations of variables, including spatial distance, temporal distance, traffic state, and the number of lanes, and they found that the primary factor in correlation is spatial distance. \citet{zeng2015application} also incorporated the spatial correlation of link travel time in finding the reliable path of stochastic networks. \citet{geroliminis2006real} estimated the variance of route travel time; they used full observations of six links to directly compute the covariance and correlation. \citet{ramezani2012estimation} applied Markov chains to estimate the route travel time distribution considering the correlation between successive links. They first established a two-dimensional (2D) diagram with data points representing travel times of two consecutive links; then used a heuristic grid clustering method to cluster the 2D diagram to different spaces (states). With a Markov chain procedure, they can integrate the correlation between states of 2D diagrams for successive links. \citet{jenelius2013travel} incorporated the spatial link travel time correlation into travel time estimation for urban road networks; they used a spatial moving average (SMA) structure to model link correlation by assuming that the stochastic component of each link is expressed as an independent term with zero mean and variance, plus a linear combination of the independent terms of the other links.  \citet{westgate2016large} proposed the method for estimating large-network travel time distribution; they model travel time at the trip level instead of the link level, but they consider the dependency between links by incorporating explanatory factors like the road class, speed limit, one-way road. \citet{rodriguez2017urban} used the multi-output
Gaussian Processes to estimate network-wide travel time distribution. They considered the squared exponential (SE) kernel to capture correlations between any pair of time points and they applied Graph/Laplacian kernel to model correlations between two link travel times. Copula functions can describe the dependence between random variables. \citet{chen2017copula} and \citet{chen2019modeling}  developed a copula-based approach to model the link travel time correlation. The approach applied a two-dimensional Gaussian copula function to fit the link travel time distribution of two adjacent links. \citet{qin2020estimation} proposed a pair-copula mixture model for estimating urban arterial travel time distribution, and it can reduce the computational complexity. The copula-based models are limited to expensive computations, especially for many links; thus, they are applied on a few links and only consider the correlation between adjacent links.

In summary, most studies about link travel time correlation are for car travel time estimation. There are only a few studies for bus travel time estimation with link travel time correlation being considered. \citet{uno2009using} estimated the variance of individual path travel time by aggregating the covariance between link travel time; thus, the travel time distribution of one path can be estimated by summing up directly observed multiple links mean travel time with their covariance based on bus probe data. \citet{dai2019bus} attempted to estimate the bus path travel time distribution using GPS probe and smart card data. They considered that path travel time distribution could be estimated by statistically summarizing link travel time distributions and dwell-time distributions at bus stops. Therefore, both studies need to obtain the correlation or covariance matrices, but they only compute the correlation matrix directly using many full observations without considering the temporal difference. In fact, the correlation matrix should be time-varying due to the temporal variations in bus operation and road traffic. To the best of our knowledge, little attention was paid to quantifying time-varying link travel time correlation in the literature. 

\section{Problem Description}
\label{sec:problem}
For a bus route with $n$ links (i.e., $n+1$ bus stops), we define a bus link as the directional road segment between two adjacent bus stops. For conciseness, we often omit the word ``bus'' and simply use ``route'' and ``link'' in the following. Denote a random variable $x_i$ to be the travel time of the $i$-th link in the route, whose value is observed by the time difference between the arrival of a bus at the two adjacent bus stops. Next, the travel time of all the links of the route can be represented by multivariate random variable $\boldsymbol{x}=\left[x_1,x_2,\cdots,x_n\right]^\top$.

\begin{figure}[!ht]
\centering
\subfigure{
    \centering
    \includegraphics[width = 0.7\textwidth]{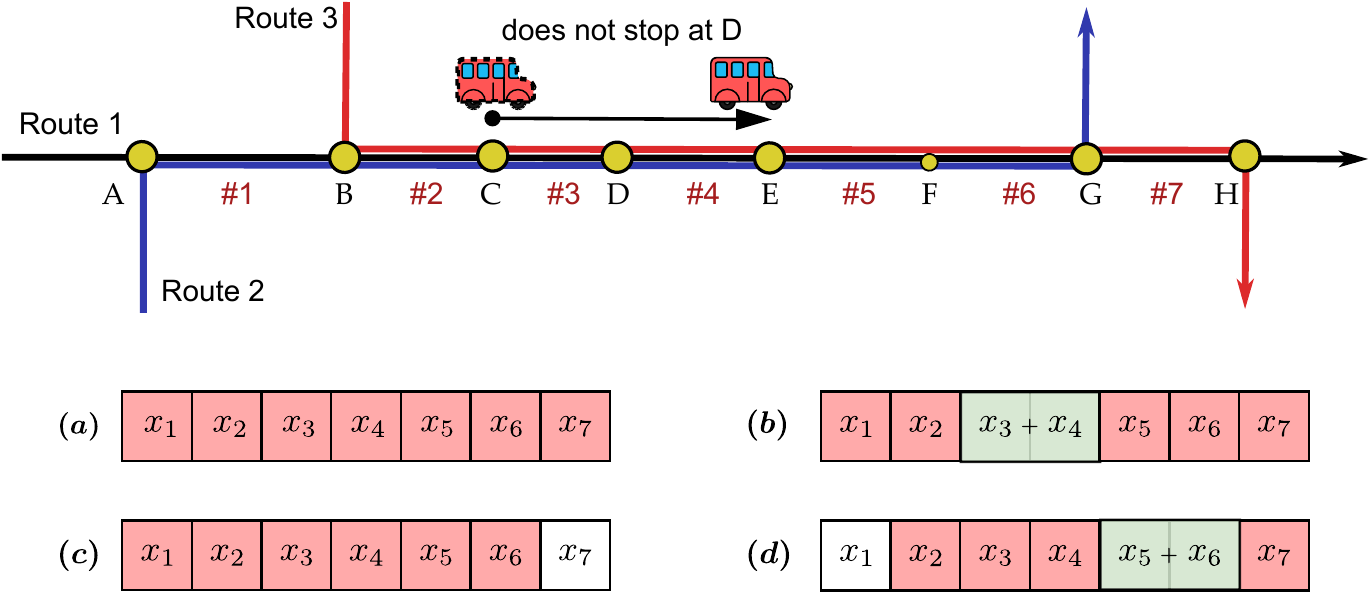}
}
\caption{Graphical illustration of full observations and incomplete observations. (a) a complete observation of Route 1; (b) an observation of Route 1 with ragged values; (c) an observation of Route 2, where $x_7$ is inaccessible (missing); (d) an observation of Route 3, where $x_1$ is inaccessible (missing), $x_5$ and $x_6$ are ragged.}
\label{links_example}
\end{figure}

This study uses the data from in-out-stop record systems. When a bus arrives at or leaves a bus stop, the system registers vehicle ID, route ID, action type (arrival/departure), together with a timestamp. We can thus calculate the link travel time from the bus in-out-stop data. Each bus has a vector for its link travel times on the route. Besides bus in-out-stop record data, smart card data---a more common type of data---can be equivalently used to obtain the link travel time. However, in practice, the link travel time vector $\boldsymbol{x}$ from a bus run on the target route or on a related route is often incomplete due to several reasons. Fig.~\ref{links_example} uses a simple bus network to illustrate the issue of missing and ragged values in the data. Route 1 is a bus route of interest with seven links. We assume all buses go through the same bus link (during the same period of the day) have the same link travel time distribution, regardless of which route they belong; this allows us to use the data from Route 2 and Route 3 to improve the estimation of link travel time correlations in Route 1. An ideal observation for Route 1 is Fig.~\ref{links_example} (a), where the travel times of all links are obtained. However, we cannot obtain the travel time of both link $\#3$ and link $\#4$ if a bus of Route 1 skipped stop D; we can instead observe the total travel time ($x_3$+$x_4$) from stop C to stop E, and this is the case of ragged value shown in Fig.~\ref{links_example} (b). Moreover, Route 2 goes through six out of seven links, resulting in observations with missing values as shown in Fig.~\ref{links_example} (c). The last example is Route 3, which goes through six links (B to H) and has no stop at F, bringing incomplete observations with both missing and ragged values as shown in Fig.~\ref{links_example} (d). Essentially, data from other relevant bus routes can be considered as a general type of incomplete observations with both missing or ragged values.

We denote $\boldsymbol{x}_i=[x_{i,1}, x_{i,2}, \cdots, x_{i,n}]^{\top}$, which is a sample of the random variable $\boldsymbol{x}$, to be the link travel time of the $i$-th bus during the study period. Since not all entries of $\boldsymbol{x}_i$ are always available from data, we denote $\boldsymbol{r}_i={\mathbf{G}}_i\boldsymbol{x}_i \in \mathbb{R}^{n_i}$ to be a vector of observed information attached with $\boldsymbol{x}_i$, where ${\mathbf{G}}_i \in \left\{0,1\right\}^{n_i \times n}$ is a binary matrix encodes the missing and ragged positions of the $i$-th bus. We call a $\boldsymbol{r}_i$ a \textit{recording vector} and ${\mathbf{G}}_i$ an \textit{alignment matrix}. For example, we have $\boldsymbol{x}_i=\boldsymbol{r}_i$ and ${\mathbf{G}}_i$ being an identity matrix for a bus with a complete observation; for a bus $i$ in the case of Fig.~\ref{links_example} (d), the link travel time and its missing and ragged values can be represented as
\begin{equation}
    \begin{matrix}
          \begin{bmatrix}
          0&		1&		0&		0&		0&		0&		0\\
          0&		0&		1&		0&		0&		0&		0\\
          0&		0&		0&		1&		0&		0&		0\\
          0&		0&		0&		0&		1&		1&		0\\
          0&		0&		0&		0&		0&		0&		1\\
    \end{bmatrix}&		\begin{bmatrix}
          x_{i,1}\\
          x_{i,2}\\
          x_{i,3}\\
          x_{i,4}\\
          x_{i,5}\\
          x_{i,6}\\
          x_{i,7}\\
    \end{bmatrix}&		=&		\begin{bmatrix}
          x_{i,2}\\
          x_{i,3}\\
          x_{i,4}\\
          x_{i,5}+x_{i,6}\\
          x_{i,7}\\
    \end{bmatrix}&		=&		\begin{bmatrix}
          r_{i,1}\\
          r_{i,2}\\
          r_{i,3}\\
          r_{i,4}\\
          r_{i,5}\\
    \end{bmatrix} \\
          {\mathbf{G}}_i&		\boldsymbol{x}_i&		&		&		&		\boldsymbol{r}_i\\
    \end{matrix}.
\end{equation}

In the real world, the number of incomplete samples can be even greater than that of complete samples. However, such ``incomplete'' observations should not be discarded as they also encode valuable information in estimating link travel time correlations. Assuming there are $m$ buses that went through a target bus route during a study period, the goal of this research is to incorporate all recording vectors $\left\{\boldsymbol{r}_i \right\}_{i=1}^{m}$ and alignment matrices $\left\{{\mathbf{G}}_i \right\}_{i=1}^{m}$ to quantify the link travel time correlation matrix $\boldsymbol{\mathrm{C}} = \operatorname{Corr}\left[\boldsymbol{x}\right]$.

\section{Methodology}
\label{methodology}
\subsection{Multivariate Gaussian Model}

Gaussian distribution offers numerous analytical and computational advantages and has been extensively used in modeling travel time distribution \citep[e.g.,][]{smeed1971variability,li2004examining,seshadri2010algorithm}. In this paper, we assume the joint probability of link travel times in a bus route follows a multivariate Gaussian distribution with the probability density function
\begin{equation}\label{Multi-variate Density}
p\left(\boldsymbol{x}\mid \boldsymbol{\mu},\boldsymbol{\Sigma}\right) = \frac{1}{\left(2\pi\right)^{n/2}\left|\boldsymbol{\Sigma}\right|^{1/2}}\exp\left[ -\frac{1}{2}\left(\boldsymbol{x}-\boldsymbol{\mu}\right)^{\top}\boldsymbol{\Sigma}^{-1}\left(\boldsymbol{x}-\boldsymbol{\mu}\right)\right], \end{equation}
where $\boldsymbol{\mu} = \mathbb{E}\left[\boldsymbol{x}\right] \in \mathbb{R}^n $ is a mean vector, and $\boldsymbol{\Sigma}=\operatorname{Cov}\left[\boldsymbol{x}\right]$ is an $n\times n$ covariance matrix, and $\left|\boldsymbol{\Sigma}\right|$ is the determinant of $\boldsymbol{\Sigma}$. The covariance matrix $\boldsymbol{\Sigma}$ and its relationship with the correlation matrix $\boldsymbol{\mathrm{C}}$ are shown in Eq.~\eqref{Covariance} and Eq.~\eqref{Covariance2Correlation}, respectively.
\begin{equation}\label{Covariance}
\boldsymbol{\Sigma} = \mathbb{E}\Big[\left(\boldsymbol{x}-\mathbb{E}\left[\boldsymbol{x}\right]\right)\left(\boldsymbol{x}-\mathbb{E}\left[\boldsymbol{x}\right]\right)^\top\Big]
= \begin{bmatrix}
\mathbb{V}\left[x_1\right] & \operatorname{Cov}\left[x_1,x_2\right] & \cdots & \operatorname{Cov}\left[x_1,x_n\right] \\
\operatorname{Cov}\left[x_2,x_1\right] & \mathbb{V}\left[x_2\right] & \cdots & \operatorname{Cov}\left[x_2,x_n\right] \\
\vdots & \vdots & \ddots & \vdots \\
\operatorname{Cov}\left[x_n,x_1\right] & \operatorname{Cov}\left[x_n,x_2\right] & \cdots & \mathbb{V}\left[x_n\right]
\end{bmatrix},\end{equation}
\begin{equation}\label{Covariance2Correlation}
    \boldsymbol{\mathrm{C}} = \left(\operatorname{diag}\left(\boldsymbol{\Sigma}\right)\right)^{-\frac{1}{2}} \boldsymbol{\Sigma} \left(\operatorname{diag}\left(\boldsymbol{\Sigma}\right)\right)^{-\frac{1}{2}}
    = \begin{bmatrix}
    1 & \frac{\operatorname{Cov}\left[x_1,x_2\right]}{\sqrt{\mathbb{V}\left[x_1\right]\mathbb{V}\left[x_2\right]}} & \cdots & \frac{\operatorname{Cov}\left[x_1,x_n\right]}{\sqrt{\mathbb{V}\left[x_1\right]\mathbb{V}\left[x_n\right]}} \\
    \frac{\operatorname{Cov}\left[x_2,x_1\right]}{\sqrt{\mathbb{V}\left[x_2\right]\mathbb{V}\left[x_1\right]}} & 1 & \cdots & \frac{\operatorname{Cov}\left[x_2,x_n\right]}{\sqrt{\mathbb{V}\left[x_2\right]\mathbb{V}\left[x_n\right]}} \\
    \vdots & \vdots & \ddots & \vdots \\
    \frac{\operatorname{Cov}\left[x_n,x_1\right]}{\sqrt{\mathbb{V}\left[x_n\right]\mathbb{V}\left[x_1\right]}} & \frac{\operatorname{Cov}\left[x_n,x_1\right]}{\sqrt{\mathbb{V}\left[x_n\right]\mathbb{V}\left[x_1\right]}} & \cdots & 1
    \end{bmatrix},
\end{equation}
where $\operatorname{Cov}\left[x_i,x_j\right] = \mathbb{E} \left[\left(x_i- \mathbb{E}\left[x_i\right]\right)\left(x_j-\mathbb{E}\left[x_j\right]\right)\right]$, $\mathbb{V}\left[x_i\right]=\operatorname{Cov}\left[x_i,x_i\right]$, and $\operatorname{diag}\left(\boldsymbol{\Sigma}\right)$ is the diagonal elements of $\boldsymbol{\Sigma}$. However, this naive approach that directly calculates sample variance and covariance using Eq.~\eqref{Covariance} and Eq.~\eqref{Covariance2Correlation} fails with the presence of incomplete observations. 

\begin{figure}[!ht]
    \centering
    \subfigure{
        \centering
        \includegraphics[width = 0.25\textwidth]{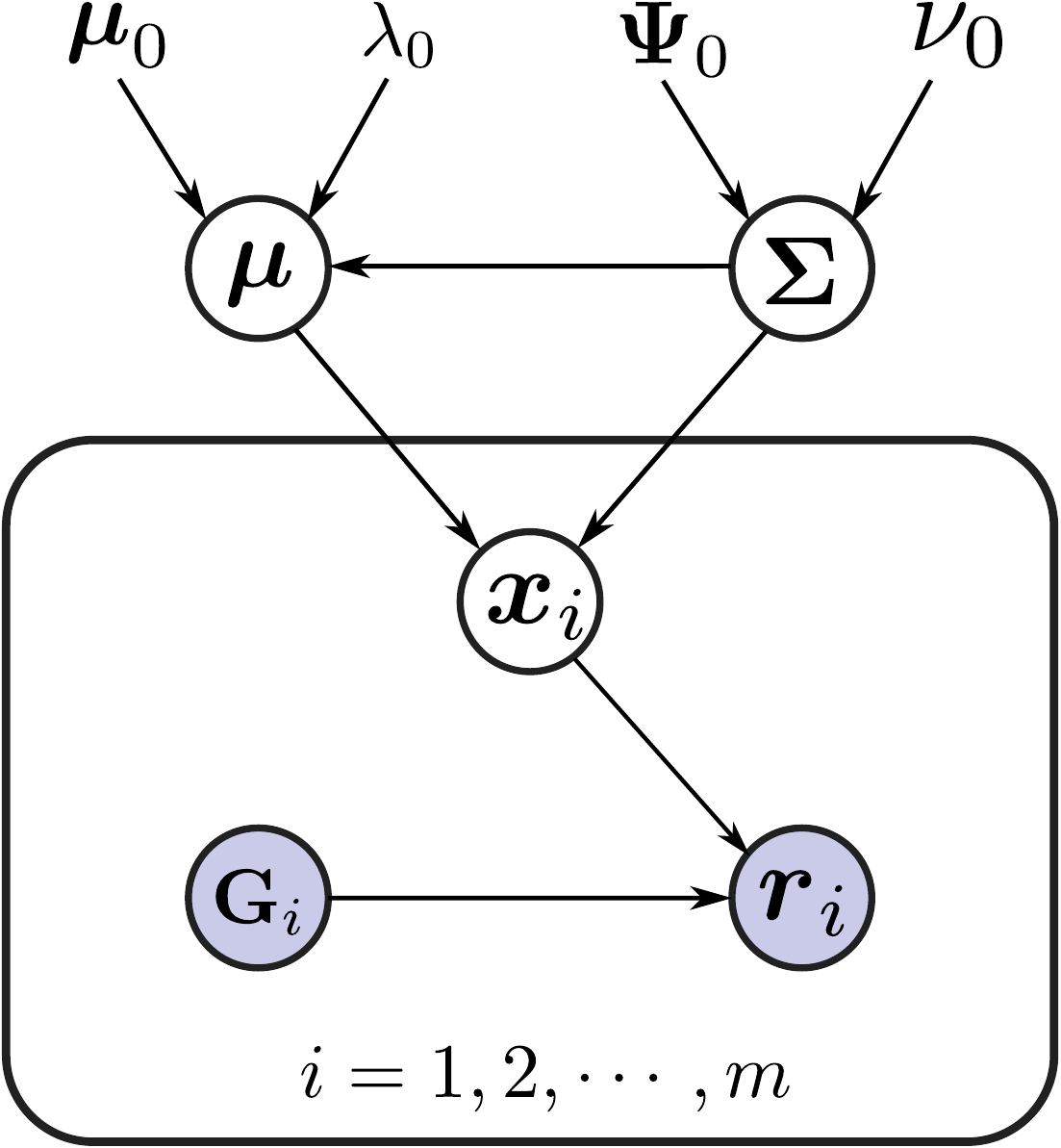}
    }
    \caption{The graphical illustration of Bayesian Gaussian model.}
    \label{fig:graphical model}
\end{figure}

Fig.~\ref{fig:graphical model} shows the overall graphical representation of our Bayesian model that can handle missing and ragged values. For a collection of $m$ ragged observations $\mathcal{R} = \left\{\boldsymbol{r}_i \right\}_{i=1}^{m}$ over a pre-defined time window, we have $\boldsymbol{r}_i = \mathbf{G}_i \boldsymbol{x}_i$ for $i=1,\ldots,m$. Next, we assume $\boldsymbol{x}_i$ is a ``latent'' realization/sample from a multivariate Gaussian distribution $\mathcal{N}\left(\boldsymbol{\mu},\boldsymbol{\Sigma}\right)$ following Eq.~\eqref{Multi-variate Density}. In a Bayesian setting, we further use a conjugate Gaussian-inverse-Wishart distribution on $\boldsymbol{\mu}$ and $\boldsymbol{\Sigma}$ \citep{gelman1995bayesian} for efficient inference. The overall data generation process is summarized as:
\begin{align}
    \boldsymbol{\Sigma}& \sim \mathcal{W}^{-1}\left(\boldsymbol{\Psi}_0,\nu_0\right), \label{eq:cov prior} \\
    \boldsymbol{\mu} &\sim \mathcal{N}\left(\boldsymbol{\mu}_{0}, \frac{1}{\lambda_{0}}\boldsymbol{\Sigma} \right), \label{eq:mean prior} \\
    \boldsymbol{x}_i &\sim \mathcal{N}\left(\boldsymbol{\mu},\boldsymbol{\Sigma}\right), \text{ for } i=1,\ldots,m, \label{xpdf}\\
    \boldsymbol{r}_i &= \mathbf{G}_i \boldsymbol{x}_i, \text{ for } i=1,\ldots,m, \label{eq:linear}
\end{align}
where $\mathcal{W}^{-1}\left(\boldsymbol{\Psi}_0,\nu_0\right)$ is the inverse-Wishart distribution with $\nu_0$ degrees of freedom ($\nu_0\ge n$), and an $n\times n$ scale matrix $\boldsymbol{\Psi}_0$; $\boldsymbol{\mu}_0$ is the prior mean. The probability density function of the inverse-Wishart distribution is
\begin{equation}
\mathcal{W}^{-1}\left(\boldsymbol{\Sigma}\mid \boldsymbol{\Psi}_0,\nu_0\right)=C\left|\boldsymbol{\Sigma}\right|^{-\left(\nu_0 + n +1\right)/2}\exp\left[-\frac{1}{2} \operatorname{Tr} \left(\boldsymbol{\Psi}_0\boldsymbol{\Sigma}^{-1}\right) \right],
\end{equation}
where $C$ is a normalizing constant and $\operatorname{Tr(\cdot)}$ is the trace of a matrix.

Based on the graphical structure presented in Fig.~\ref{fig:graphical model}, we can derive an efficient MCMC scheme using Gibbs sampling. For simplicity, we denote by $\mathcal{G}=\left\{{\mathbf{G}}_i \right\}_{i=1}^{m}$ the set of alignment matrices corresponding to observations $\mathcal{R} = \left\{\boldsymbol{r}_i \right\}_{i=1}^{m}$,  by $\mathcal{X} = \left\{\boldsymbol{x}_i \right\}_{i=1}^{m}$ the set of ``full'' link travel time for the $m$ bus runs, and by $\Theta=\left\{\boldsymbol{\mu}_0,\lambda_0, \boldsymbol{\Psi}_0, \nu_0\right\}$ the set of hyperparameters for the Gaussian-inverse-Wishart prior distribution in Eqs.~\eqref{eq:cov prior} and \eqref{eq:mean prior}. We start the Gibbs sampling with randomly initialized values for all variables and then iteratively sample each variable from its conditional distribution on other variable:
\begin{itemize}
    \item Sample $(\boldsymbol{\mu},\boldsymbol{\Sigma})$ from $p\left(\boldsymbol{\mu}, \boldsymbol{\Sigma} \given \mathcal{X}, \Theta \right)$. Because of the conjugate prior distribution, the conditional distribution of the mean vector and the covariance matrix
$p\left(\boldsymbol{\mu}, \boldsymbol{\Sigma} \given \mathcal{X}, \Theta \right)$ is a Gaussian-inverse-Wishard distribution:
\begin{equation}\label{mu123}
p\left(\boldsymbol{\mu},\boldsymbol{\Sigma} \mid \mathcal{X}, \Theta \right) \sim \mathcal{N}\left(\boldsymbol{\mu}\mid\boldsymbol{\mu}_0^*,\frac{1}{\lambda_0^*}\boldsymbol{\Sigma}\right)\mathcal{W}^{-1}\left(\boldsymbol{\Sigma}\mid\boldsymbol{\Psi}_0^*,\nu_0^*\right),
\end{equation}
where
\begin{equation}\label{l123}
\begin{aligned}
\boldsymbol{\mu}_0^* = \frac{\lambda_0\boldsymbol{\mu}_0+m\bar{\boldsymbol{x}}}{\lambda_0+m}, \qquad \lambda_0^* = \lambda_0+m, \qquad \nu_0^*=\nu_0+m, \qquad \bar{\boldsymbol{x}}=\frac{1}{m}\sum_{i=1}^{m}\boldsymbol{x}_i,\\
\boldsymbol{\Psi}_0^* = \boldsymbol{\Psi}_0+\boldsymbol{\mathrm{S}}+\frac{\lambda_0m}{\lambda_0+m}\left(\bar{\boldsymbol{x}}-\boldsymbol{\mu_0}\right)\left(\bar{\boldsymbol{x}}-\boldsymbol{\mu_0}\right)^{\top}, \qquad \boldsymbol{\mathrm{S}}=\sum_{i=1}^{m}\left(\boldsymbol{x}_i-\bar{\boldsymbol{x}}\right)\left(\boldsymbol{x}_i-\bar{\boldsymbol{x}}\right)^{\top}.
\end{aligned}
\end{equation}
    \item Sample $\mathcal{X}$ from $p\left(\mathcal{X} \given \boldsymbol{\mu}, \boldsymbol{\Sigma}, \mathcal{R}, \mathcal{G}  \right)$. For this step, we no longer have a simple analytical formulation to sample $\mathcal{X}$ due to the linear constraints in Eq.~\eqref{eq:linear}. We next introduce an effective solution to sample $\boldsymbol{x}_i$ from its conditional distribution in Section~\ref{sec:imputation}.
\end{itemize}

\subsection{Sampling Link Travel Time}\label{sec:imputation}
Assuming the link travel times of different buses are independent. Next, we can factorize the conditional distribution of link travel time as
\begin{equation}\label{eq:factor}
    p\left(\mathcal{X} \given \boldsymbol{\mu}, \boldsymbol{\Sigma}, \mathcal{R}, \mathcal{G}  \right) = \prod_{i=1}^{m}{p\left(\boldsymbol{x}_i \given \boldsymbol{\mu}, \boldsymbol{\Sigma}, \boldsymbol{r}_i, {\mathbf{G}}_i \right)}.
\end{equation}
Therefore, we can draw sample of the bus-specific link travel time vector $\boldsymbol{x}_i$ independently. The conditional distribution of $\boldsymbol{x}_i$ in Eq.~\eqref{eq:factor} can be viewed as a multivariate Gaussian distribution truncated on the intersection with a hyperplane, i.e.,
\begin{equation}
    \boldsymbol{x}_i \sim \mathcal{N}_{\mathcal{S}_i}\left(\boldsymbol{\mu},\boldsymbol{\Sigma}\right), \quad \mathcal{S}_i=\left\{\boldsymbol{x}_i: {\mathbf{G}}_i \boldsymbol{x}_i = \boldsymbol{r}_i \right\}.
\end{equation}
The probability density function of the hyperplane-truncated multivariate Gaussian is
\begin{equation}\label{eq:truncated_pdf}
    p(\boldsymbol{x}_i \mid \boldsymbol{\mu}, \boldsymbol{\Sigma}, \boldsymbol{r}_i, \mathbf{G}_i)=\frac{1}{Z_i} \exp \left[-\frac{1}{2}(\boldsymbol{x}_i-\boldsymbol{\mu})^{T} \boldsymbol{\Sigma}^{-1}(\boldsymbol{x}_i-\boldsymbol{\mu})\right] \delta(\mathbf{G}_i \boldsymbol{x}_i=\boldsymbol{r}_i),
\end{equation}
where $Z_i$ is a normalizing constant; $\delta(*)$ is a function whose value is 1 only if the condition $*$ holds, and 0 otherwise.

There are several available algorithms for efficient sampling over hyperplane-truncated Gaussian distributions \citep[e.g.,][]{cong2017fast, botev2017normal}. We apply a fast sampling algorithm developed by \citet{cong2017fast} for this problem. For a given mean vector and covariance matrix, the algorithm for sampling the link travel time vector of the $i$-th bus is described in Algorithm~\ref{imputation}.

\begin{algorithm}
\caption{Sampling from a hyperplane-truncated multivariate Gaussian distribution \citet{cong2017fast}.}
\label{imputation}
\begin{algorithmic}[1]
\STATE Sample $\boldsymbol{y} \sim \mathcal{N}\left(\boldsymbol{\mu},\boldsymbol{\Sigma}\right)$;
\STATE Return $\boldsymbol{x}_i = \boldsymbol{y} + {\boldsymbol{\Sigma}}{\mathbf{G}}_i^{\top}\left({\mathbf{G}}_i{\boldsymbol{\Sigma}}{\mathbf{G}}_i^{\top}\right)^{-1}\left(\boldsymbol{r}_i-{\mathbf{G}}_i\boldsymbol{y}\right)$, which can be more efficiently and accurately calculated by\\
\begin{itemize}
    \item Solve $\boldsymbol{\alpha}$ such that $\left({\mathbf{G}}_i{\boldsymbol{\Sigma}}{\mathbf{G}}_i^{\top}\right)\boldsymbol{\alpha}=\boldsymbol{r}-{\mathbf{G}}_i\boldsymbol{y}$;
    \item Return $\boldsymbol{x}_i=\boldsymbol{y}+{\boldsymbol{\Sigma}}{\mathbf{G}}_i^{\top}\boldsymbol{\alpha}$.
\end{itemize}
\end{algorithmic}
\end{algorithm}

\subsection{Overall Gibbs Sampling Algorithm}

Having obtained the two conditional distributions in Eqs.~\eqref{mu123} and \eqref{eq:factor}, we summarize the overall Gibbs sampling procedure for estimating the correlation matrix in Algorithm~\ref{alg:gibbs}. We drop the first $k_1$ iterations as burn-in and estimate the correlation matrix $\hat{\boldsymbol{\mathrm{C}}}$ as the average of samples from the last $k_2$ iterations. Besides, we store samples of correlation matrices $\{\boldsymbol{\mathrm{C}}^{(i)}\}_{i=1}^{k_2}$ and covariance matrices $\{\boldsymbol{\Sigma}^{(i)}\}_{i=1}^{k_2}$, which are critical ingredients for deriving the Bayesian credible interval for each entry in the correlation matrix and performing probabilistic forecasting of bus travel time. For hyperparameters $\Theta$, we set $\boldsymbol{\mu}_0 = \boldsymbol{0}_n$, $\lambda_0=10$,  $\boldsymbol{\Phi}_0 = \boldsymbol{I}_n$, $\nu_0 = n+2$, where $n$ is number of links.

\begin{algorithm}
\renewcommand{\algorithmicrequire}{\textbf{Input:}}
\renewcommand{\algorithmicensure}{\textbf{Output:}}
\caption{Gibbs sampling for correlation estimation.}\label{alg:gibbs}
\begin{algorithmic}[1]
\REQUIRE Recording vectors $\mathcal{R}$, alignment matrices $\mathcal{G}$, initial values for hyperparameters $\Theta$, the number of iterations $k_1$, $k_2$.
\ENSURE Estimated correlation matrix $\hat{\boldsymbol{\mathrm{C}}}$, a set of samples for correlation matrices $\{\boldsymbol{\mathrm{C}}^{(i)}\}_{i=1}^{k_2}$, a set of samples for covariance matrices $\{\boldsymbol{\Sigma}^{(i)}\}_{i=1}^{k_2}$.
\FOR{$\mathrm{iter}=1$ to $k_1+k_2$}
\STATE Draw $\boldsymbol{\Sigma}$ and $\boldsymbol{\mu}$ according to Eq.~\eqref{eq:cov prior} and Eq.~\eqref{eq:mean prior}.
\IF{$\mathrm{iter}>k_1$}
    \STATE Calculate $\boldsymbol{\mathrm{C}}$ by Eq.~\eqref{Covariance2Correlation}, collect $\boldsymbol{\mathrm{C}}$ and $\boldsymbol{\Sigma}$ to the output sets.
\ENDIF
\FOR{$i$ = $1$ to $m$}
    \STATE Draw $\boldsymbol{x}_i$ by Algorithm~\ref{imputation}.
\ENDFOR
\STATE Update the parameters $\Theta=\{\boldsymbol{\mu}_0,\lambda_0, \boldsymbol{\Psi}_0, \nu_0\}$ by Eq.~\eqref{l123}.
\ENDFOR
\STATE Compute $\hat{\boldsymbol{\mathrm{C}}}$ as the average of samples in $\{\boldsymbol{\mathrm{C}}^{(i)}\}_{i=1}^{k_2}$.
\RETURN{$\hat{\boldsymbol{\mathrm{C}}}$, $\{\boldsymbol{\mathrm{C}}^{(i)}\}_{i=1}^{k_2}$, $\{\boldsymbol{\Sigma}^{(i)}\}_{i=1}^{k_2}$.}
\end{algorithmic}
\end{algorithm}

\section{Case Study}
\label{casestudy}
This section provides three numerical case studies using both synthetic data and real-world data. First, we use a synthetic experiment to test the accuracy of the proposed correlation estimation method and the improvement brought by incorporating missing/ragged values. Next, we apply our model on bus in-out-stop record data to quantify link travel time correlation in a transit corridor in Guangzhou. Finally, we demonstrate an application of using our model in probabilistic forecasting of bus link/trip travel time.

\subsection{Case 1: Synthetic Data}
\begin{figure}[!ht]
\centering
\subfigure{
    \centering
    \includegraphics[width = 0.7\textwidth]{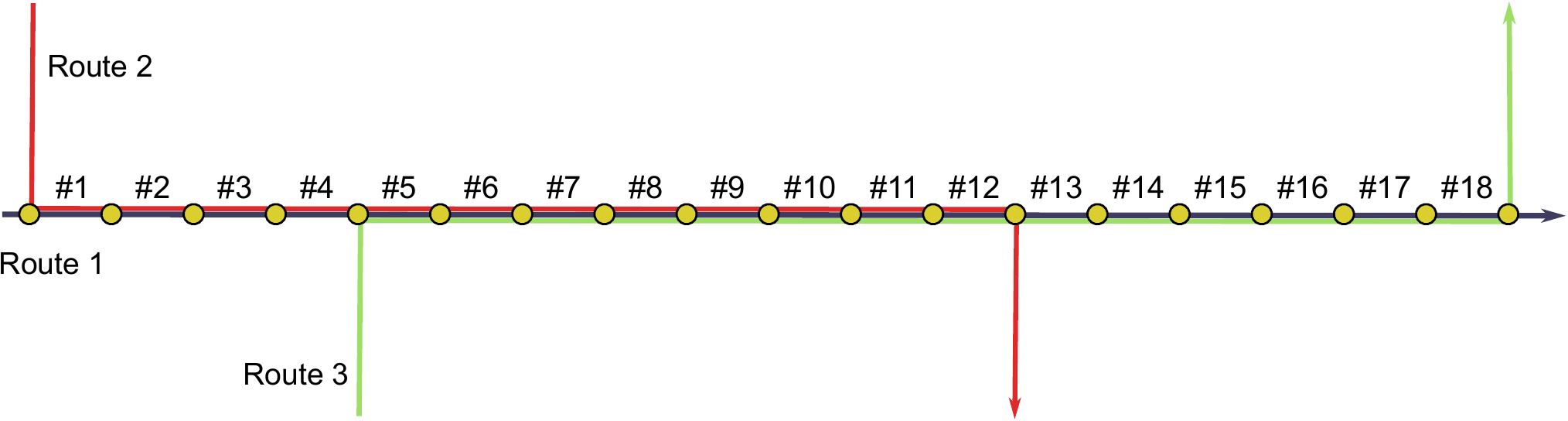}
}
\caption{The bus network of synthetic data.}
\label{Bus_synthesis}
\end{figure}

\begin{figure*}[!ht]
\centering
\subfigure{
    \centering
    \includegraphics[width=0.42
    \textwidth]{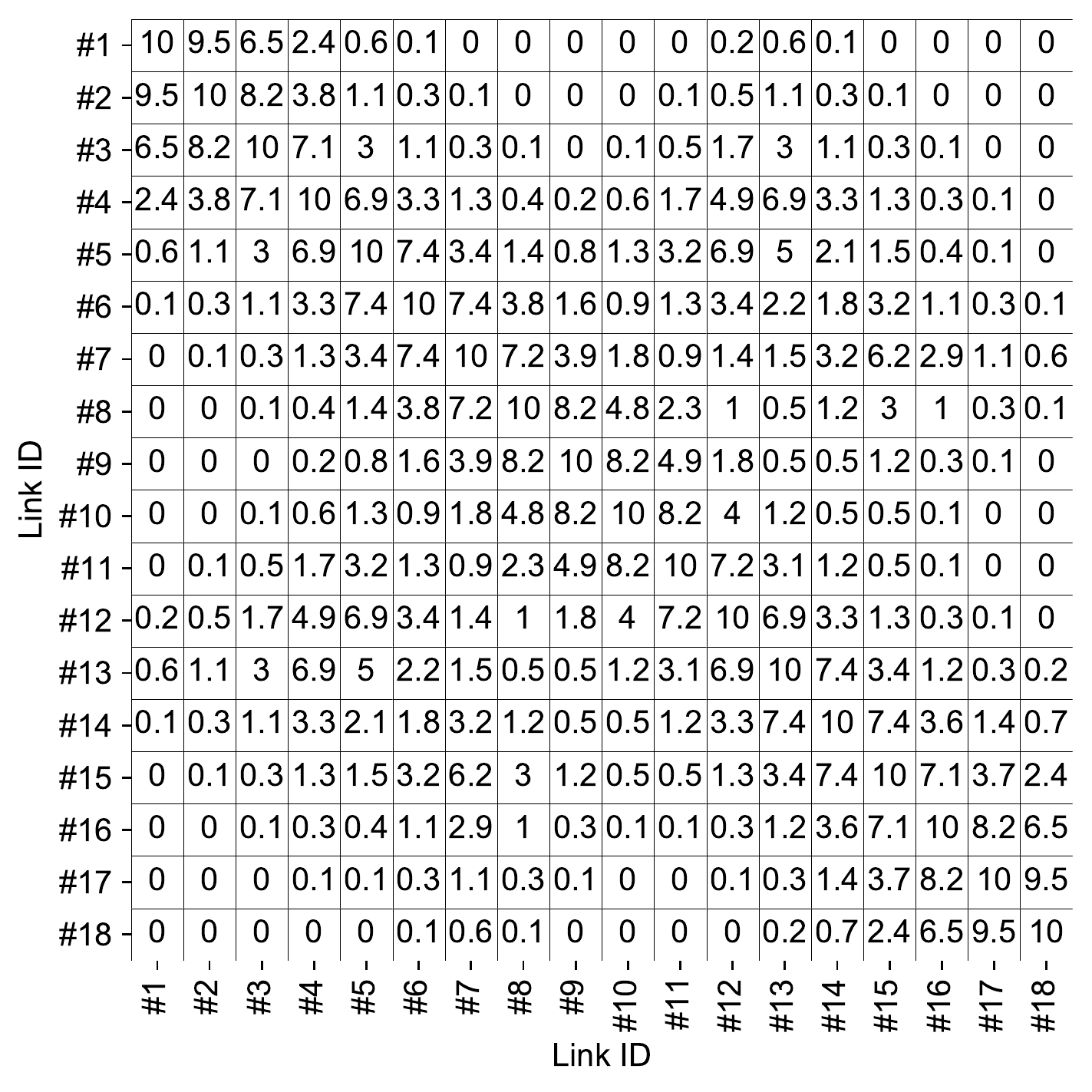}
}
\caption{Generated covariance matrix $\boldsymbol{\mathrm{\Sigma}}$.}
\label{cov_parameters}
\end{figure*}

We design a simple bus network with 18 links as shown in Fig.~\ref{Bus_synthesis} to test the performance of the proposed correlation estimation method. In this bus network, The target Route 1 has 18 links; Route 2 shares 12 links with Route 1 (from link $\#1$ to link $\#12$); Route 3 shares 14 links with Route 1 (from link $\#5$ to link $\#18$). We use a multivariate Gaussian model with pre-defined mean and covariance to synthesize a link-travel-time data set. We set the mean vector to be $\boldsymbol{\mu} = \left[14,15,18,13,17,15,10,24,15,11,12,15,9,13,17,15,19,21\right]$. We use the Graph kernel to set the covariance matrix by the following steps: (1) The route's structure has local correlations, and we assume the following link pairs are virtually adjacent: (link $\#4$, link $\#13$), (link $\#5$, link $\#12$), (link $\#7$, link $\#15$) to simulate long-range correlations. (2) From the structure, we can obtain the degree matrix $\boldsymbol{\mathrm{D}}$ and the adjacency matrix $\boldsymbol{\mathrm{A}}$, then we can get the symmetrically normalized Laplacian matrix $\boldsymbol{\mathrm{L}}=\boldsymbol{\mathrm{D}}^{-\frac{1}{2}}\left(\boldsymbol{\mathrm{D}}-\boldsymbol{\mathrm{A}}\right)\boldsymbol{\mathrm{D}}^{-\frac{1}{2}}$. (3) Next, we can compute the kernel matrix $\boldsymbol{\mathrm{K}}$ using the kernel function $\boldsymbol{\mathrm{K}} = \exp{\left(\beta\boldsymbol{\mathrm{L}}\right)}$. (4) Finally, we can get the correlation matrix $\boldsymbol{\mathrm{Corr}}=\left(\operatorname{diag}\left(\boldsymbol{\mathrm{K}}\right)\right)^{-\frac{1}{2}} \boldsymbol{\mathrm{K}} \left(\operatorname{diag}\left(\boldsymbol{\mathrm{K}}\right)\right)^{-\frac{1}{2}}$, and the covariance matrix $\boldsymbol{\mathrm{\Sigma}}=\sigma\boldsymbol{\mathrm{Corr}}$. Here, we set $\beta=3,\sigma=10$ to generate the covariance shown in Fig.~\ref{cov_parameters}.

We draw 240 samples of link travel time vectors from the multivariate Gaussian distribution with the above parameters $\{\boldsymbol{\mu}$, $\boldsymbol{\Sigma}\}$. Next, we assign 160 samples to Route 1, 80 samples to Route 2, and 80 samples to Route 3. The travel time values for links that do not belong to Route 2 or Route 3 are dropped and regarded as missing values. Among the 160 samples for Route 1, we keep 80 samples as full observations, and set ragged values to the rest 80 samples by adding the travel time of link $\#5$ and link $\#6$.

We applied Algorithm~\ref{alg:gibbs} to estimate the correlation matrix from the above synthetic data. For convenience, we refer to  incomplete observations with missing values (not including ragged values) as ``missing observations'' and incomplete observations with ragged values as ``ragged observation''. Three experiments are designed to compare the estimation accuracy with different types of observations: (1) only uses full observations; (2) uses full and missing observations; (3) uses all observations. The numbers of MCMC iterations are $k_1=10000$ and $k_2=5000$. 

With samples drawn in the last $k_2$ iterations, the posterior mean of mean vectors estimated with different types of observations are shown in Table~\ref{mean_vector}, and the posterior mean of covariance matrices are shown in Fig.~\ref{cov_estimation}. Next, we use Kullback–Leibler (KL) divergence \citep{kullback1951information} to evaluate measure how our estimated distribution $q\left(\boldsymbol{x}\right)=\mathcal{N}(\hat{\boldsymbol{\mu}}, \hat{\boldsymbol{\Sigma}})$ is different from the true distribution  $p\left(\boldsymbol{x}\right)=\mathcal{N}(\boldsymbol{\mu},\boldsymbol{\Sigma})$.
Specifically, the KL divergence from $q(\boldsymbol{x})$ to $p(\boldsymbol{x})$ is defined as:
\begin{equation}\label{eq:kl}
\begin{aligned}
D_{KL}\left(p\left(\boldsymbol{x}\right)\parallel q\left(\boldsymbol{x}\right)\right) = \int p\left(\boldsymbol{x}\right)\ln\frac{p\left(\boldsymbol{x}\right)}{q\left(\boldsymbol{x}\right)}d\boldsymbol{x}.
\end{aligned}
\end{equation}
A smaller KL divergence indicates that the distribution $q\left( x \right)$ is closer to the reference distribution $p\left( x \right)$. In the experiment, both $p\left(\boldsymbol{x}\right)$ and $q\left(\boldsymbol{x}\right)$ are multivariate Gaussian distributions. We can derive the KL divergence formulation for multivariate Gaussian distributions as:
\begin{equation}\label{eq:kl_multi}
\begin{aligned}
D_{KL}\left(p\left(\boldsymbol{x}\right)\parallel q\left(\boldsymbol{x}\right)\right) &= \frac{1}{2}\left[\ln\frac{|\hat{\boldsymbol{\Sigma}}|}{|\boldsymbol{\Sigma}|} - N + \text{Tr} \{ \hat{\boldsymbol{\Sigma}}^{-1}\boldsymbol{\Sigma} \} + (\hat{\boldsymbol{\mu}} - \boldsymbol{\mu})^{\top} \hat{\boldsymbol{\Sigma}}^{-1}(\hat{\boldsymbol{\mu}} - \boldsymbol{\mu})\right].
\end{aligned}
\end{equation}
The KL divergences for the distributions estimated with different observations are shown in Table~\ref{performance}. We can see the KL divergence of using all observations is the lowest, indicating that using missing/ragged values can improve the accuracy of the estimated distribution.

\begin{table}[!ht]
\caption{Posterior mean of mean vectors estimated with different types of observations.}
\label{mean_vector}
\centering
\footnotesize
\begin{tabular}{c|c}
\toprule
 & Mean vector \\
\midrule
Full observations & $[14.1,15.2,18.5,13.6,17.6,15.3,9.9,23.8,15.2,11.5,12.4,15.5,9.6,13.4,17.1,15.0,19.1,21.3]$\\
Full, missing observations & $[14.2,15.2,18.3,13.2, 17.2,15.2,10.0,24.0,15.1,11.1,12.0,15.0,9.3,13.3,17.1,14.7,18.7,20.8]$\\
All observations & $[14.1,15.1,18.2,13.3,17.1,15.0,10.0,24.1,15.1,11.1,12.0,15.0,9.2,13.2,17.2,14.9,18.8,20.8]$\\
\bottomrule
\end{tabular}
\end{table}

\begin{table}[!ht]
\caption{KL divergence of distributions estimated with different types of observations.}
\label{performance}
\centering
\footnotesize
\begin{tabular}{c|c|c|c}
\toprule
 & Full observations & Full, missing observations & Full, missing, ragged observations\\
\midrule
  $D_{KL}$ & 0.2502 & 0.0748 & \textbf{0.0565}\\

\bottomrule
\end{tabular}
\end{table}

\begin{figure}[!ht]
\centering

\subfigure[Markov sampling with full observations.]{
    \centering
    \includegraphics[width = 0.9\textwidth]{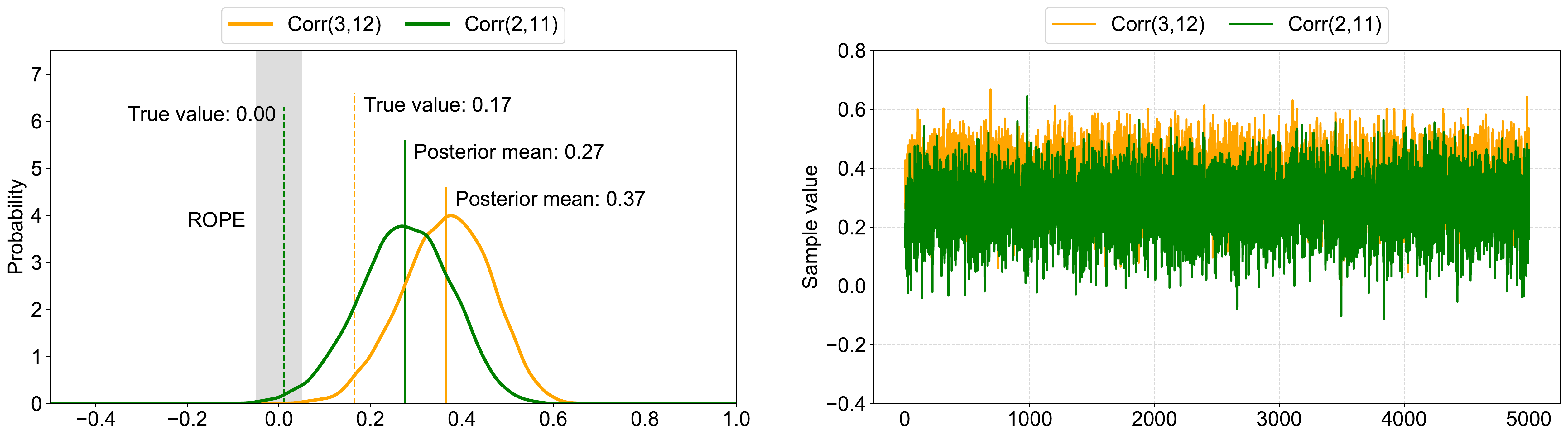}
}
\subfigure[Markov sampling with full and missing observations.]{
    \centering
    \includegraphics[width = 0.9\textwidth]{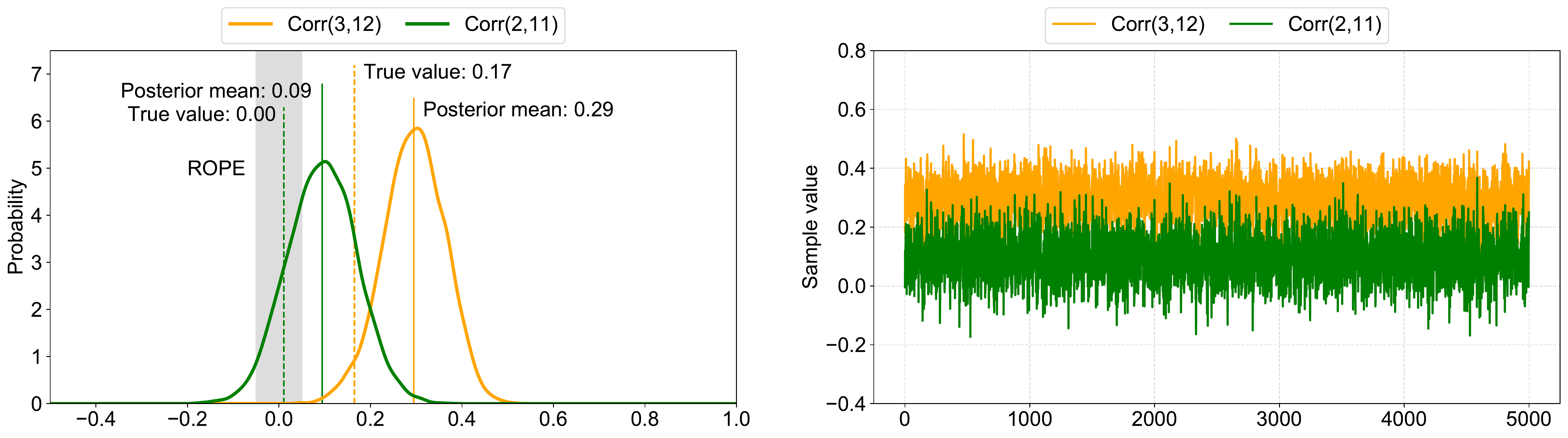}
}
\subfigure[Markov sampling with full, missing, and ragged observations.]{
    \centering
    \includegraphics[width = 0.9\textwidth]{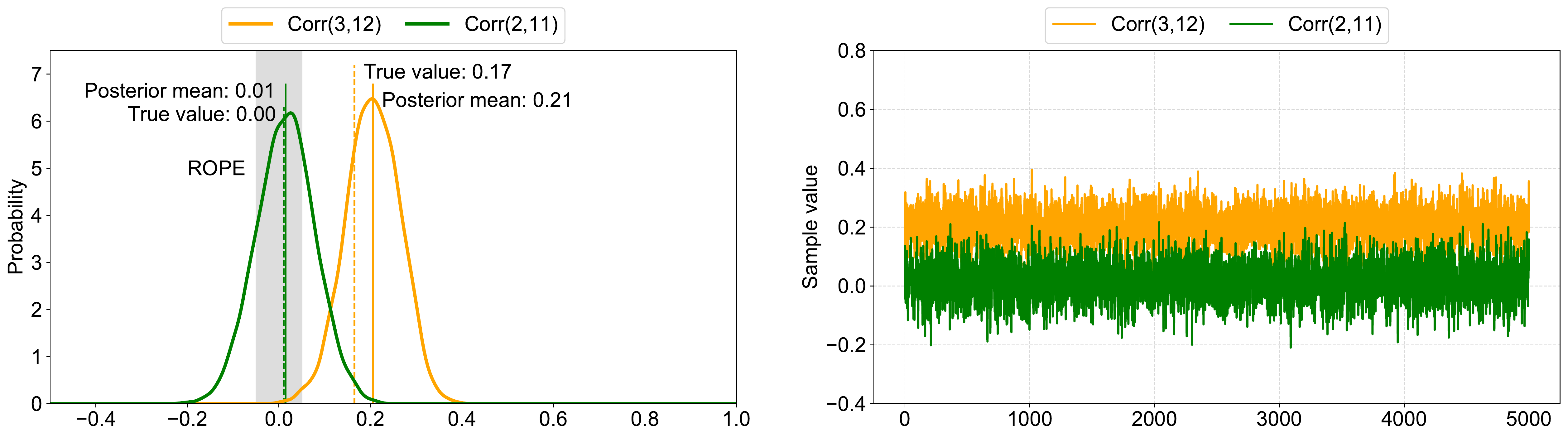}
}

\caption{The estimated posterior distributions over two entries of the covariance matrix with different observations.}
\label{iteration}
\end{figure}

Bayesian approach has the advantage that we can estimate the posterior distributions over covariance/correlation matrices. We use credible interval (CI) to measures the uncertainty of parameters. A CI is an interval with a particular probability to contain an unknown parameter value, and throughout this paper use 95\% equal-tailed interval interval as CI unless stated otherwise. Moreover, we want to determine whether a particular parameter value is credible, such as the ``null'' value of 0 for the correlation. For making decisions about the null value, we can use the equivalence test based on the full posterior distribution and region of practical equivalence (ROPE) \citep{kruschke2018bayesian}. The equivalence test checks the percentage of full posterior that falls inside the ROPE. The null value is declared to be rejected when the percentage is sufficiently low; the null value is considered to be accepted if the percentage is sufficiently high. Throughout this paper, we set the ROPE range with $(-0.05,0.05)$ and the rejected-threshold with 5\%. Fig.~\ref{iteration} presents the estimated posterior distributions over two entries ($\operatorname{Corr}(2, 11)$ and $\operatorname{Corr}(3, 12)$) of correlation matrices with different observations. The true value of  $\operatorname{Corr}(2, 11)$ is zero. From Fig.~\ref{iteration} (a), we can see the posterior mean of $\operatorname{Corr}(2, 11)$ estimated with full observations is 0.27 (CI: $[0.07, 0.46]$), and the posterior distribution shows that the estimated $\operatorname{Corr}(2, 11)$ is much larger than zero, and the Bayesian credible interval is largely outside the ROPE. From Fig.~\ref{iteration} (b), we can see the posterior mean of $\operatorname{Corr}(2, 11)$ estimated with full and missing observations is 0.09 (CI: $[-0.05, 0.24]$), which is more accurate than only using only full observations and we fail to reject the value zero. In Fig.~\ref{iteration} (c), the posterior mean of $\operatorname{Corr}(2, 11)$ estimated by using all observations is 0.01 (CI: $[-0.11, 0.14]$), and the percentage of the credible interval that falls in the ROPE is larger than 5\%, meaning that we cannot reject the value zero. The true value of $\operatorname{Corr}(3, 12)$ is 0.17, and the posterior mean values estimated with full observations, full and missing observations, and all observations are 0.31 (CI: $[0.17, 0.54]$), 0.24 (CI: $[0.16, 0.42]$), and 0.19 (CI: $[0.09, 0.31]$), respectively. All the credible intervals of $\operatorname{Corr}(7, 14)$ largely fall outside the ROPE, indicating that we can reject the value zero. With the use of missing/ragged observations, the posterior mean of covariance becomes more accurate, and the posterior probability density becomes thinner, indicating a smaller standard deviation.

\begin{figure*}[!ht]
\centering
\subfigure[True correlation matrix.]{
    \centering
    \includegraphics[width=0.47
    \textwidth]{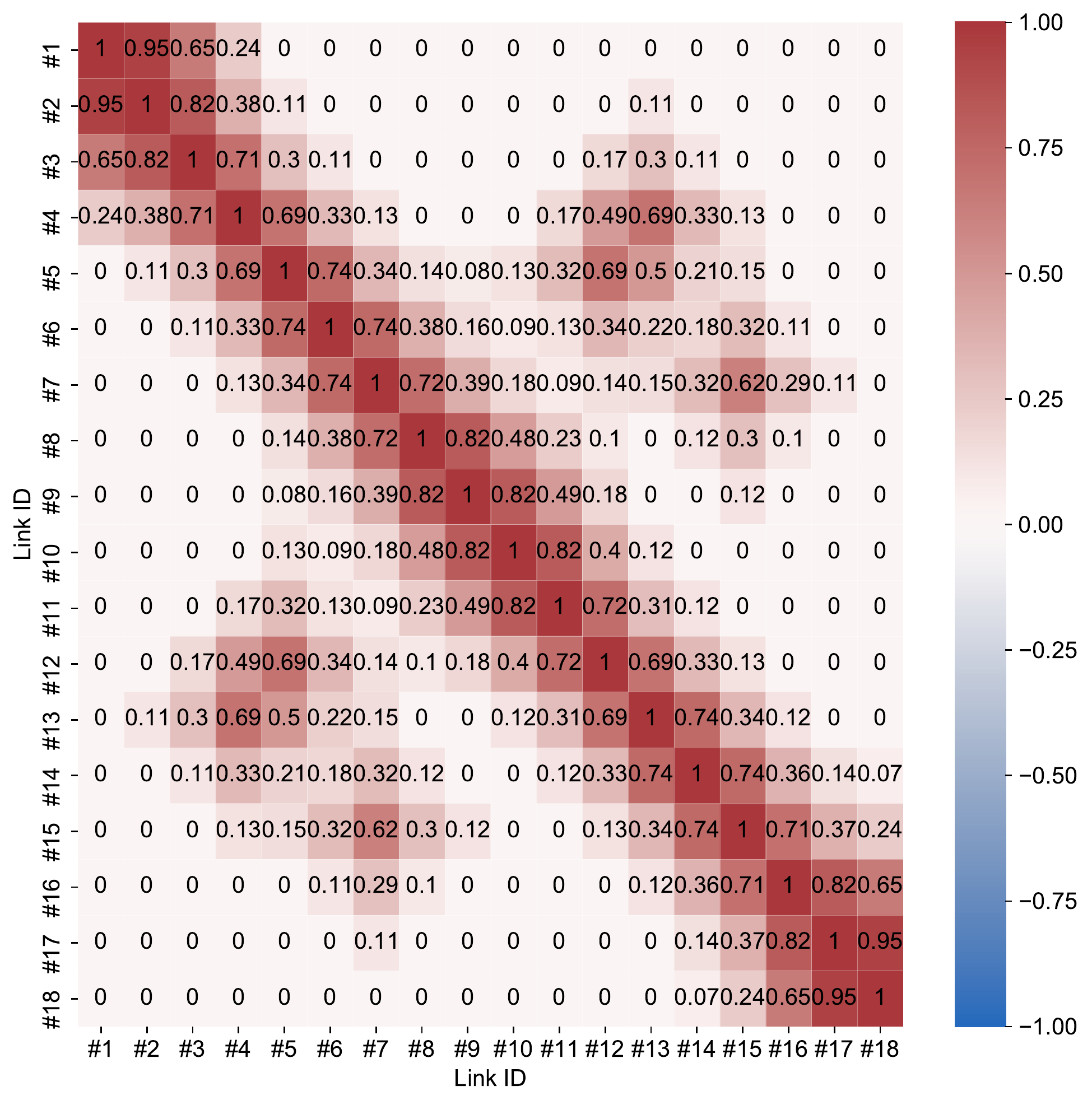}
}
\subfigure[Estimated correlation matrix with full observations.]{
    \centering
    \includegraphics[width=0.47
    \textwidth]{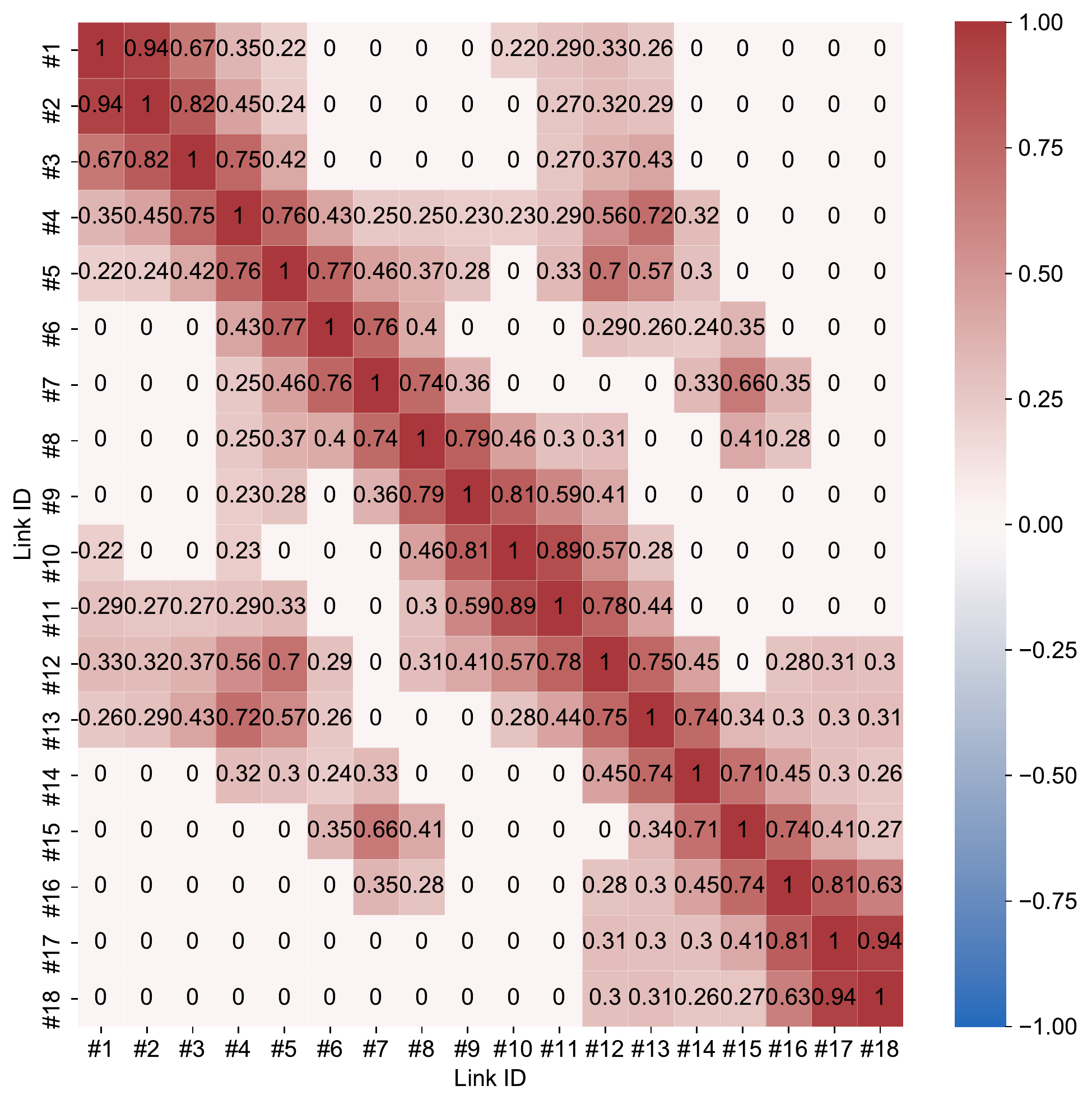}
}
\subfigure[Estimated correlation with full and missing observations.]{
    \centering
    \includegraphics[width=0.47
    \textwidth]{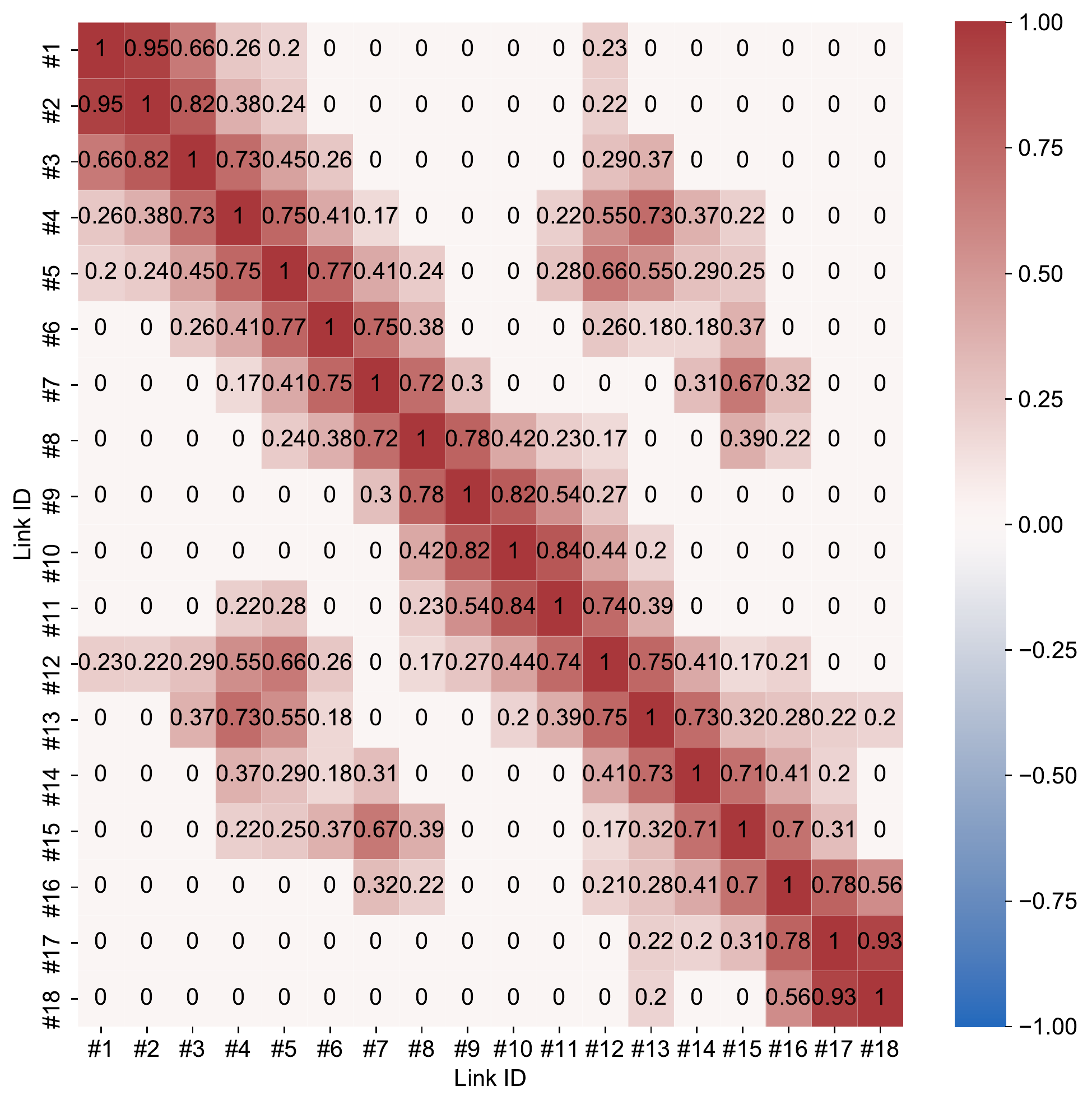}
}
\subfigure[Estimated correlation with all observations.]{
    \centering
    \includegraphics[width=0.47
    \textwidth]{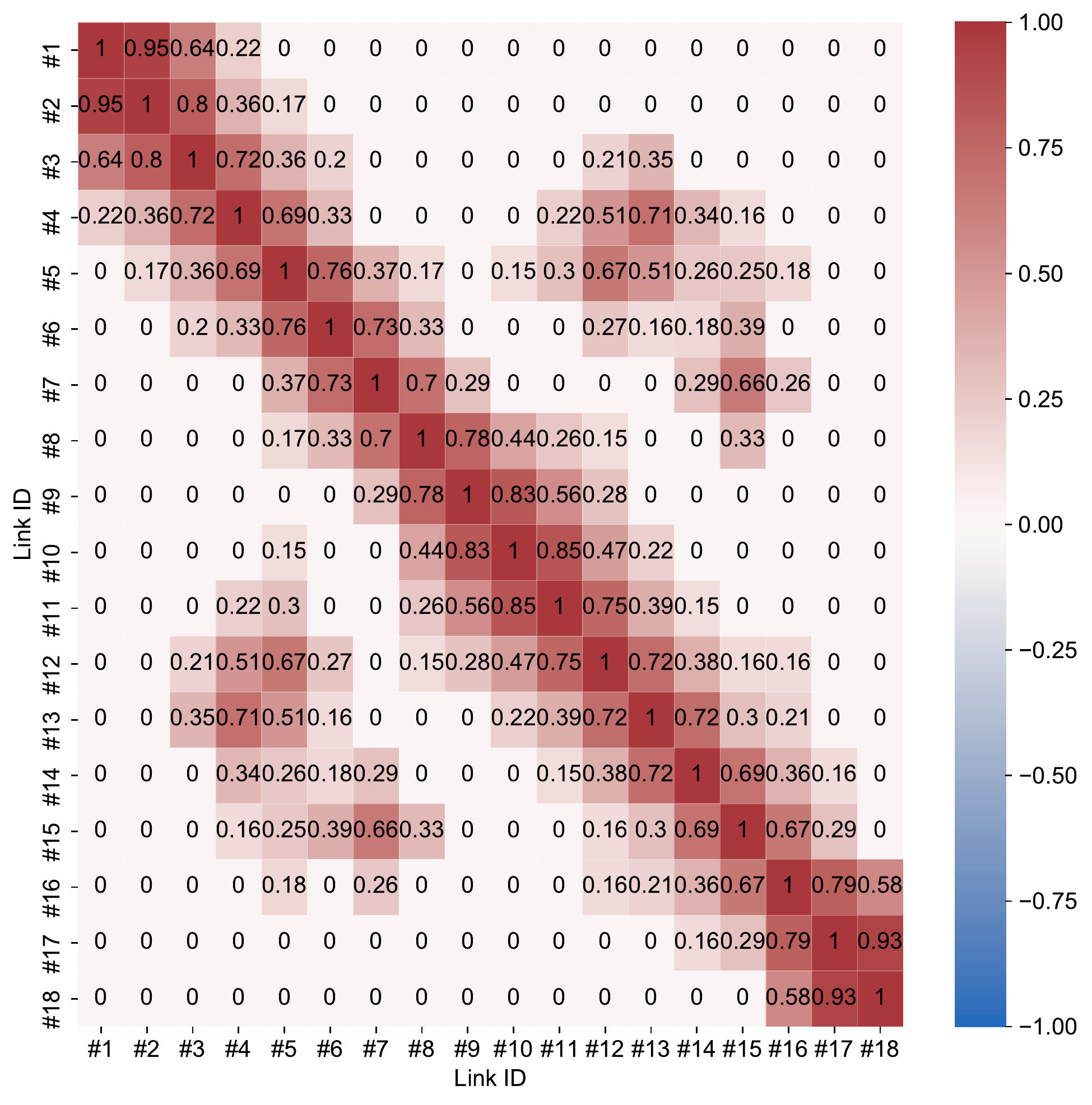}
}
\caption{The true correlation matrix and estimated correlation matrices.}
\label{syn_est}
\end{figure*}

Finally, we do the equivalence tests for entries of estimated correlation matrices and we set zero for correlations that fail to reject the value zero for better visualization. Moreover, we set zeros for the true correlations which are lower than 0.05 for convenient comparison considering that we use the ROPE with $(-0.05, 0.05)$. The estimated and true correlation matrices are shown in Fig.~\ref{syn_est}. Fig.~\ref{syn_est} (a) presents the true correlation matrix. Fig.~\ref{syn_est} (b)-(c) show the correlation matrices estimated with full observations, full and missing observations, all observations, respectively. We can see that using all observations can obtain the most accurate estimated correlation matrix, which agrees with the evaluation using KL divergences.

\subsection{Case 2: Guangzhou Bus Data}

\begin{figure*}[!ht]
\centering
\subfigure{
    \centering
    \includegraphics[width=0.6\textwidth]{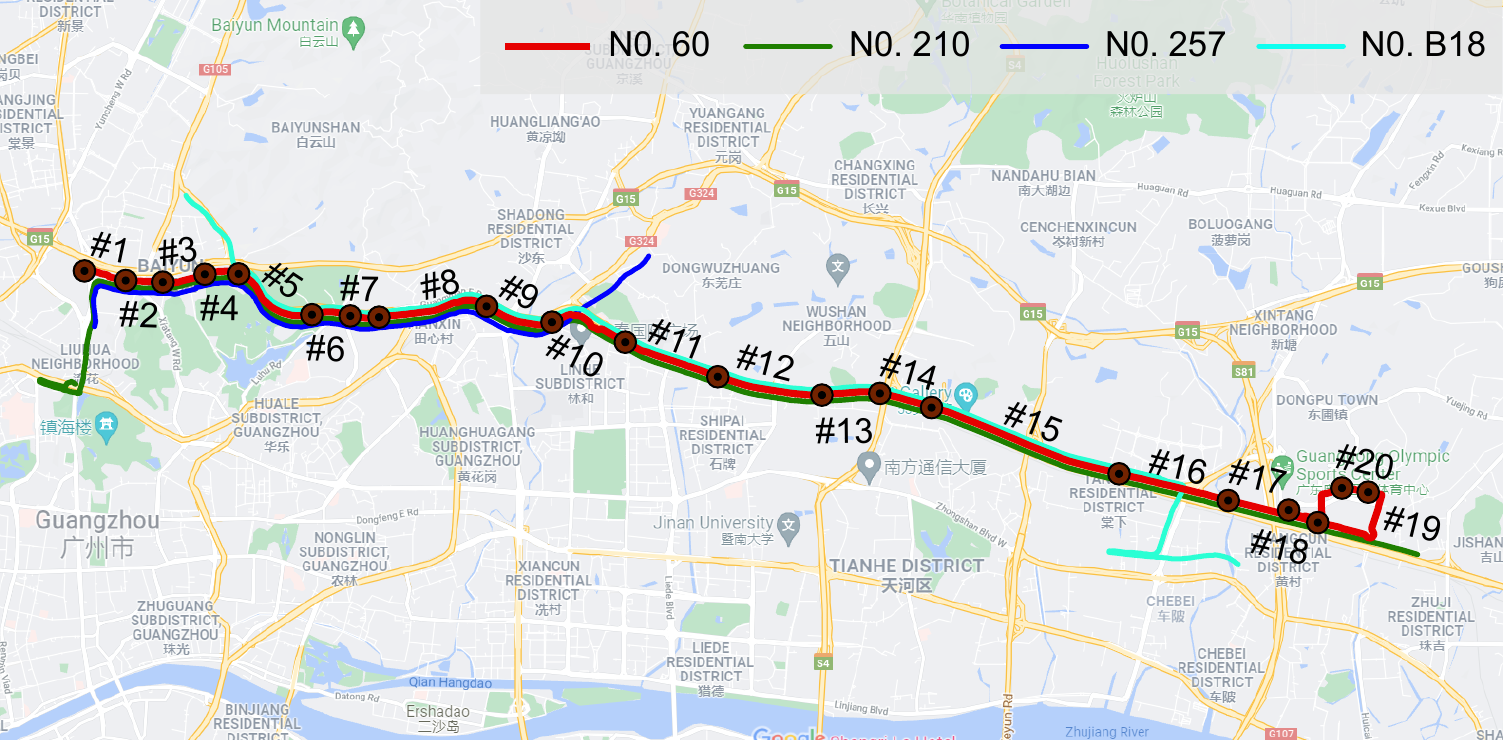}
}
\caption{Bus route No. 60 and the related routes in Guangzhou bus network.}
\label{Bus_network}
\end{figure*}

\begin{figure*}[!ht]
\centering
\subfigure{
    \centering
    \includegraphics[width=1\textwidth]{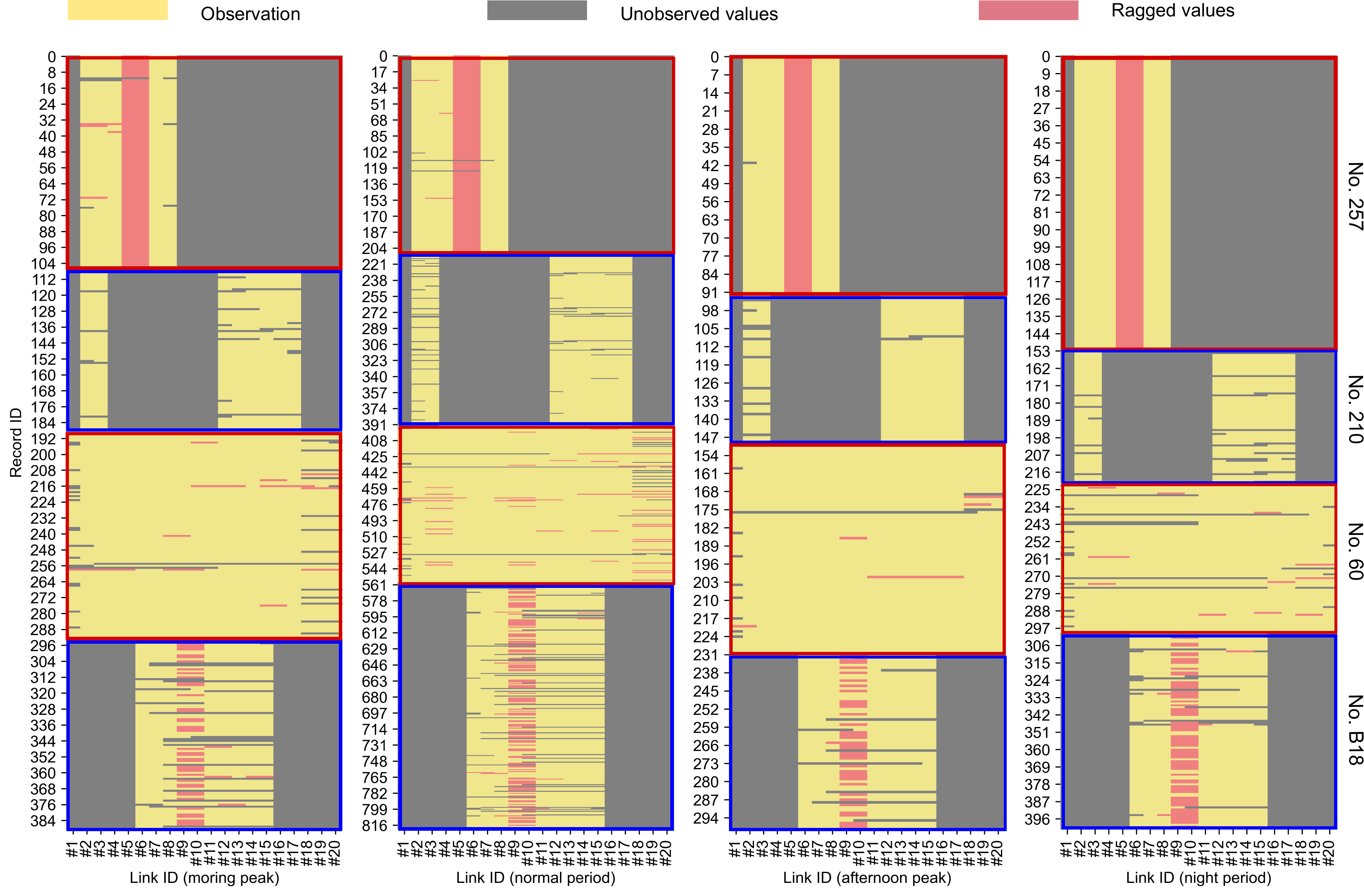}
}
\caption{Data overview.}
\label{missing_data}
\end{figure*}

In this section, we apply the proposed Bayesian model to real-world data to empirically quantify the link travel time correlation of a bus route. The data used in this paper are the bus in-out-stop record data collected in Guangzhou, China, during the weekdays from December 8, 2016 to December 15, 2016. The information of the data is outlined in Table~\ref{desciption_data}. These data were collected by the bus in-out-stop record system, i.e., the automatic bus announcing system. When a bus enters or exits a bus stop, the system reports the arrival or departure information and records the time stamp accordingly. Thus we can easily obtain the link travel times from the data. We take bus route No. 60 as a case and aim to quantify this bus route's link travel time correlation. First, we select the other three bus routes related to route No. 60. All the studied bus routes are displayed in Fig.~\ref{Bus_network}, and they are in the CBD of Guangzhou. Route No. 60 has 20 links; route No. 257 shares 7 links with route No. 60 (from link $\#2$ to link $\#8$); route No. B18 shares 10 links with route No. 60 (from link $\#6$ to link $\#15$); route No. 210 shares 17 links with route No. 60, but the buses of route No. 210 do not stop after entering link $\#3$ until leaving link $\#11$. As our defined link travel time considers the dwell time, these long-ragged data may have a negative impact on the estimation; we thus only used the 8 shared links (link $\#2$, $\#3$, $\#12$-$\#17$). We divide one day into four periods: morning peak $(7:00-10:00)$, normal period $(10:00-17:00)$, afternoon peak $(17:00-20:00)$, and night period $(20:00-7:00)$. The overview of all used data is shown in Fig.~\ref{missing_data}. We can see that all the bus routes have many missing and ragged values.

\begin{table}[!ht]
\caption{Description of Data.}
\label{desciption_data}
\centering
\footnotesize
\begin{tabular}{l|l|l}
\toprule
Variable & Description & Example\\
\midrule
ID & Identity for bus data record & 1612020547101390 \\
OBUID & Identity for bus & 911721 \\
TRIP\_ID & Identity for bus trip & 1612012250030880 \\
ROUTE\_ID & Identity for bus route & 201 \\
ROUTE\_NAME & Bus route name & No. 24 \\
ROUTESUB\_ID & Identity of bus route direction & 502669 \\
ROUTE\_STA\_ID & Identity of bus stop & 84279 \\
STOP\_NAME & Bus stop name & Dunhe Stop \\
AD\_FLAG & Bus state: arrival (1) or departure (0) & 1\\
AD\_TIME & The time bus reported arriving at/leaving a bus stop & 20161202, 05:47:08\\
\bottomrule
\end{tabular}
\end{table}

\begin{figure}[!ht]
\centering
\subfigure[Markov sampling for morning peak.]{
    \centering
    \includegraphics[width = 0.9\textwidth]{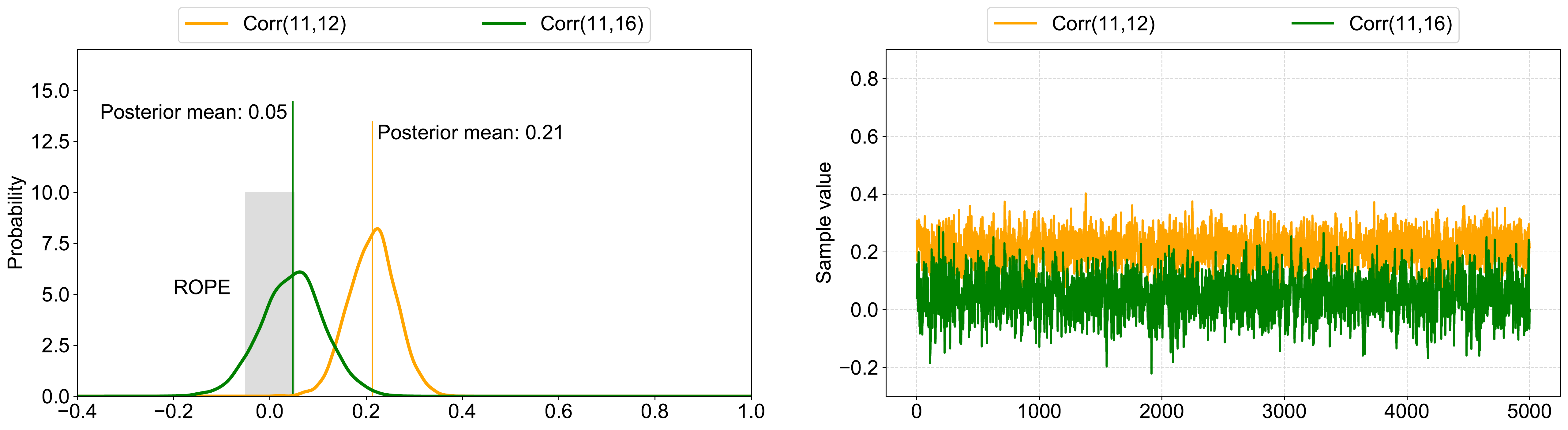}
}

\subfigure[Markov sampling for normal period.]{
    \centering
    \includegraphics[width = 0.9\textwidth]{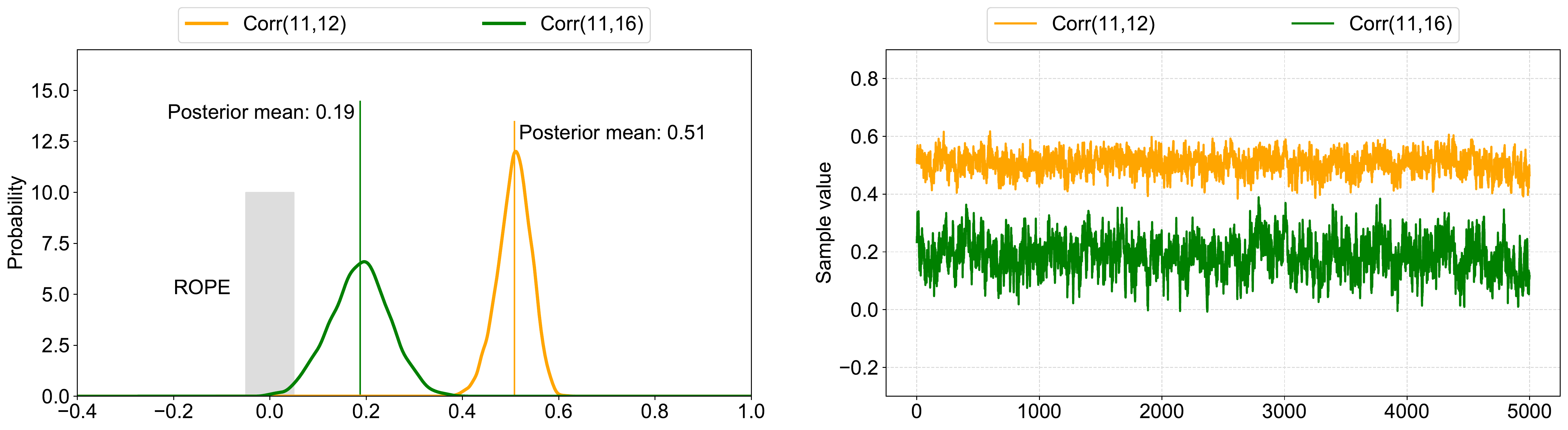}
}

\subfigure[Markov sampling for afternoon peak.]{
    \centering
    \includegraphics[width = 0.9\textwidth]{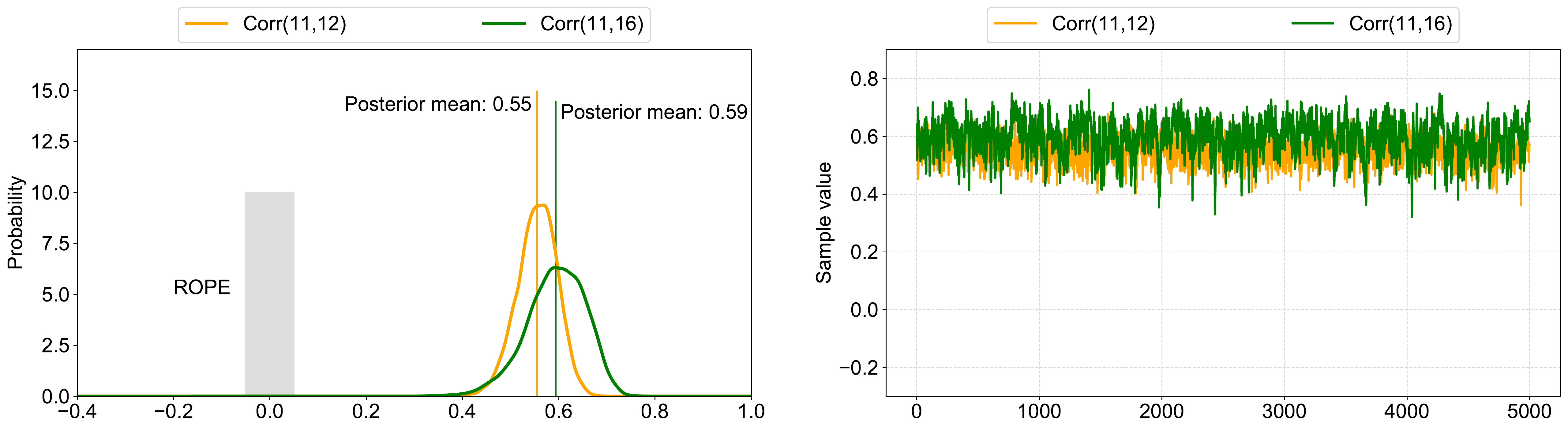}
}

\subfigure[Markov sampling for night period.]{
    \centering
    \includegraphics[width = 0.9\textwidth]{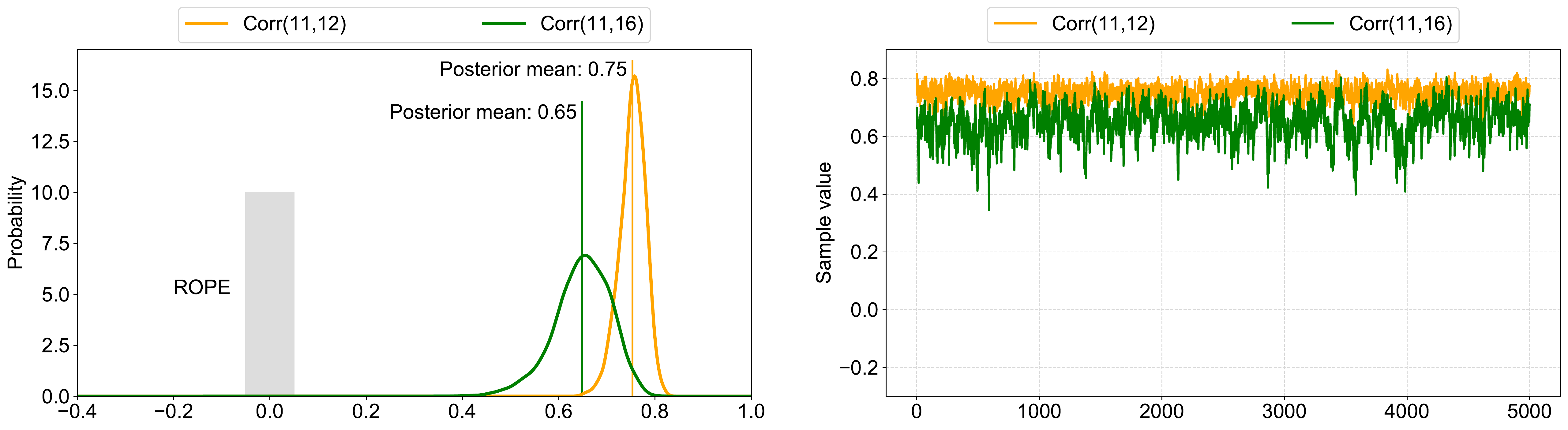}
}

\caption{The estimated posterior distributions over two entries of the correlation matrices.}
\label{sampling2}
\end{figure}

The numbers of MCMC iterations are $k_1=10000$ and $k_2=5000$, respectively. Fig.~\ref{sampling2} presents the estimated posterior distributions over two entries ($\operatorname{Corr}(11, 12)$ and $\operatorname{Corr}(11, 16)$) of
correlations matrices for different periods. We can find that the distributions over a correlation are distinct for different periods. For example, the posterior mean values of $\operatorname{Corr}(11, 16)$ are 0.05 (CI: $[-0.08, 0.18]$), 0.19 (CI: $[0.07, 0.31]$), 0.59 (CI: $[0.46, 0.7]$), 0.65 (CI: $[0.52, 0.75]$) for morning peak, normal period, afternoon peak, night period, respectively. The equivalence test of $\operatorname{Corr}(11, 16)$ for the morning peak fails to reject the value zero, while all the 95\% CI of $\operatorname{Corr}(11, 16)$ for the other periods largely fall outside ROPE, indicating that travel times on these two links are positively correlated.

\begin{figure*}[!ht]
\centering
\subfigure[Estimated correlation matrix for morning peak.]{
    \centering
    \includegraphics[width=0.45
    \textwidth]{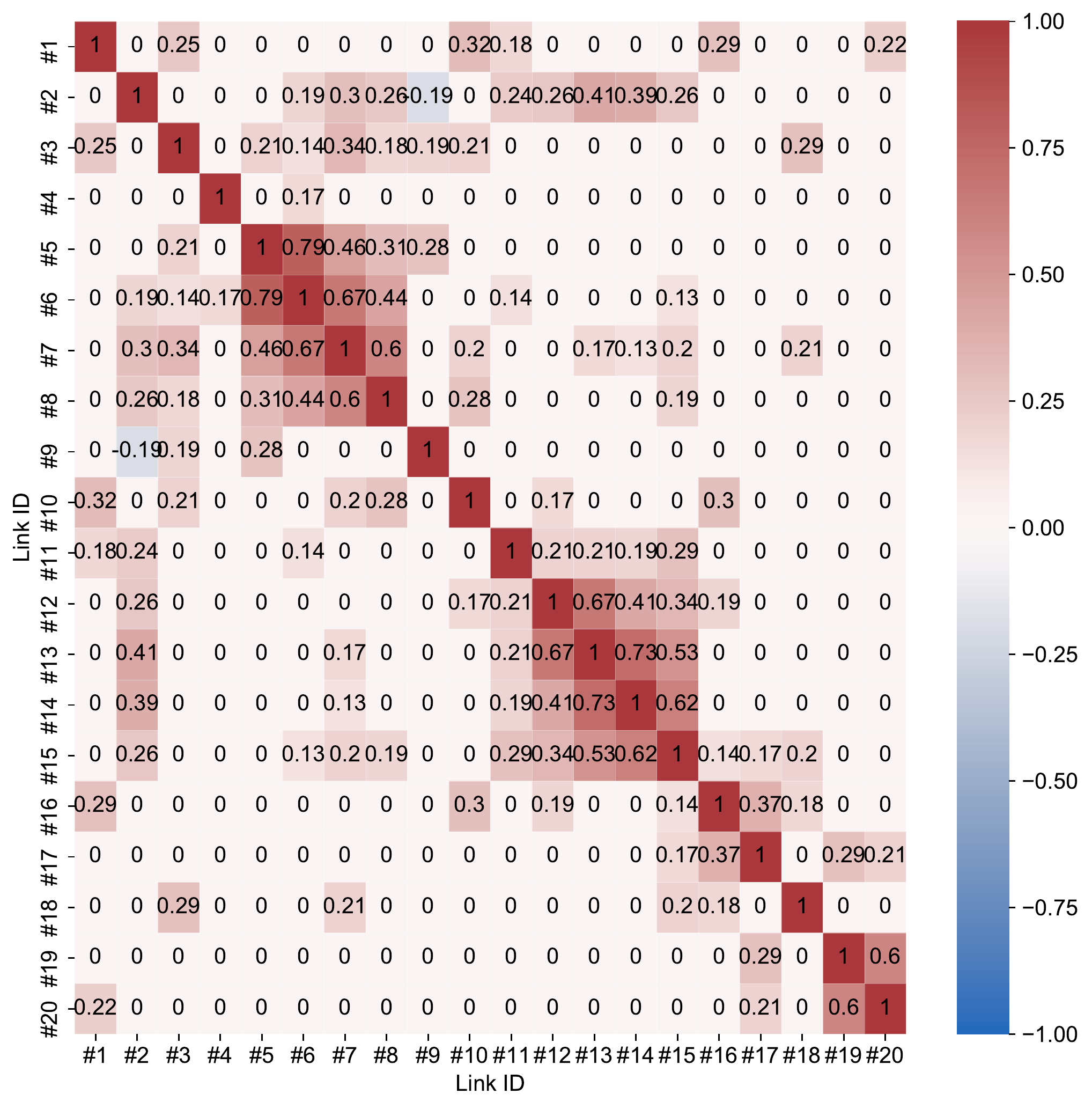}
}
\subfigure[Estimated correlation matrix for normal period.]{
    \centering
    \includegraphics[width=0.45
    \textwidth]{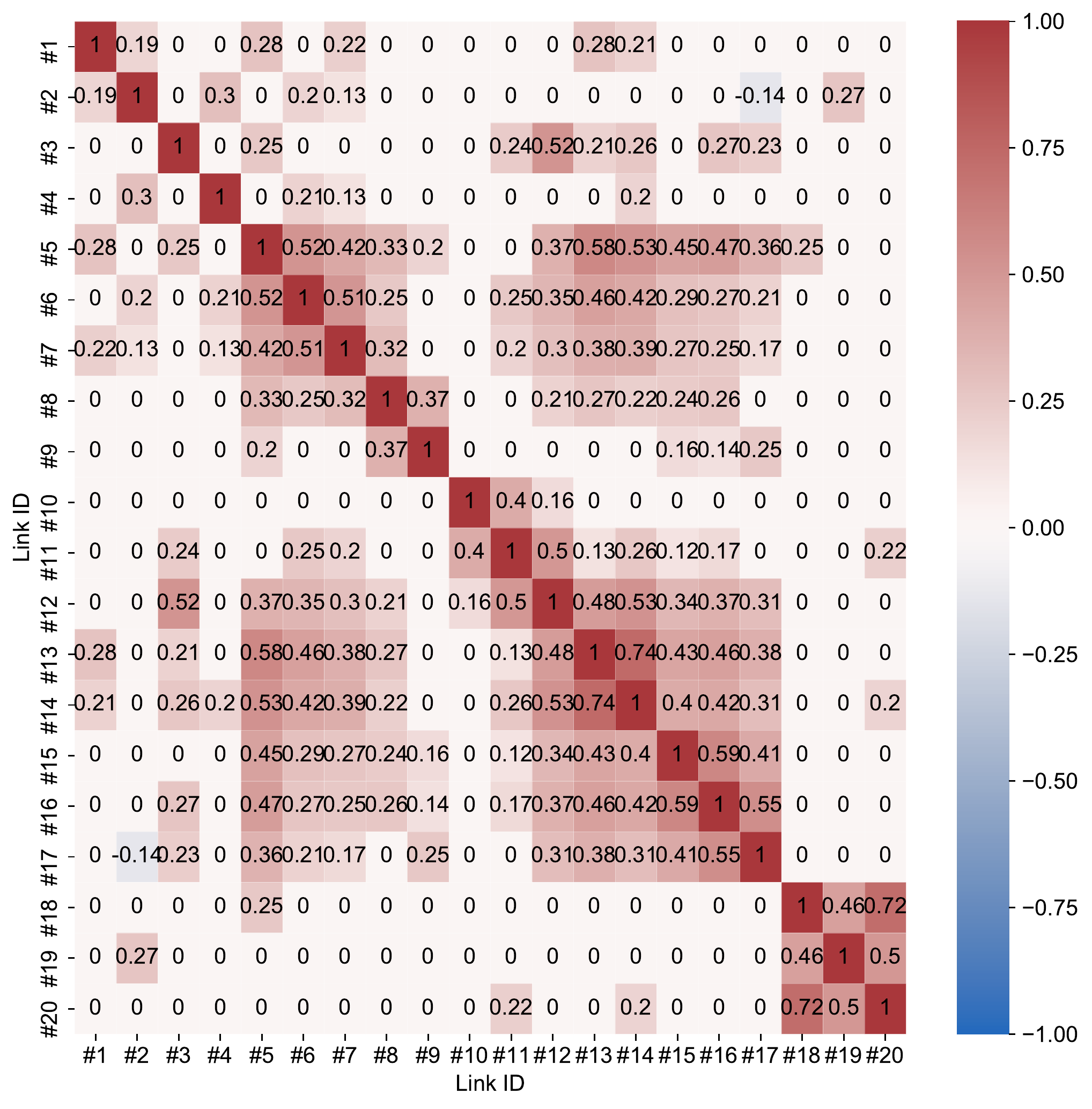}
}
\subfigure[Estimated correlation matrix for afternoon peak.]{
    \centering
    \includegraphics[width=0.45
    \textwidth]{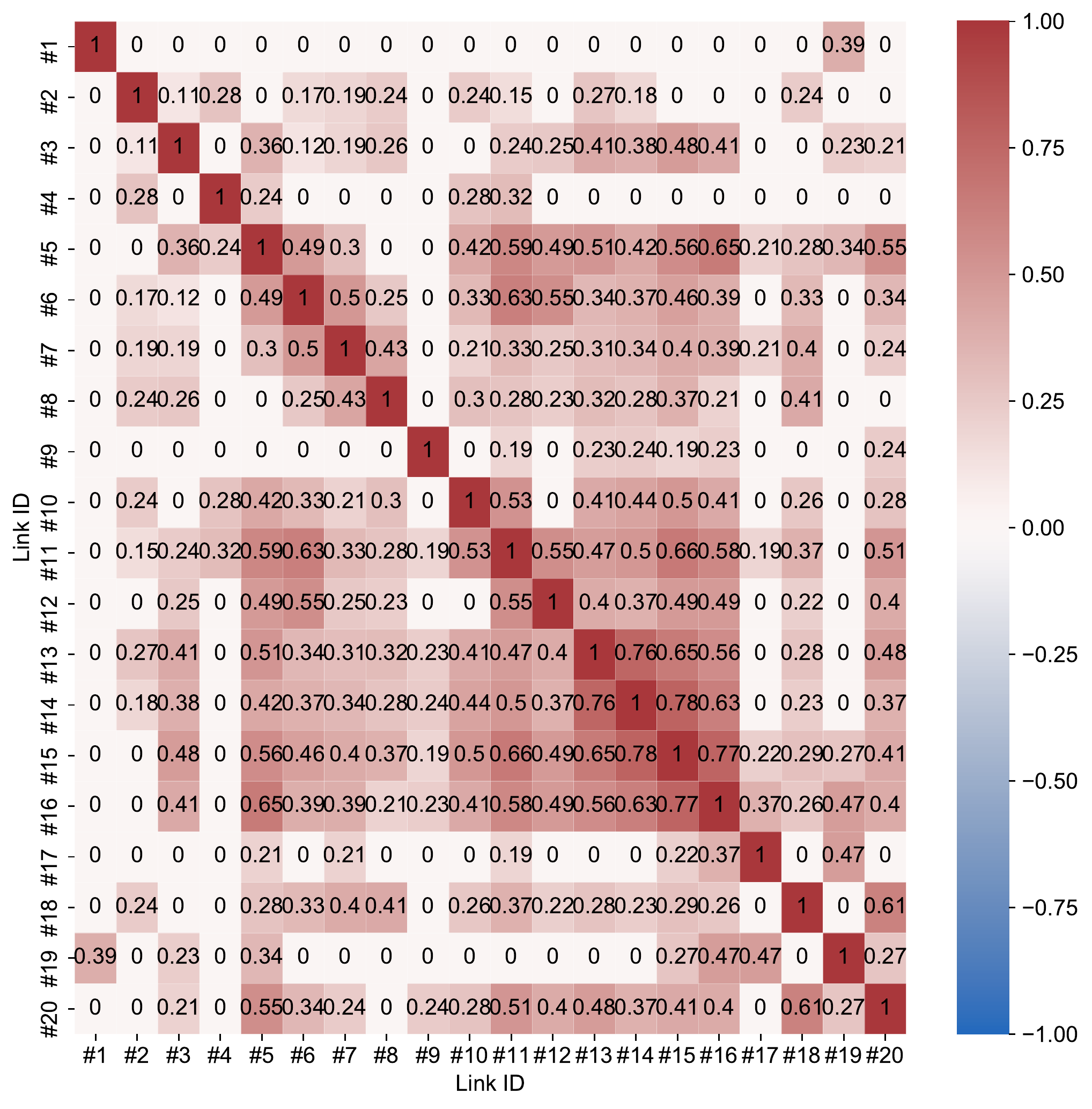}
}
\subfigure[Estimated correlation matrix for night period.]{
    \centering
    \includegraphics[width=0.45
    \textwidth]{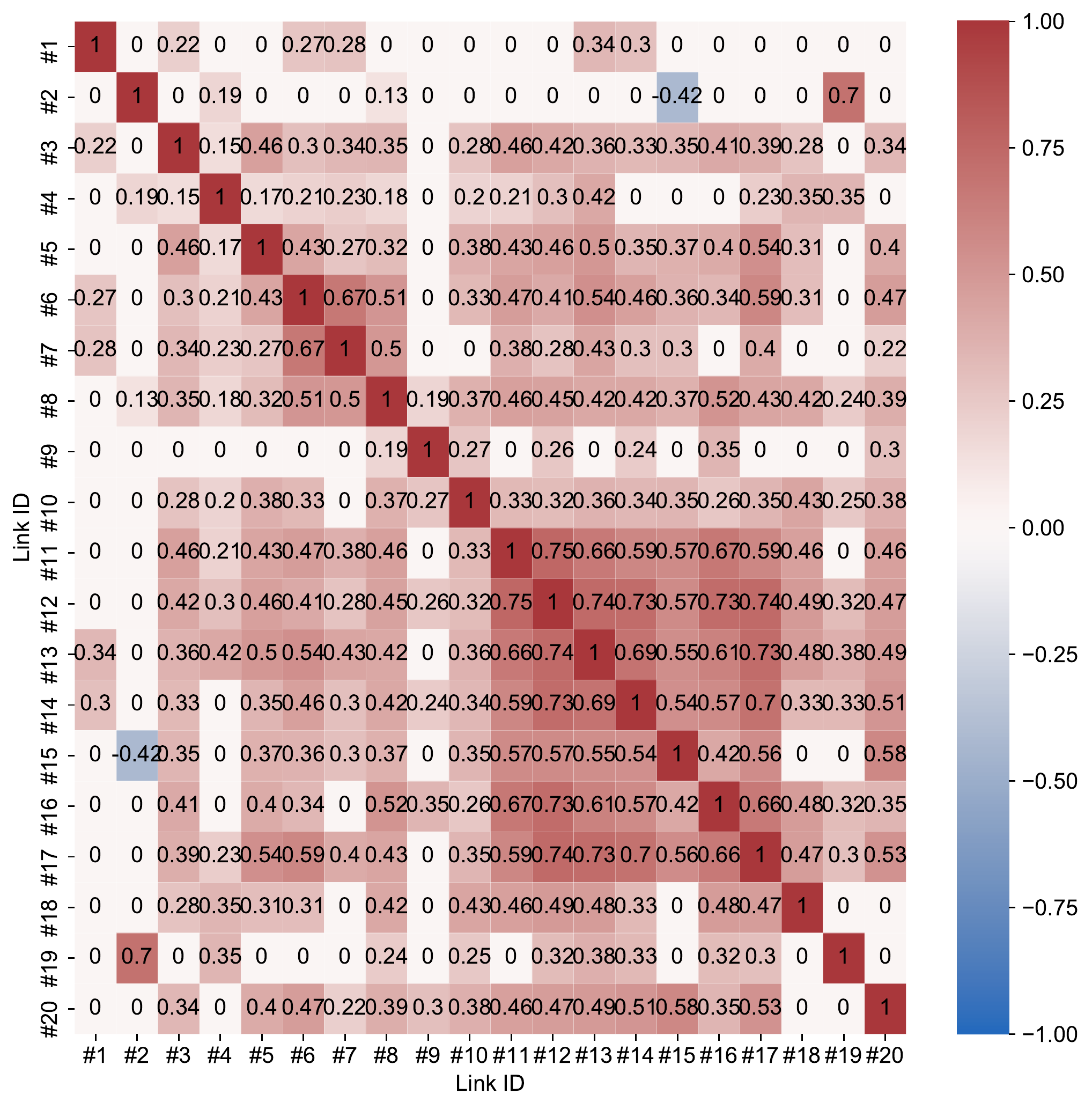}
}
\caption{The estimated correlation matrix for different periods.}
\label{Estimate4}
\end{figure*}

We calculate credible intervals for entries of estimated correlation matrices using drawn samples and we set zero for correlations that cannot reject the value zero for better visualization. Fig.~\ref{Estimate4} shows the estimated correlation matrices for four different periods. Each cell in the correlation matrix shows the correlation between two variables. Essentially, this kind of correlation matrix is Pearson's Product-Moment Correlation. The cell number can help to understand how strong a relationship is between two variables. The further away the cell value is from zero, the stronger the relationship between the two variables. Generally, when the cell's absolute value of correlation matrix is zero, the relationship between the corresponding variables will be considered as no relationship; when the absolute value is lower than 0.25, the relationship will be considered as a weak correlation; when the absolute value is between 0.25 and 0.5, the relationship will be regarded as a medium relationship; when the absolute value is larger than 0.5, it indicates these two variables are strongly correlated. Furthermore, the sign of the cell value also means a different correlation. A positive value indicates the positive correlation between two variables, while a negative value represents the negative correlation. For the positive correlation, when the value of one variable increases, the value of the other variable increases in a similar way. For the negative correlation, when the value of one variable increases, the value of the other variable tends to decrease.

Fig.~\ref{Estimate4} reveals some characteristics of link travel time correlation of the bus route. First, we can find that the estimated correlation matrices vary for four periods, indicating that the link travel time correlation is time-varying. Overall, more correlated link pairs exist during the afternoon peak and the night period, while fewer correlated link pairs exist during the morning peak. The directional bus route stretches from urban business districts to suburban areas; thus, the traffic conditions/passenger flows are different for the morning and afternoon peaks. The better traffic condition and the small passenger flow exist in the morning because few people go to suburban areas on weekdays; instead, many people go to urban business districts to work. On the contrary, traffic congestions and large passenger flow happen in the afternoon peak as more people go home from urban to suburban areas. Traffic congestions and large passenger flow can cause the bus bunching phenomenon: a lagging bus must collect more passengers and, therefore, needs more travel time; on the other hand, a subsequent bus of the lagging bus will have fewer passengers, and its travel time will be shorter. We conclude that during the afternoon peak, bus bunching can make more link pairs correlated. Second, most link pairs do not have strong correlations as most cell values are lower than 0.5 for these four time periods. Meanwhile, the values of cells with strong correlations are positive, which means link travel time variables of a bus route are more likely to have a positive correlation if they have a strong relationship. A possible reason for these positive correlations is the bus bunching phenomenon. Few negative correlations exist in link pairs though they are weak or medium correlation. Third, both local and long-range correlations exist on the bus route. Many strong correlations exist in local link pairs. For example, adjacent link pairs  (link $\#5$, link $\#6$), (link $\#6$, link $\#7$), (link $\#7$, link $\#8$) during the morning peak have strong correlations. Apart from the adjacent link pairs, strong correlations exist in link pairs with long distances. In Fig.~\ref{Estimate4} (c), (link $\#5$, link $\#11$), (link $\#5$, link $\#13$), (link $\#6$, link $\#11$) are long-range correlations. Finally, we plot the scatters of some strongly correlated link pairs for different periods shown in Fig.~\ref{Scatter_Morning} - Fig.~\ref{Scatter_Night}. We can find that all of them have intuitively positive correlations between link pairs, indicating that the estimated correlation matrices are reasonable.

\subsection{Case 3: Link/Trip Travel Time Forecasting}

In this section, we extend to applying covariance estimation for link/trip travel time probabilistic forecasting. The proposed Bayesian model can estimate the posterior mean vectors and covariance matrices for different periods; we thus can obtain the conditional posterior distribution over forecasting links conditioned on observed links. We have estimated the mean vectors and covariance matrices of four periods in Case 2. Here, the experiment uses the following five weekdays' full observations of route No. 60 from December 16 to December 22, 2016 to test the forecasting performance. As a simple experiment, the task is to forecast the link travel times of the last nine links (from link $\#12$ to link $\#20$) given the link travel times of the first eleven links (from link $\#1$ to link $\#11$).

We select the historical average (HA) as the benchmark model. For the link travel time at a certain period of the day, HA uses the average link travel time at that period in the training set as the forecast value. Then we compare the performance of these two methods, which are evaluated by the root mean square error (RMSE) and the mean absolute percentage error (MAPE):
\begin{equation}
\begin{aligned}
     \text{RMSE}&=\sqrt{\frac{1}{n}\sum_{i=1}^{n}(y_i-\hat{y}_{i})^2}, \\
     \text{MAPE}&=\frac{1}{n}\sum_{i=1}^{n}\left|\frac{y_i-\hat{y}_i}{y_i}\right|,
\end{aligned}
\end{equation}
where $y_i,\hat{y}_{i},i=1,\ldots,n$ are the true values and forecasts, respectively. Table~\ref{performance_four} presents the forecasting performance using the Bayesian model and historical average. We can find Bayesian model performs better than the historical average method for all periods.

\begin{table}[!ht]
\caption{The forecasting performance of two methods for different periods.}
\label{performance_four}
\centering
\footnotesize
\begin{tabular}{c|c|c|c|c|c}
	\toprule
	\multicolumn{2}{c|}{} & Morning peak & Normal period & Afternoon peak & Night period \\
	\midrule
	\multirow{2}*{RMSE} & Bayesian model & \textbf{27.74} & \textbf{35.45} & \textbf{61.20} & \textbf{20.95}\\
	\cmidrule{2-6}
	~ & Historical average & 32.54 & 36.07 & 67.13 & 25.86 \\
	\midrule
	\multirow{2}*{MAPE} & Bayesian model & \textbf{0.1193} & \textbf{0.1186} & \textbf{0.1638} & \textbf{0.0885}\\
	\cmidrule{2-6}
	~ & Historical average & 0.1563 & 0.1188 & 0.1742 & 0.1116 \\
	\bottomrule
\end{tabular}
\end{table}

\begin{figure}[!ht]
\centering
\subfigure[Forecasting for observation 1.]{
    \centering
    \includegraphics[width = 1\textwidth]{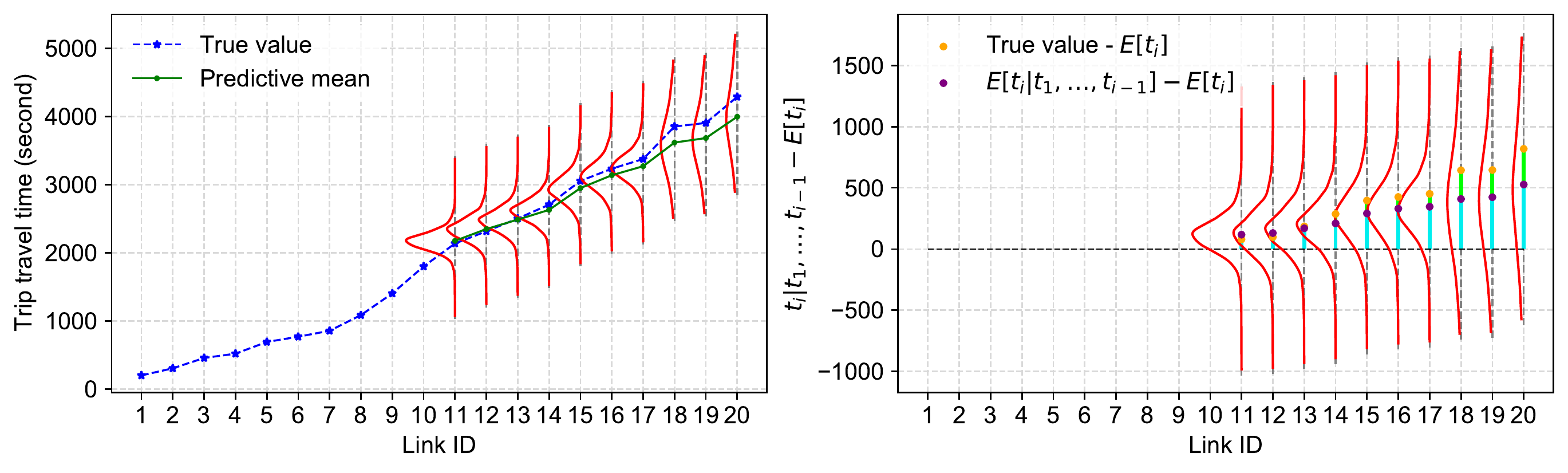}}
\subfigure[Forecasting for observation 2.]{
    \centering
    \includegraphics[width=1
    \textwidth]{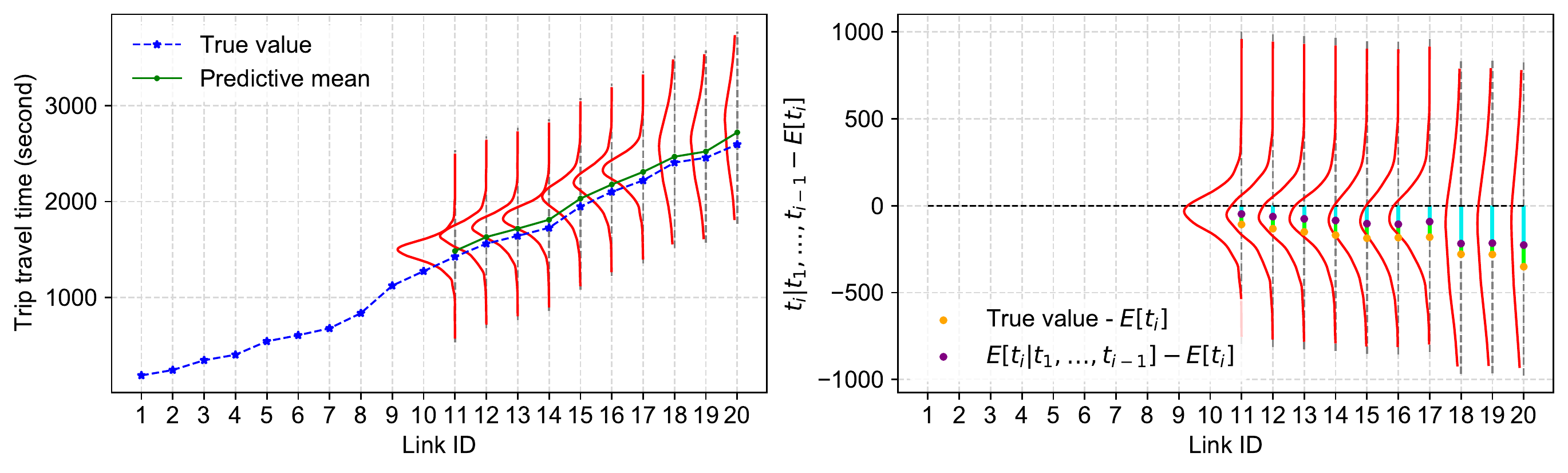}
}
\caption{The probabilistic forecasting for trip travel time.}
\label{probability_forecasting}
\end{figure}

This model can make probabilistic forecasting for bus trip travel time. As an example, we use two piece of test data in the afternoon peak to show probabilistic forecasting. Assume we have observed the first ten links' travel times, and the goal is to forecast the trip travel time distributions. Fig.~\ref{probability_forecasting} shows the probabilistic forecasting results. In the left panel, the blue points are the true trip travel times, and the green points are the predictive mean values. We can see that the predictive mean values fit the actual values, indicating the Bayesian model can make good forecasting. Moreover, the red bell curves are the trip travel time distributions, and we can see that the red bell curves are fatter with the increasing number of links in a trip, indicating the variance is increasing. The right panels present the mean corrected estimation, and the purple points (we refer to them as corrected mean values) are computed by posterior conditional mean values minus model mean values; the orange points are the difference between true values and model mean values. We can find that the posterior conditional mean can make a more accurate prediction than the model mean. If we do not use the information of the observed link travel times, the forecasting mean vectors should be equivalent to the model mean vectors. As we can see, the corrected mean values for observation 1 shown in Fig.~\ref{probability_forecasting} (a) are larger than zero, while the corrected mean values for observation 2 shown in Fig.~\ref{probability_forecasting} (b) are lower than zero, indicating the link travel time observations from link $\#1$ to $\#10$ indeed help update the forecasting values for the following links.

\section{Conclusion}
\label{conclusion}

In this paper, we have proposed a Bayesian Gaussian model to quantify the link travel time correlation of a bus route. The approach overcomes the issue of small sample sizes on a single bus route by incorporating data from other relevant bus routes. The proposed model can also impute those missing and ragged values in an incomplete link travel time vector. Three experiments are conducted in this paper. The first experiment is conducted on synthetic data with known covariance, and our result shows that the proposed Bayesian model can accurately recover the underlying mean and covariance from corrupted link travel time observations. In the second empirical experiment, we used real-world bus in-out-stop record data to quantify link travel time correlation. Our empirical analysis shows that (1) link travel times are clearly not independent on a bus route, and the estimated correlations vary substantially for different time periods of a day; (2) most link pairs are not strongly correlated, and most correlations are positive while negative correlations also exist; (3) both local and long-range correlations could exist on a bus route. Our results also suggest that simplified covariance assumptions (e.g., local spatial correlation) might be inappropriate for modeling travel time on a bus route. Finally, we applied the estimated covariance matrices to forecast link/trip travel time. An additional test data set during five weekdays is used to verify the forecasting performance, and the results show that the proposed model clearly outperforms a historical average baseline.

Our approach has potential implications for both practice and research. First, the proposed Bayesian model can estimate the covariance matrices essential to performing probabilistic forecasting of bus travel time. Second, the imputation method can also handle ragged values in other fields, such as economics, medicine, and social sciences. The ragged definition can be used to model link travel time from   origin-destination-based trip travel time observations; in this case, ${\mathbf{G}}_i$ becomes a row vector encoding the linear transformation to obtain the total travel time for (a single) trip $i$. Third, this approach can be used in estimating automobile's link travel time correlation in a small network.

Our proposed Bayesian Gaussian model has several limitations. First, it is challenging to infer the high-dimensional covariance structure (e.g., $n>100$). When the dimension of the covariance is large, the computation is highly expensive, and small-size observations constrain the estimation accuracy. Second, this model does not consider the influence of dwell time from multiple bus routes. Our defined link travel time includes dwelling time. However, different bus routes have distinct characteristics of dwell time due to factors including passenger flow/demand, bus schedule, and bus types. For example, bus routes with lower passengers flow will have shorter link travel times, while a larger passenger flow will cause longer travel times. In this case, our assumption that related bus routes share the same link travel time distribution may no longer hold. The influence of dwell time from multiple bus routes could be studied in further research. Third, the way we model the covariance structure of different time periods is by dividing samples into several periods and estimating the proposed model independently. Although simple, this approach ignores the temporal dynamic of the covariance structure---the covariance structure may vary smoothly and continuously over time. Our further research is to develop new models to characterize time-varying link travel time correlation.

\section*{Acknowledgements}

This research is supported in part by the Natural Sciences and Engineering Research Council (NSERC) of Canada. X. Chen acknowledges funding support from the China Scholarship Council (CSC).

\bibliographystyle{elsarticle-harv}
\bibliography{references}

\pagebreak
\appendix

\counterwithin{figure}{section}

\section{Figure}

\begin{figure*}[!ht]
\centering

\subfigure[Estimated with full observations.]{
    \centering
    \includegraphics[width=0.42
    \textwidth]{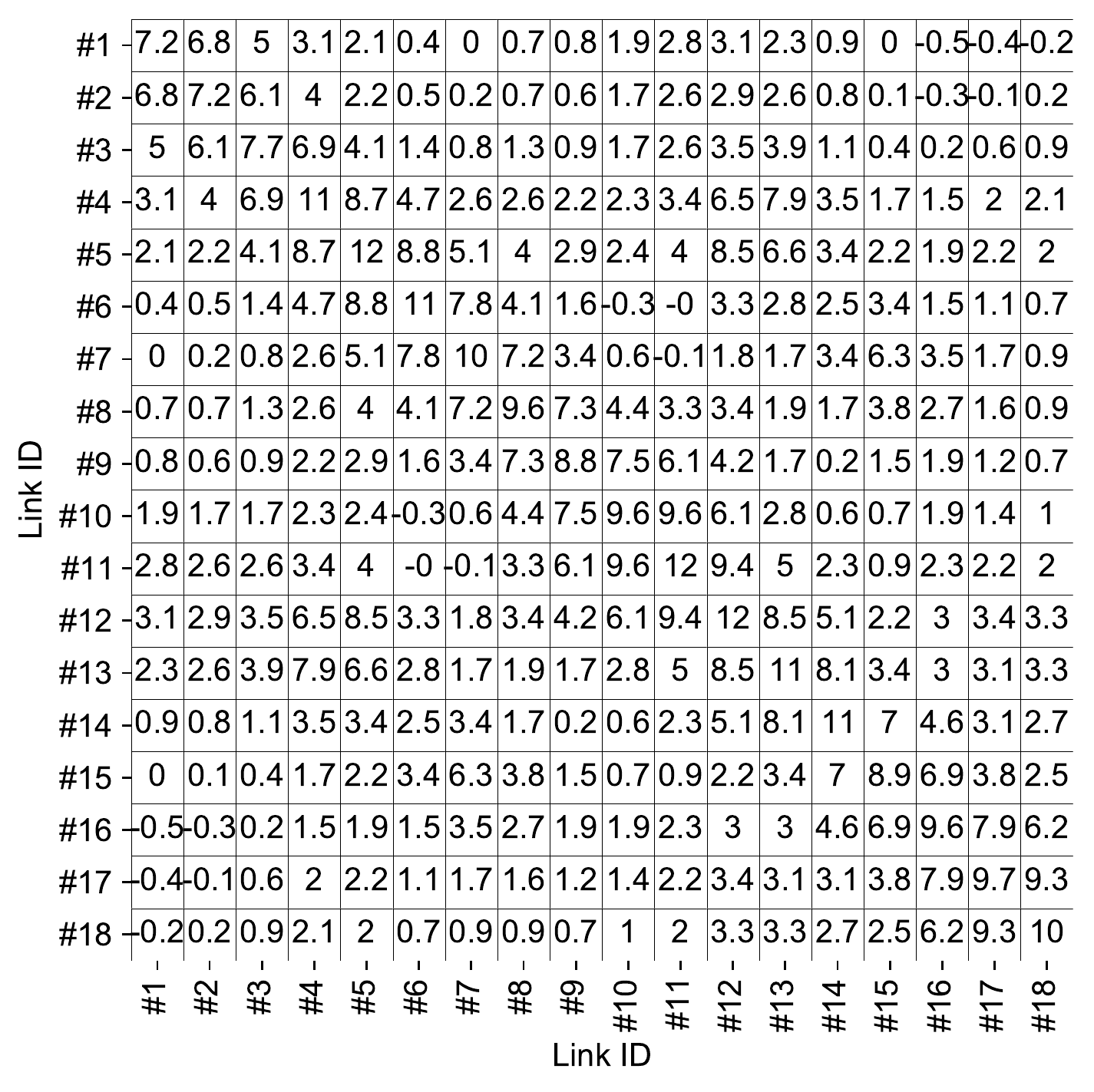}
}
\subfigure[Estimated with full, missing observations.]{
    \centering
    \includegraphics[width=0.42
    \textwidth]{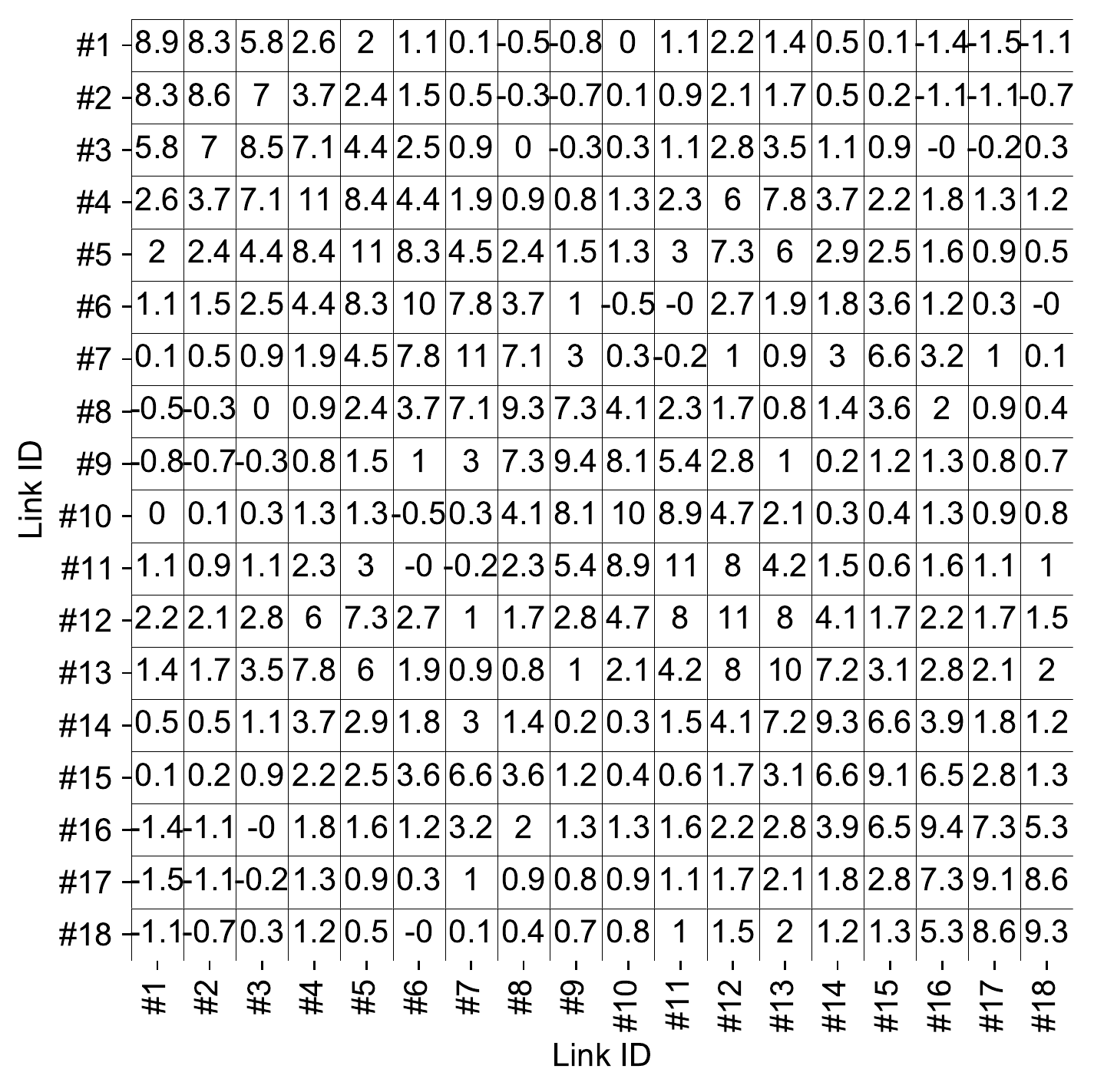}
}
\subfigure[Estimated with all observations.]{
    \centering
    \includegraphics[width=0.42
    \textwidth]{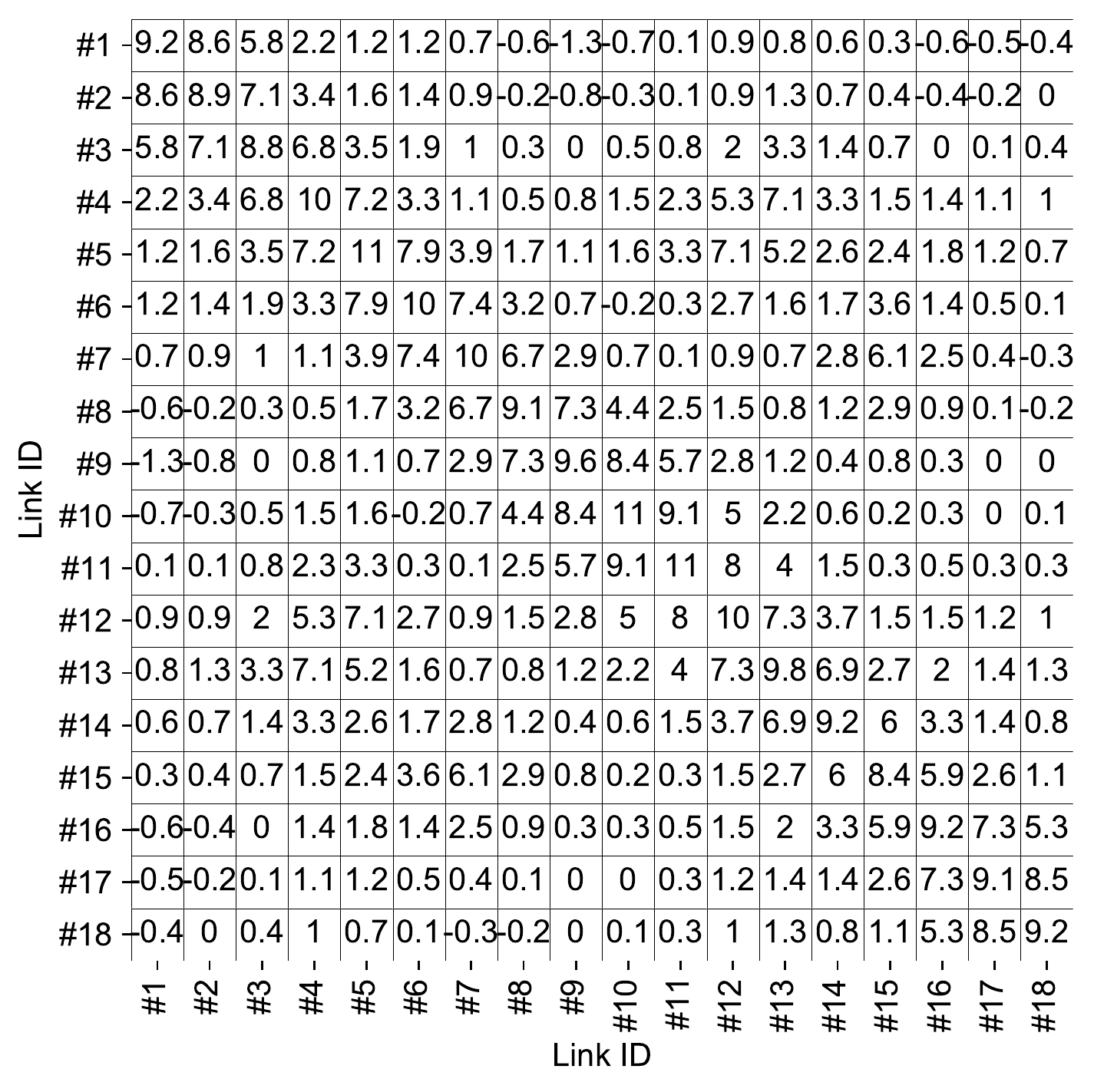}
}
\caption{Posterior mean of covariance matrices estimated with different types of observations.}
\label{cov_estimation}
\end{figure*}

\begin{figure}[htp]
\centering

\subfigure[Travel time on links \#5 and \#6.]{
    \centering
    \includegraphics[scale=0.33]{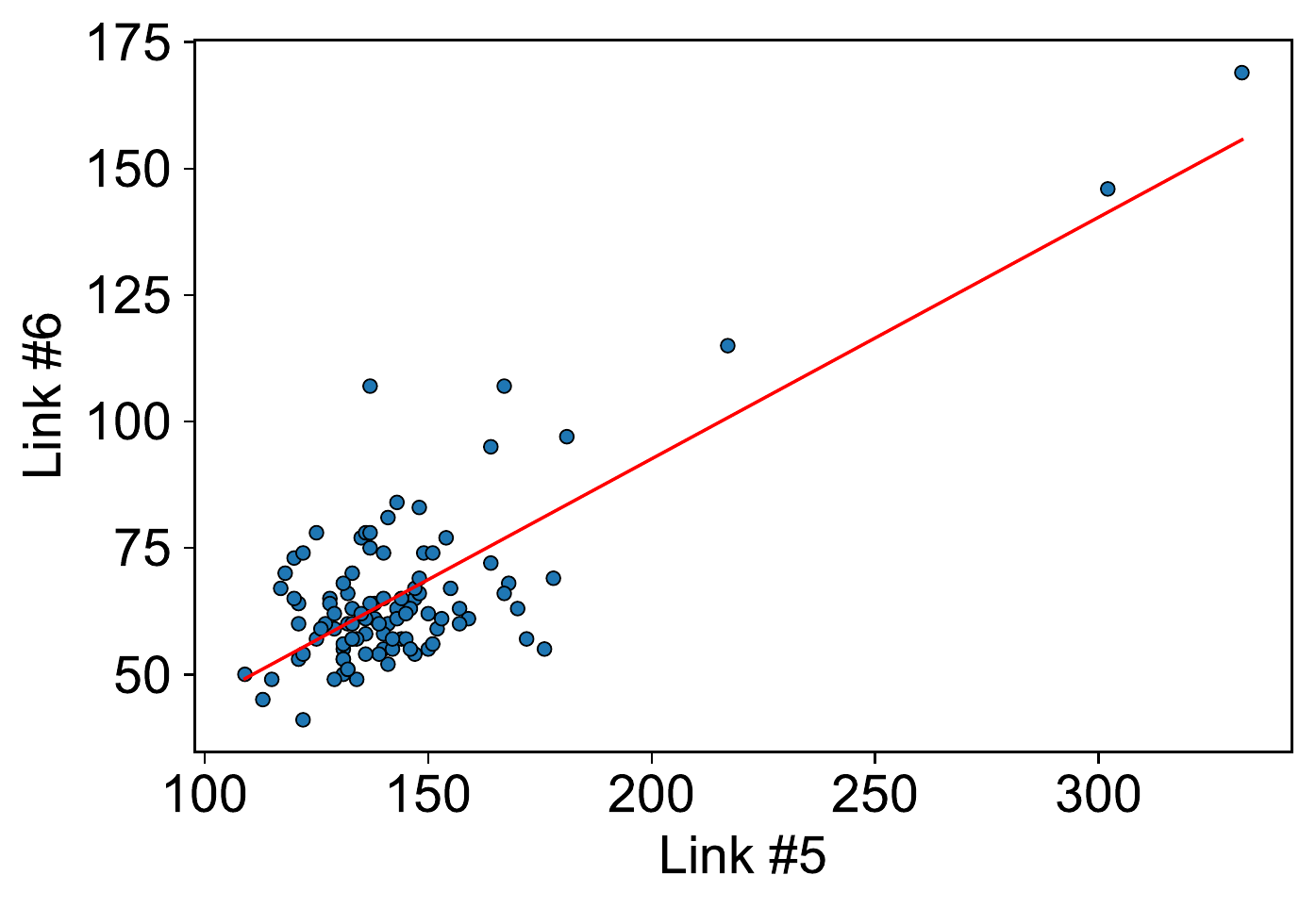}
}
\subfigure[Travel time on links \#6 and \#7.]{
    \centering
    \includegraphics[scale=0.33]{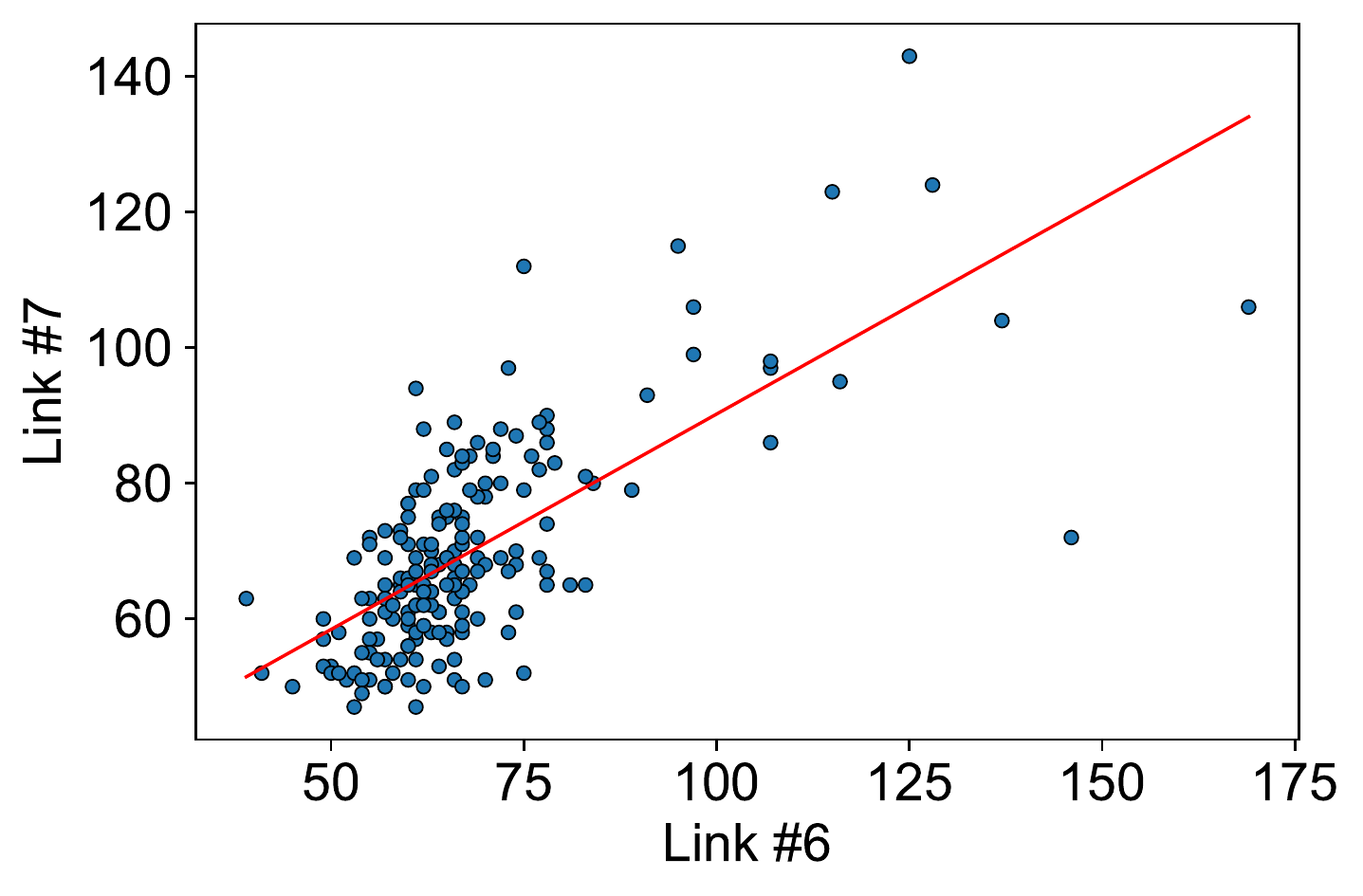}
}
\subfigure[Travel time on links \#7 and \#8.]{
    \centering
    \includegraphics[scale=0.33]{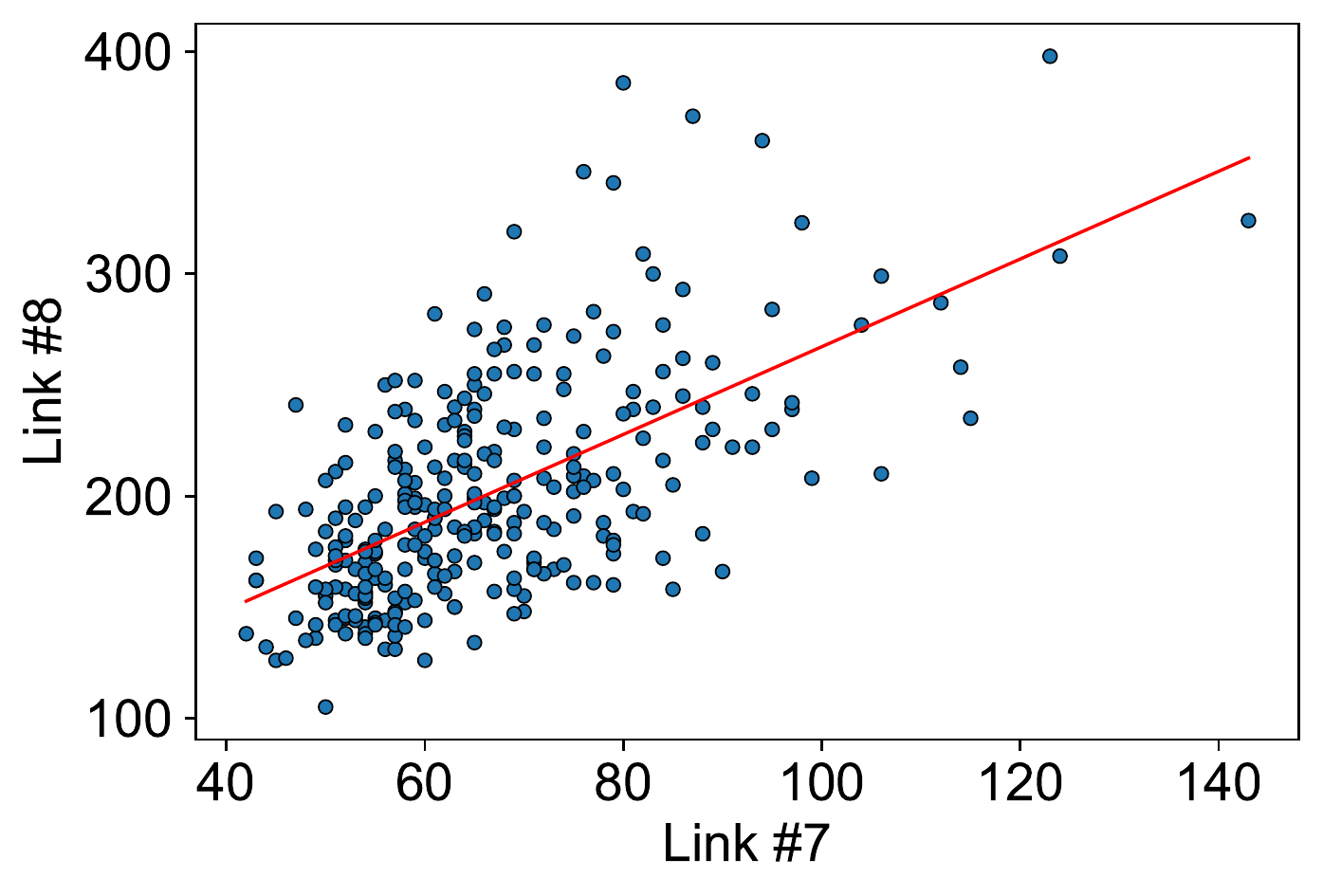}
}
\subfigure[Travel time on links \#12 and \#13.]{
    \centering
    \includegraphics[scale=0.33]{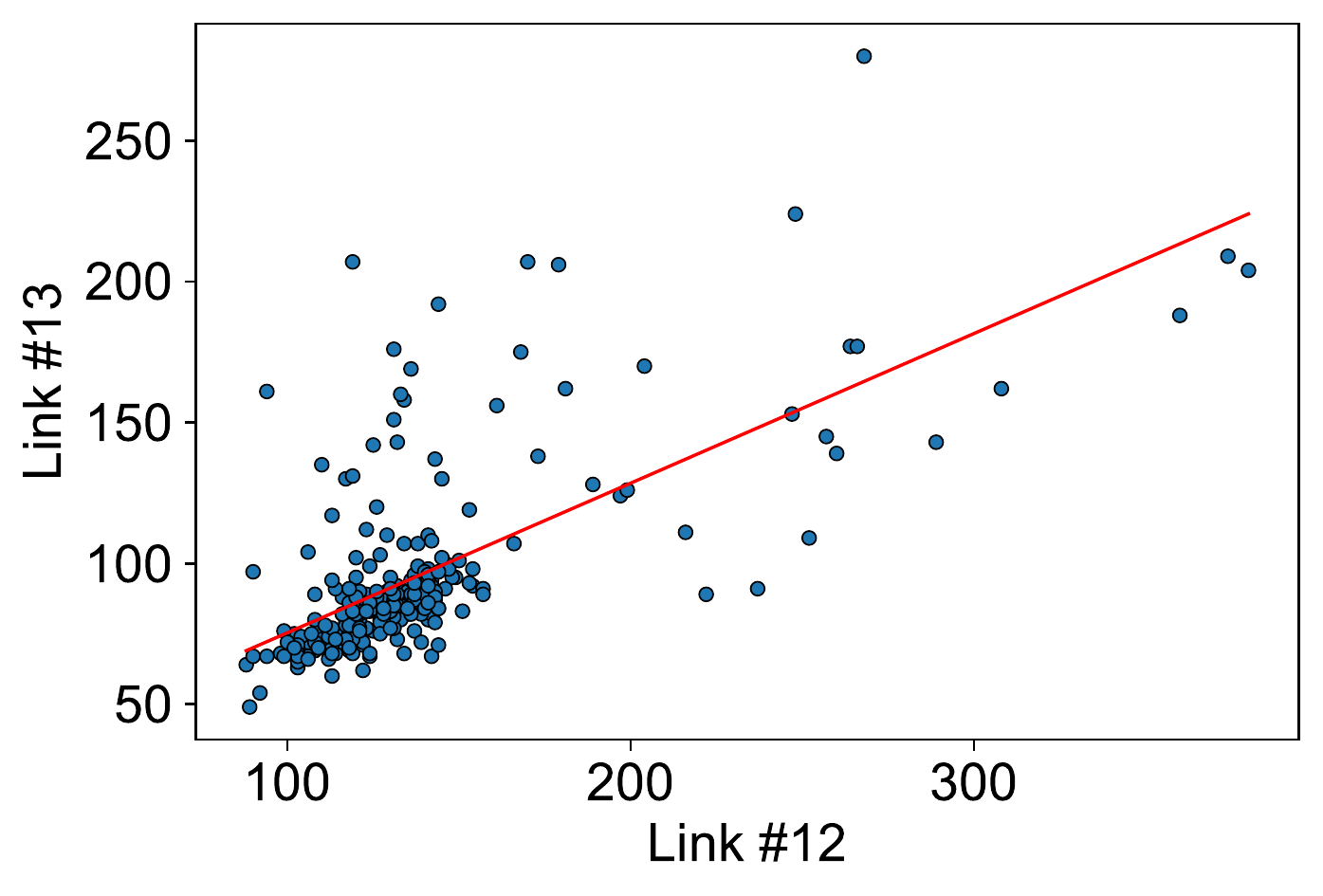}
}
\subfigure[Travel time on links \#13 and \#14.]{
    \centering
    \includegraphics[scale=0.33]{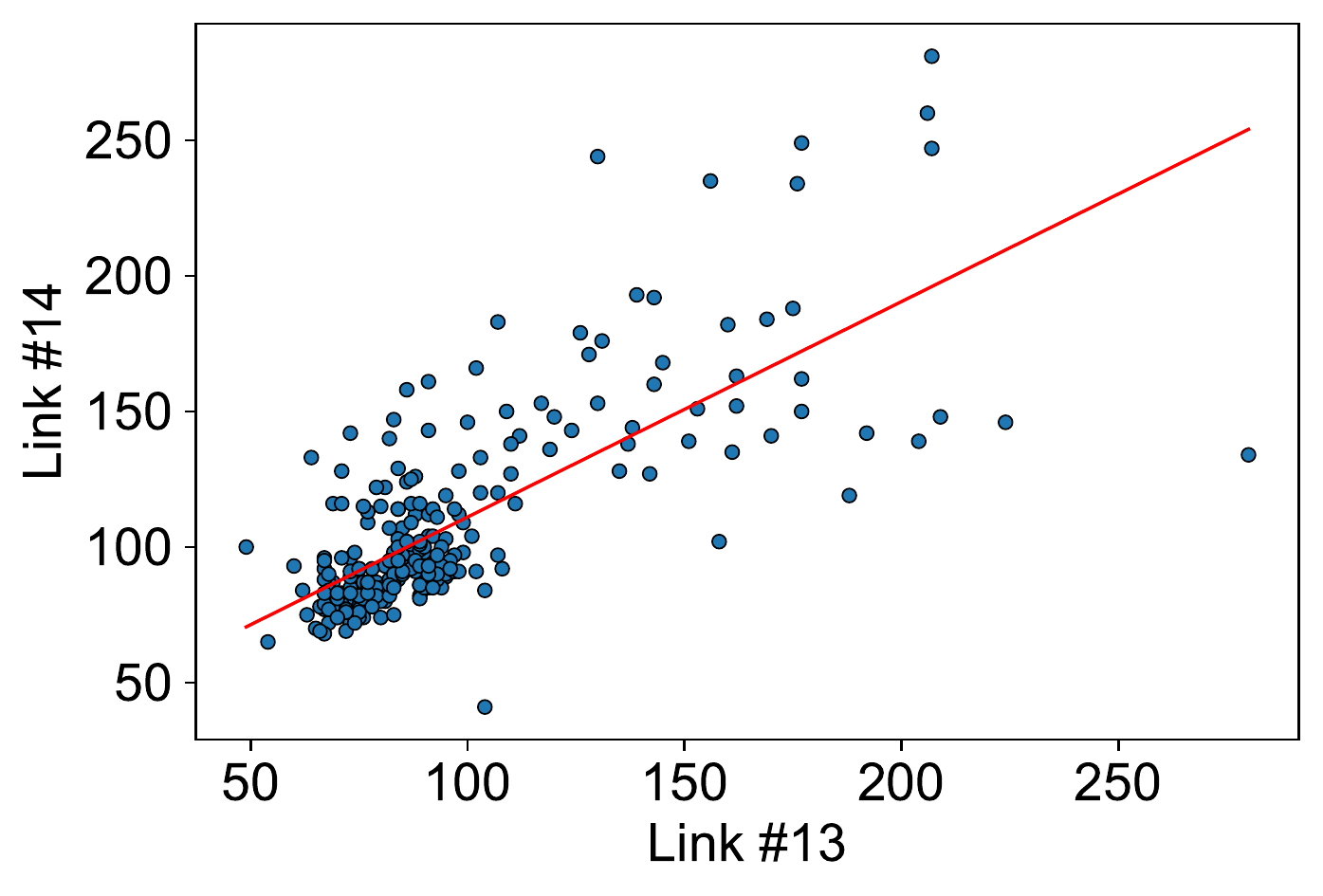}
}
\subfigure[Travel time on links \#13 and \#15.]{
    \centering
    \includegraphics[scale=0.33]{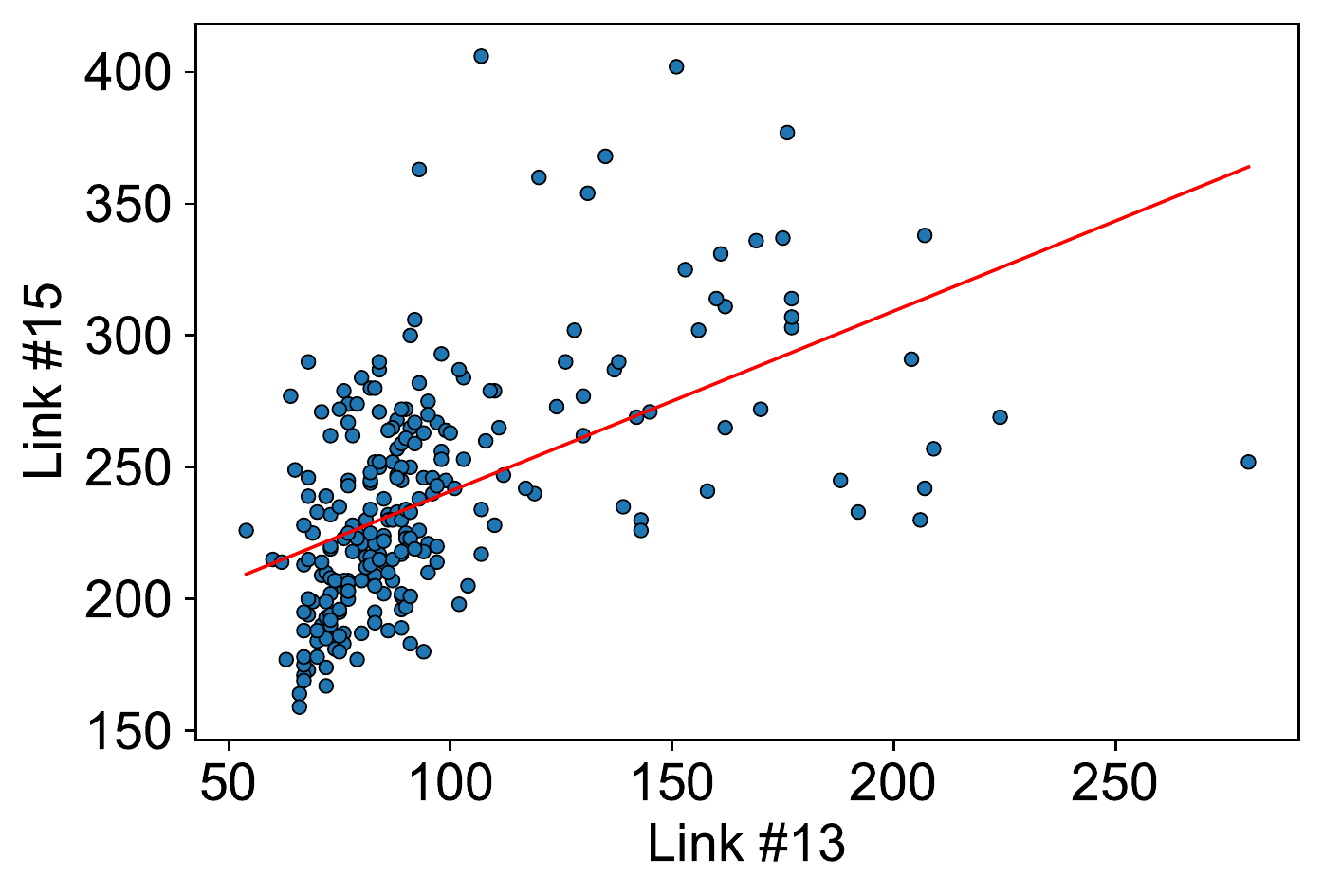}
}
\subfigure[Travel time on links \#14 and \#15.]{
    \centering
    \includegraphics[scale=0.33]{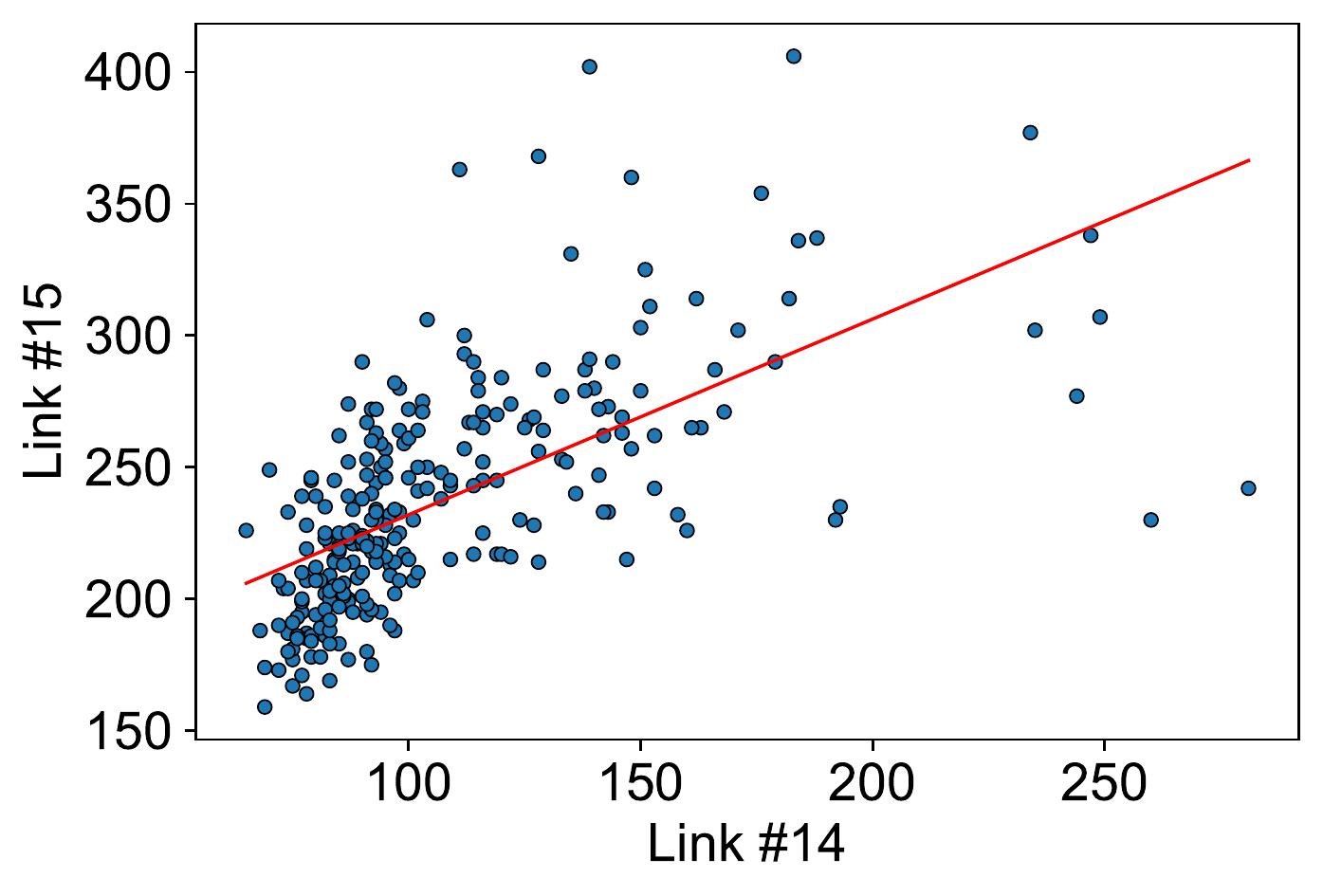}
}
\subfigure[Travel time on links \#19 and \#20.]{
    \centering
    \includegraphics[scale=0.33]{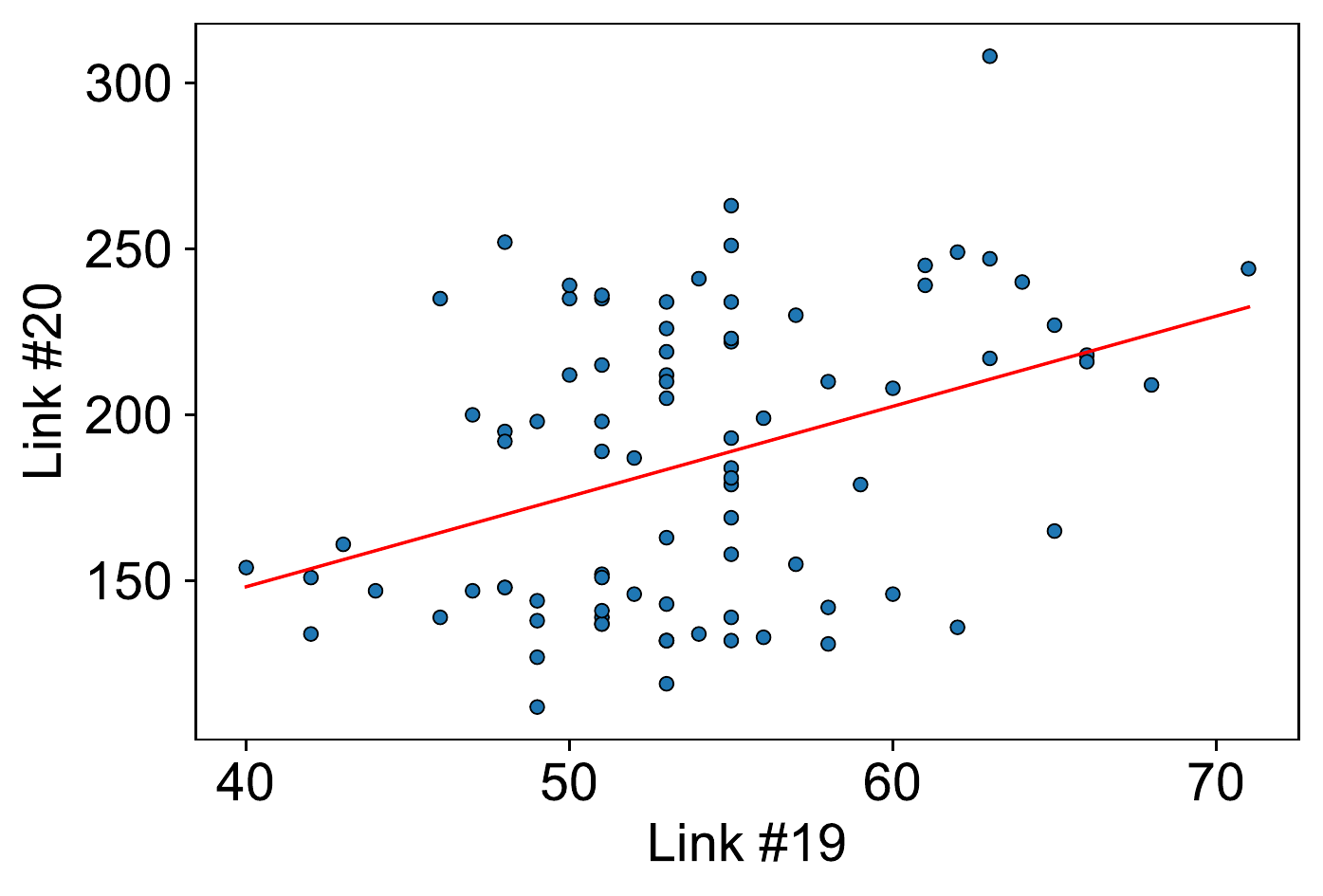}
}
\caption{The travel time scatter plots of some strong correlated link pairs during morning peak.}
\label{Scatter_Morning}
\end{figure}

\begin{figure*}[!ht]
\centering

\subfigure[Travel time on links \#5 and \#6.]{
    \centering
    \includegraphics[scale=0.33]{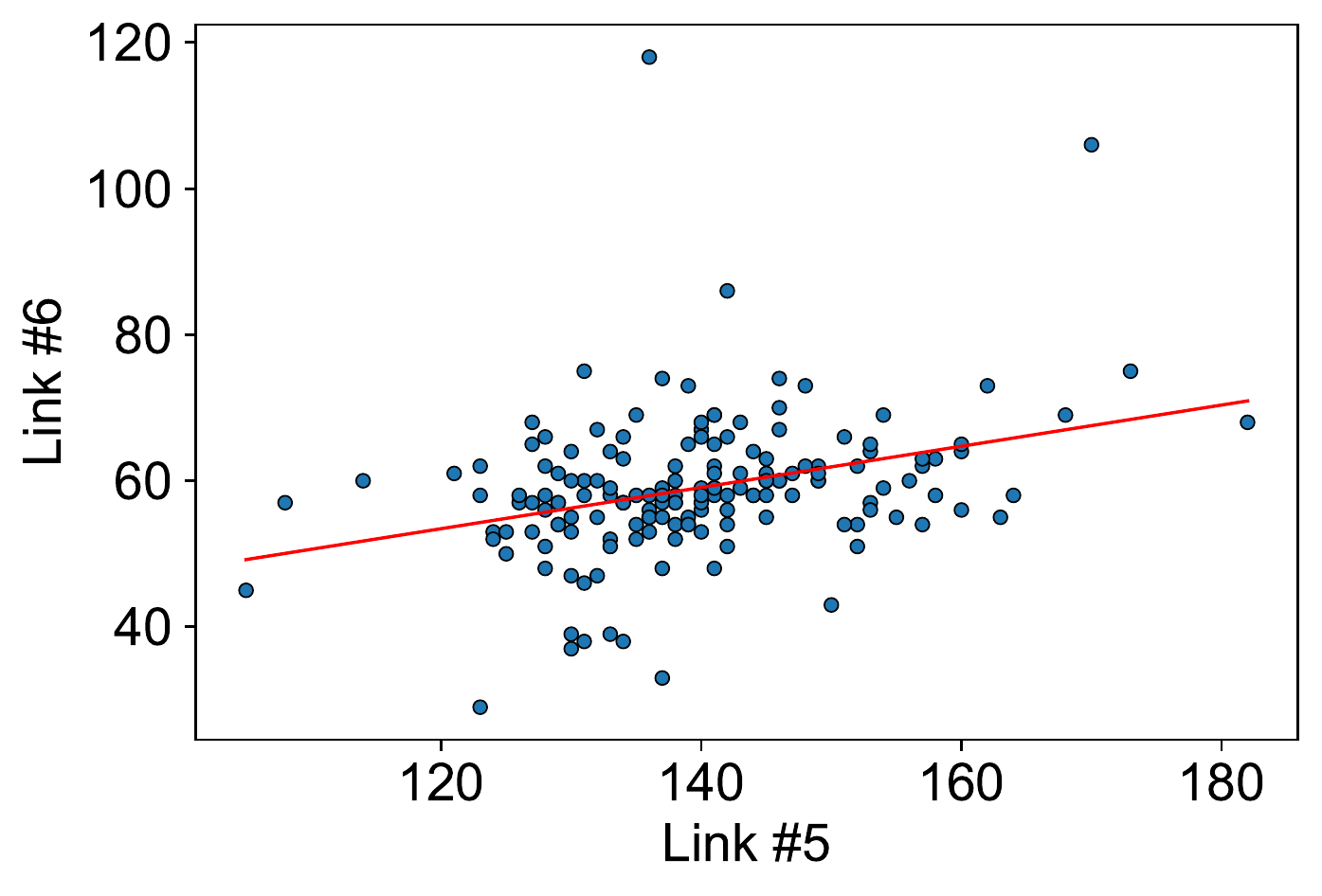}
}
\subfigure[Travel time on links \#5 and \#13.]{
    \centering
    \includegraphics[scale=0.33]{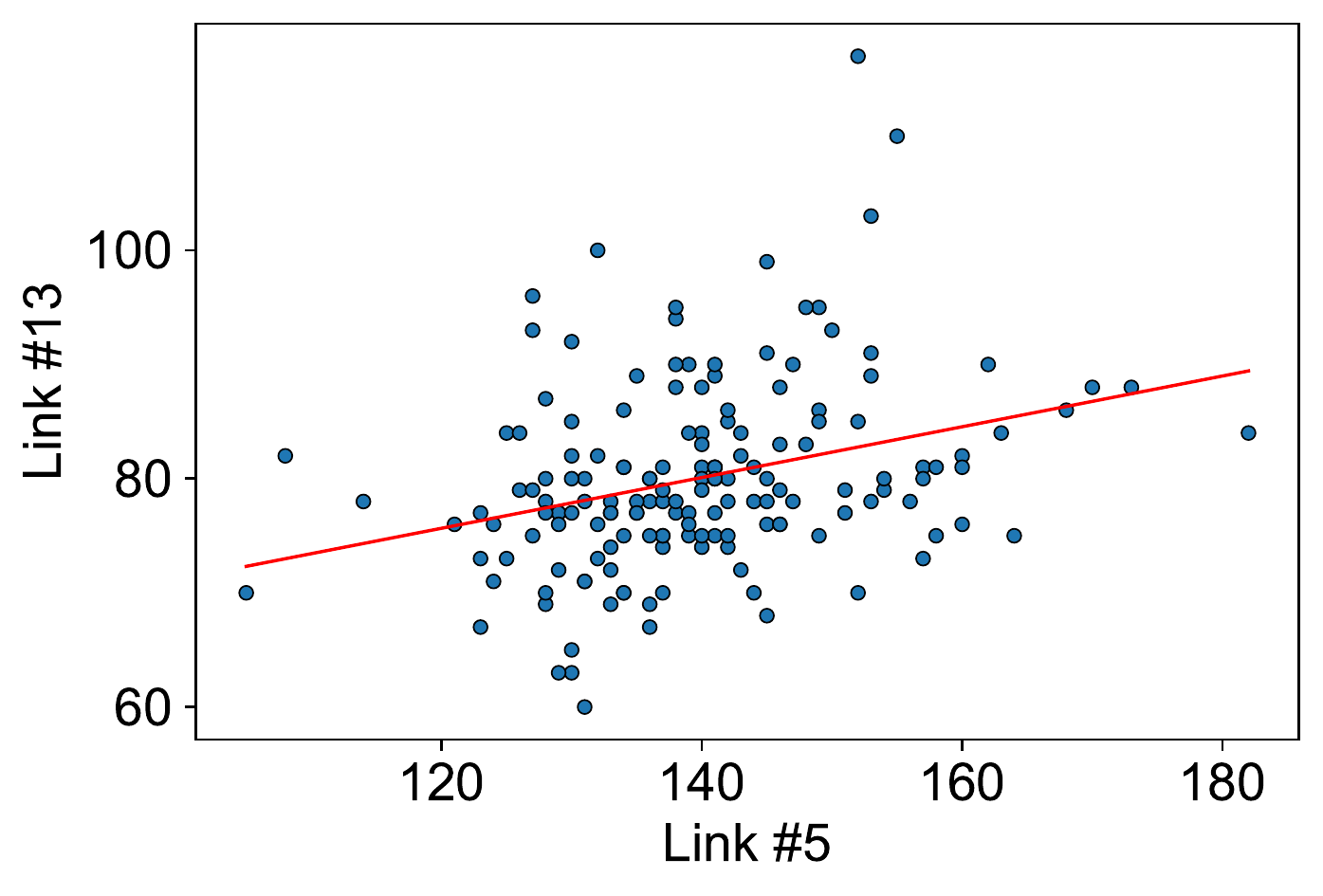}
}
\subfigure[Travel time on links \#5 and \#14.]{
    \centering
    \includegraphics[scale=0.33]{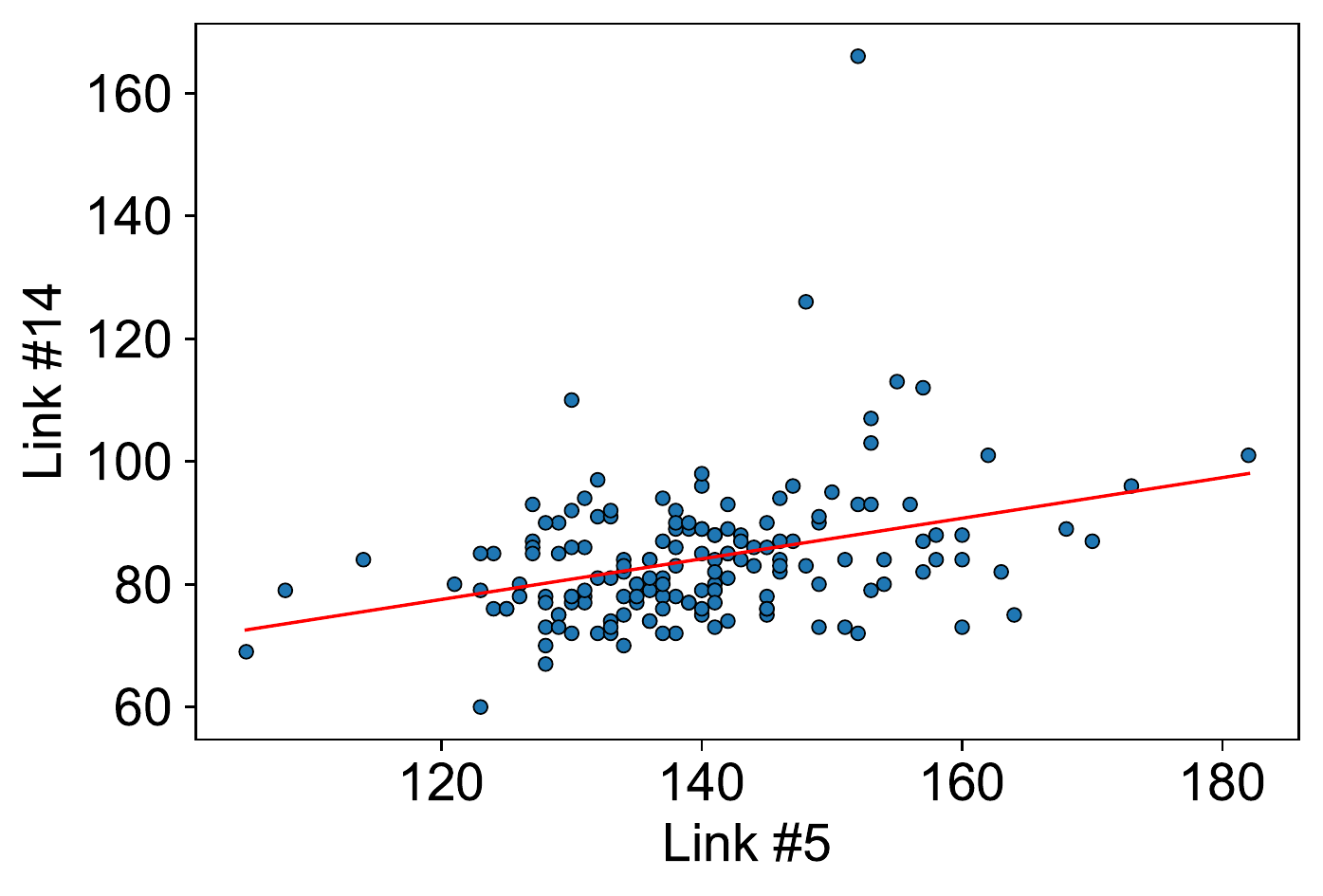}
}
\subfigure[Travel time on links \#6 and \#7.]{
    \centering
    \includegraphics[scale=0.33]{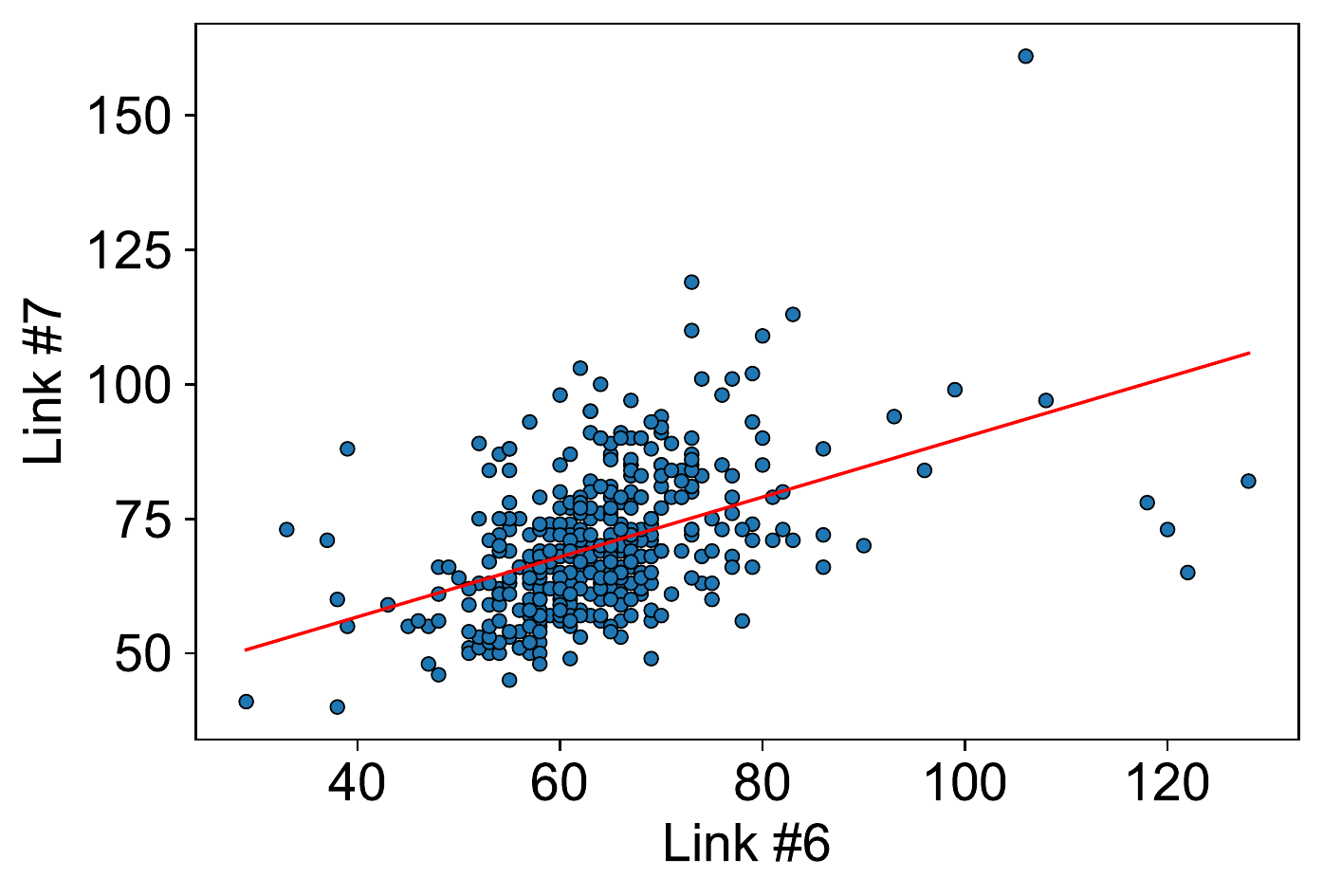}
}
\subfigure[Travel time on links \#11 and \#12.]{
    \centering
    \includegraphics[scale=0.33]{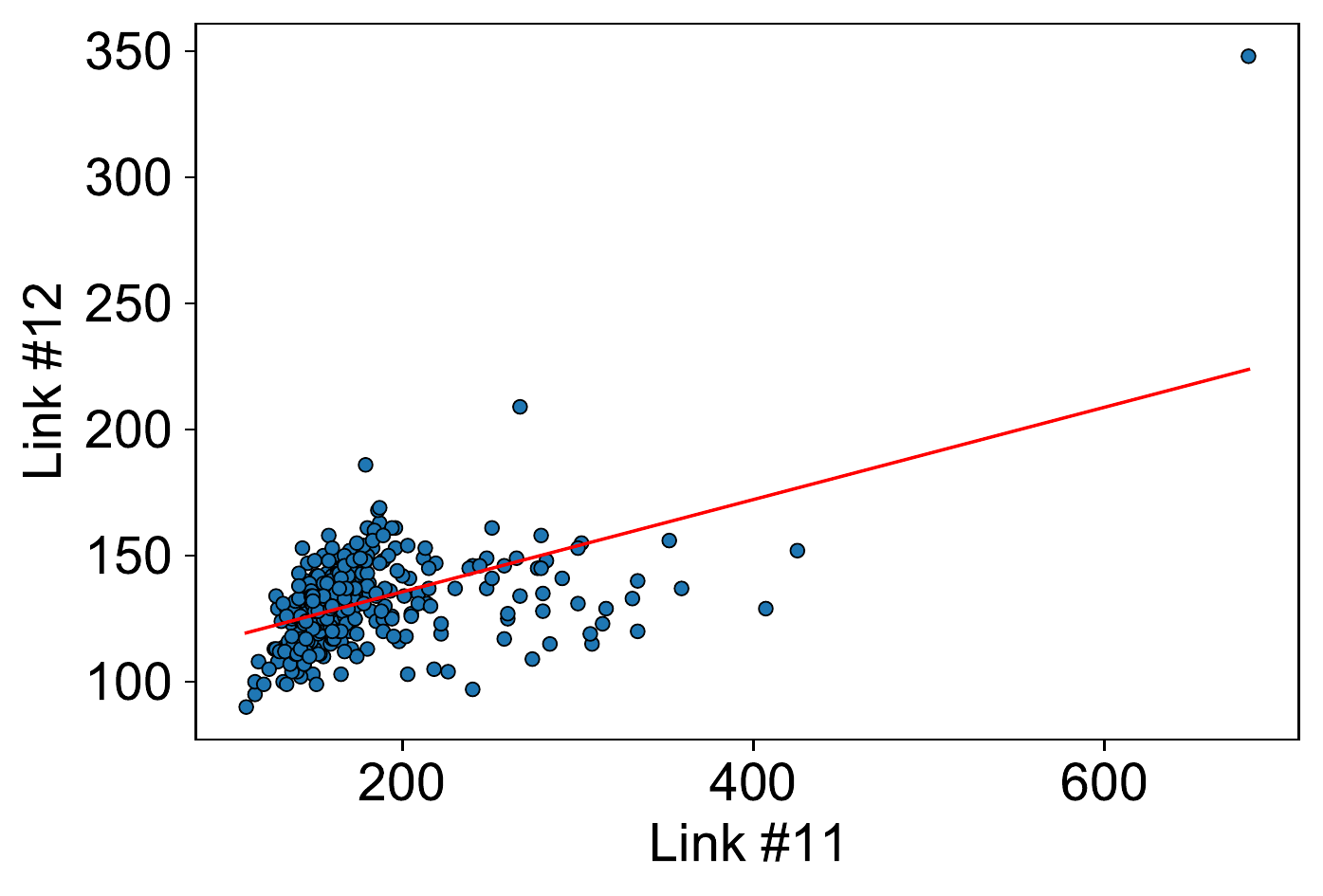}
}
\subfigure[Travel time on links \#12 and \#14.]{
    \centering
    \includegraphics[scale=0.33]{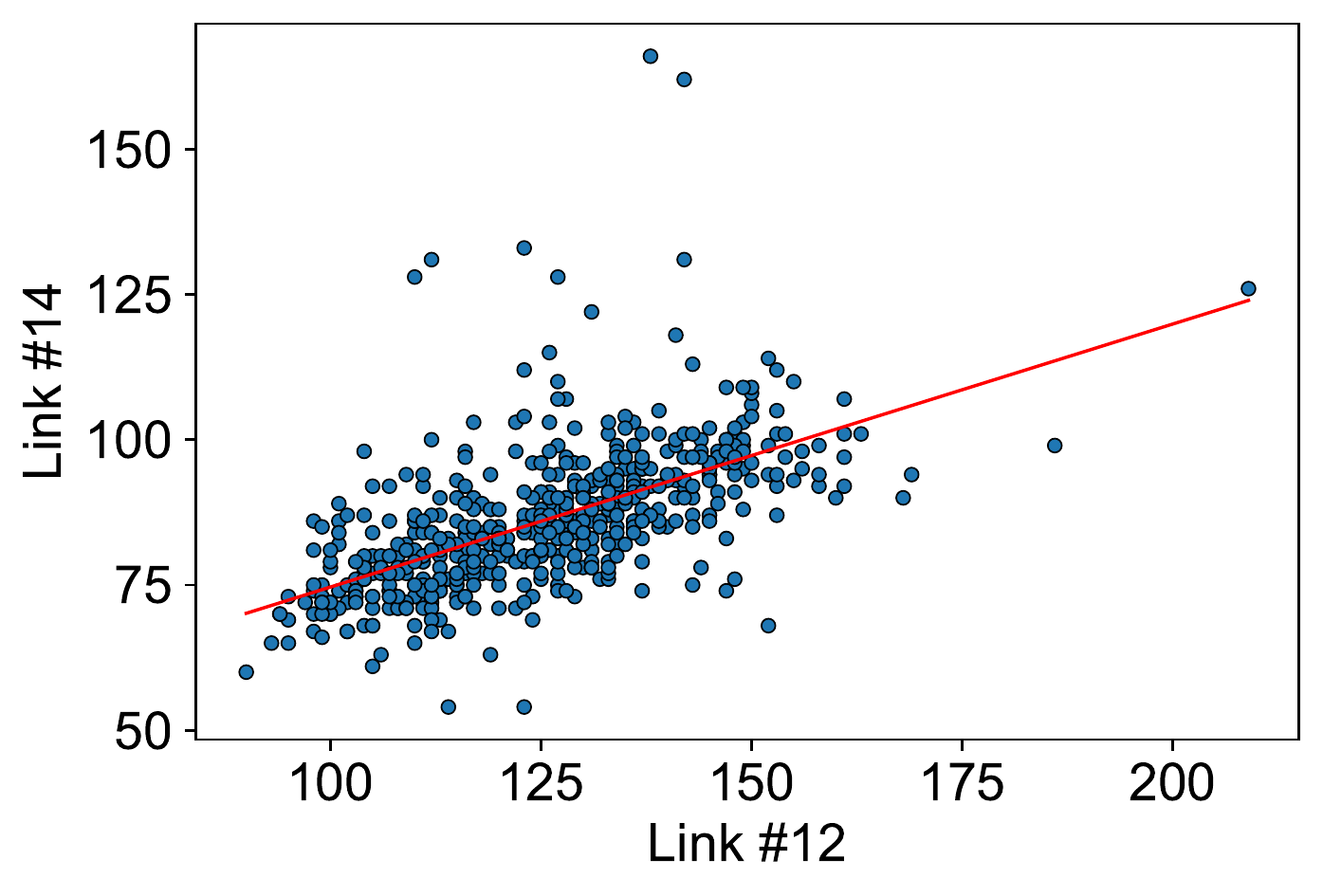}
}
\subfigure[Travel time on links \#13 and \#14.]{
    \centering
    \includegraphics[scale=0.33]{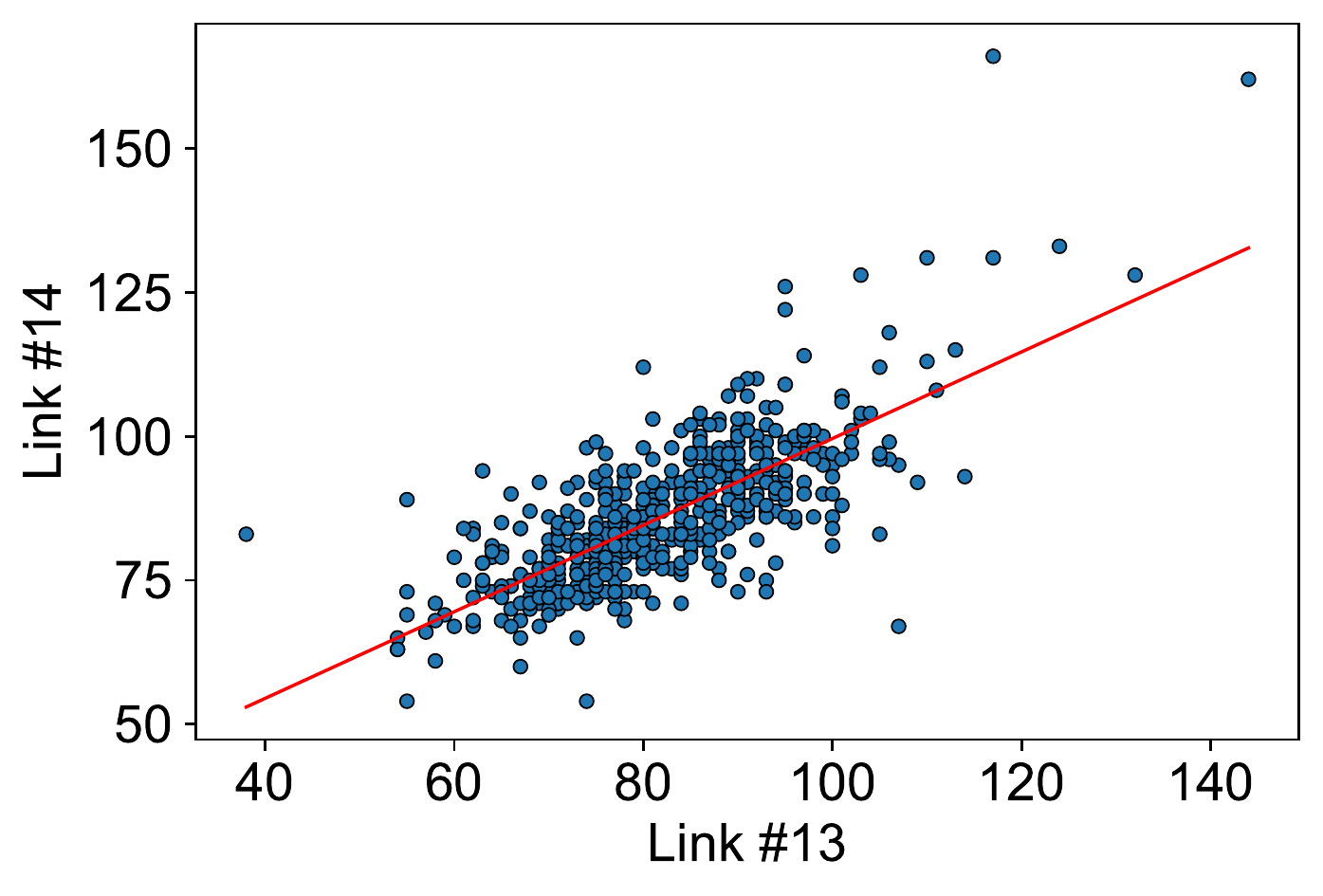}
}
\subfigure[Travel time on links \#15 and \#16.]{
    \centering
    \includegraphics[scale=0.33]{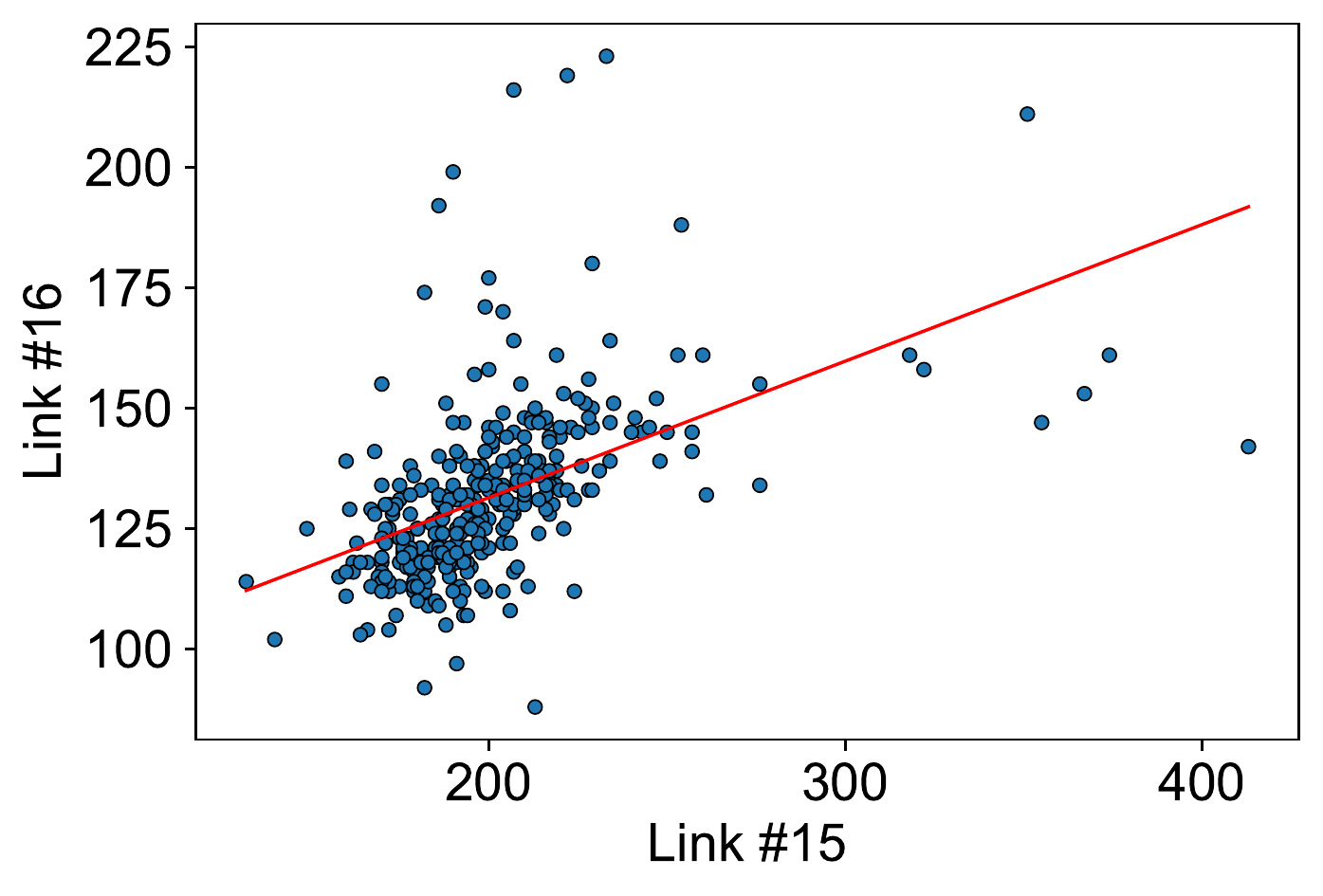}
}
\subfigure[Travel time on links \#16 and \#17.]{
    \centering
    \includegraphics[scale=0.33]{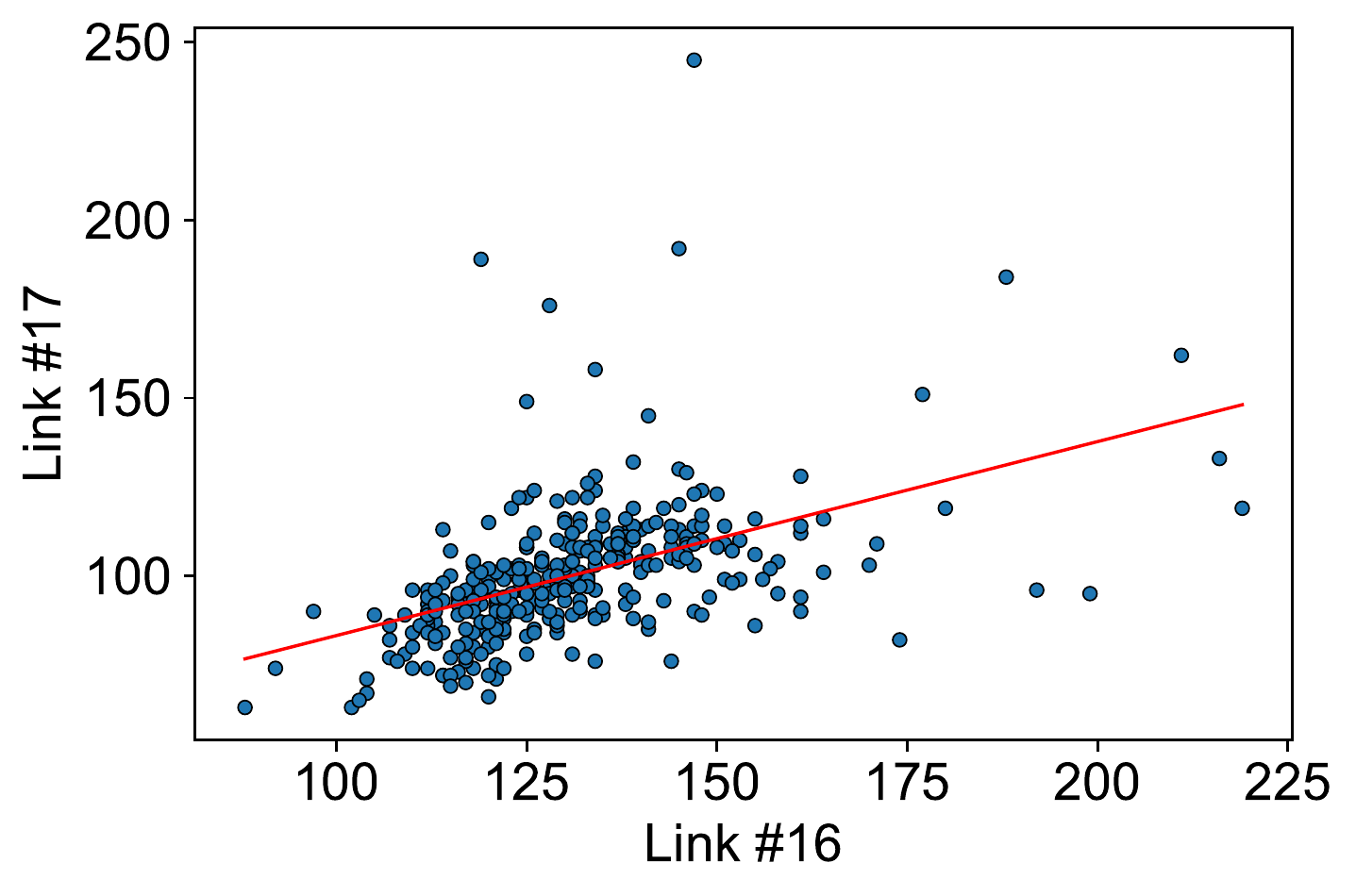}
}
\caption{The travel time scatter plots of some strong correlated link pairs during normal period.}
\label{Scatter_Normal}
\end{figure*}

\begin{figure*}[!ht]
\centering

\subfigure[Travel time on links \#5 and \#16.]{
    \centering
    \includegraphics[scale=0.33]{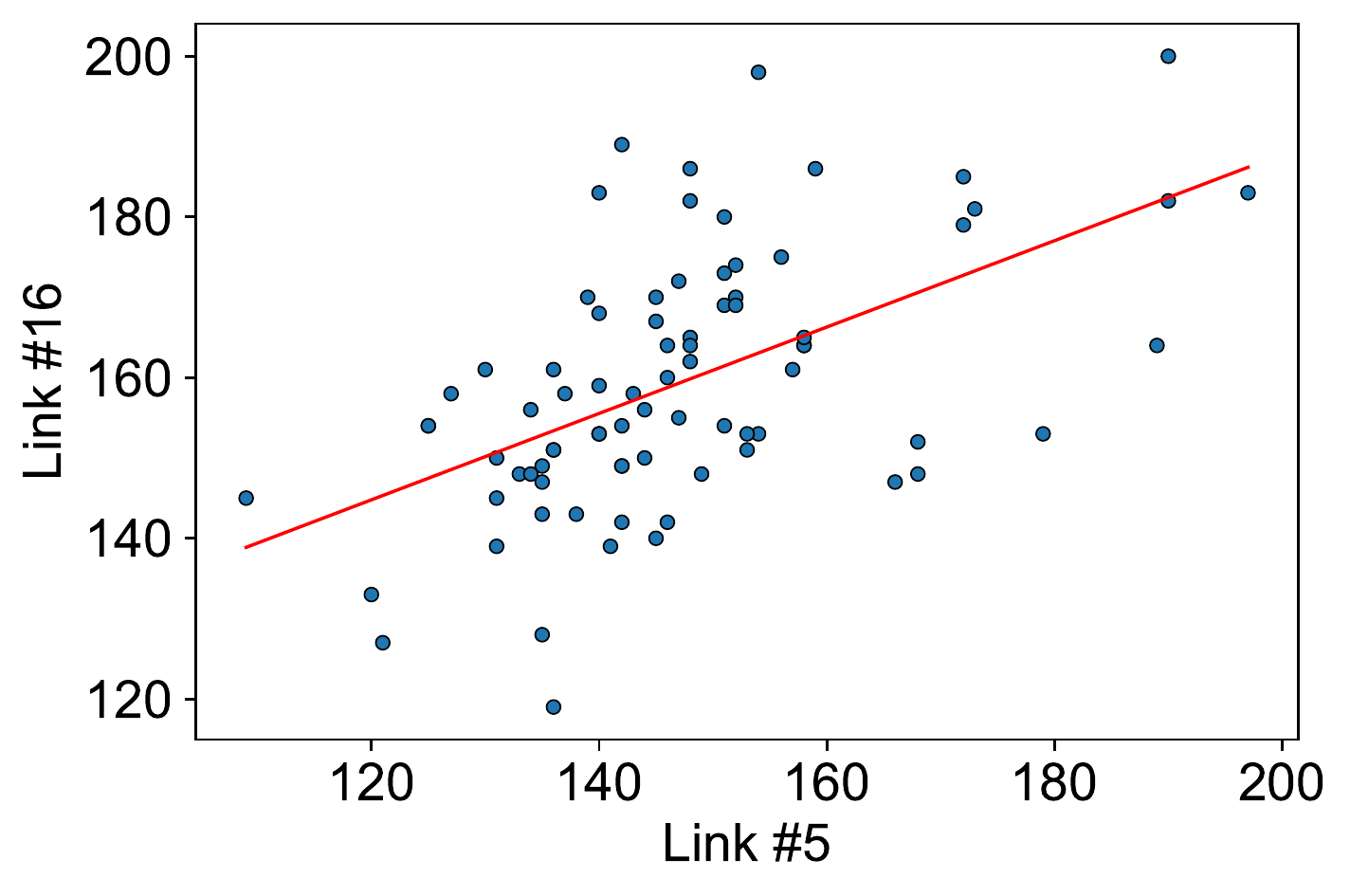}
}
\subfigure[Travel time on links \#11 and \#12.]{
    \centering
    \includegraphics[scale=0.33]{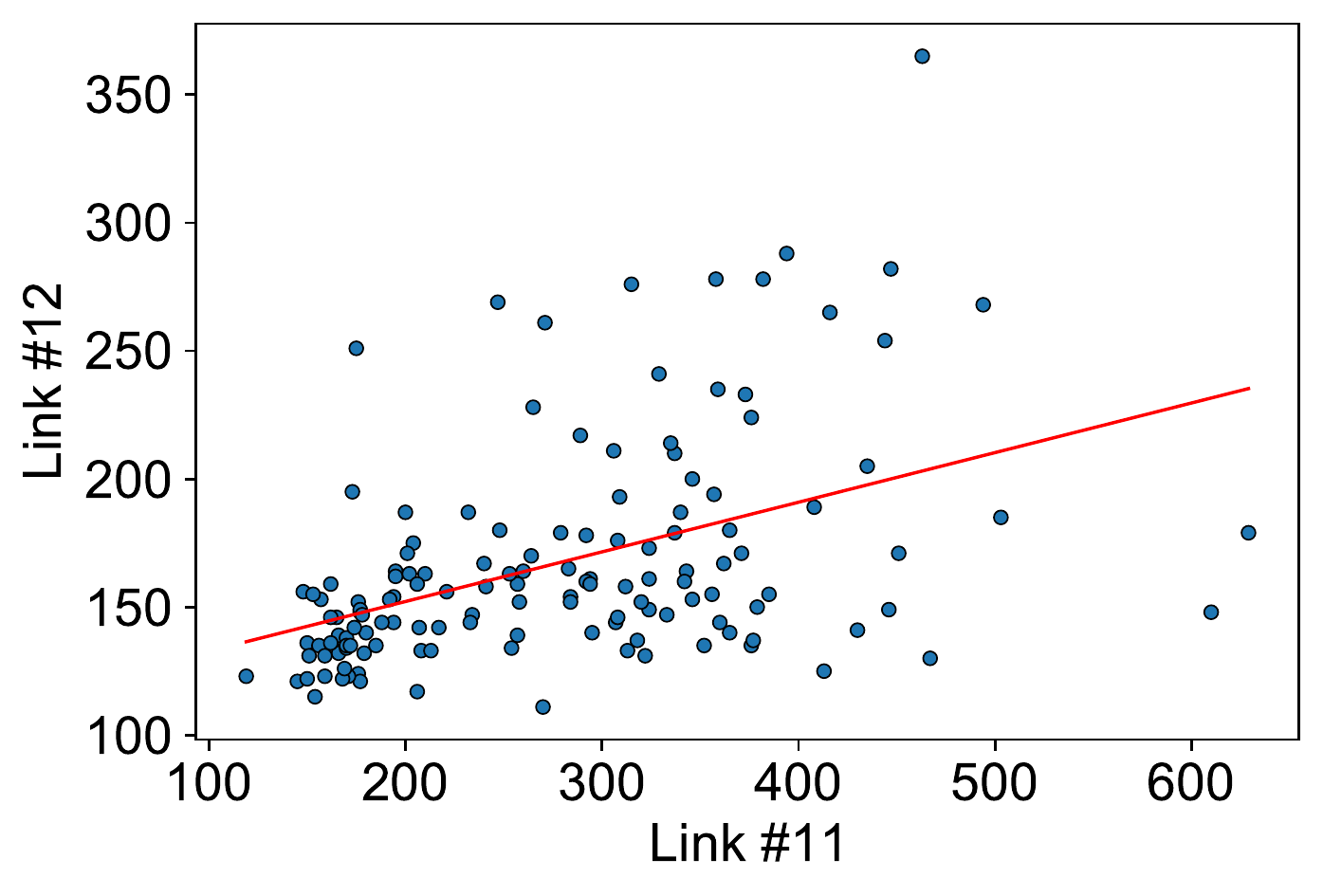}
}
\subfigure[Travel time on links \#11 and \#15.]{
    \centering
    \includegraphics[scale=0.33]{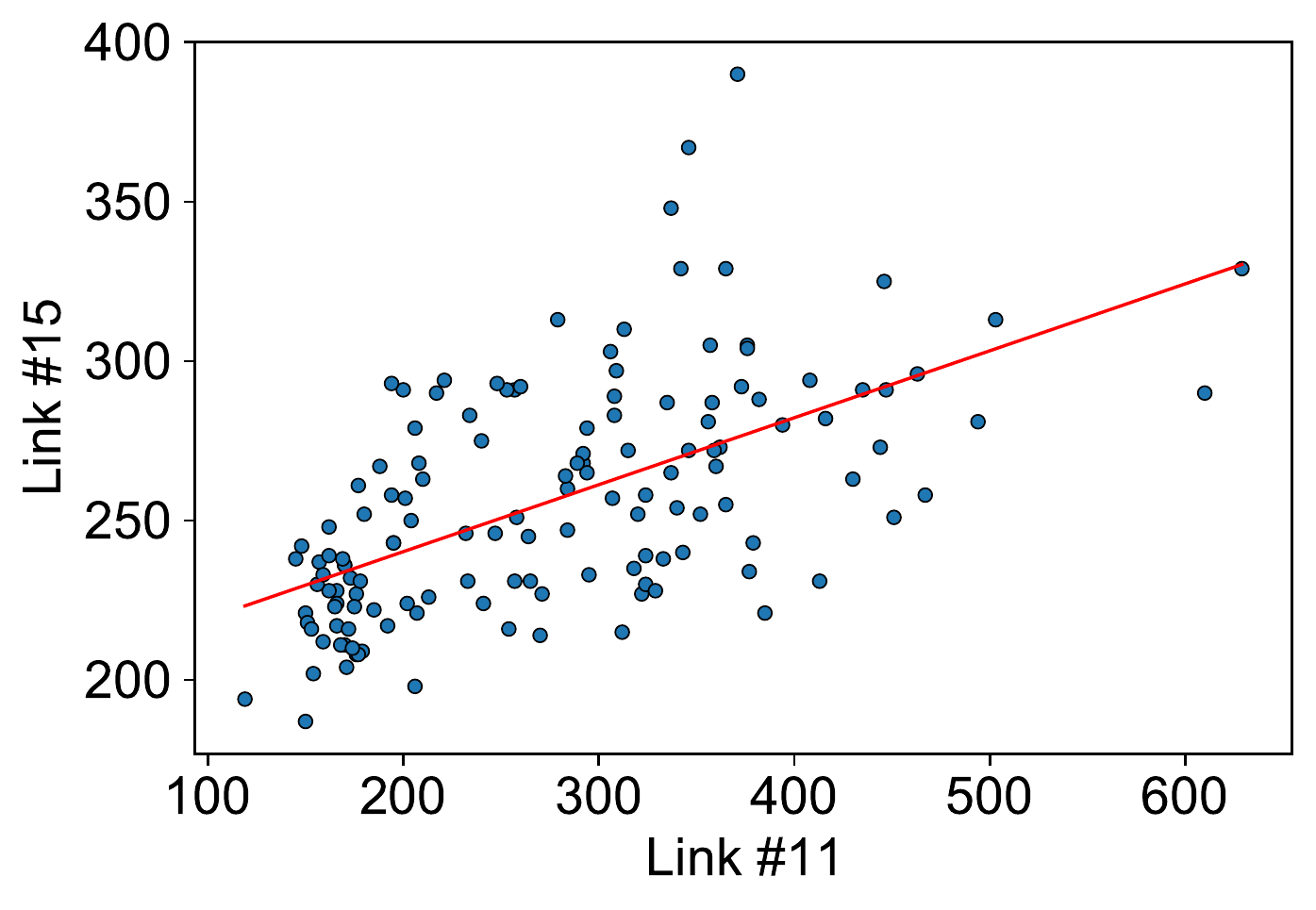}
}
\subfigure[Travel time on links \#13 and \#14.]{
    \centering
    \includegraphics[scale=0.33]{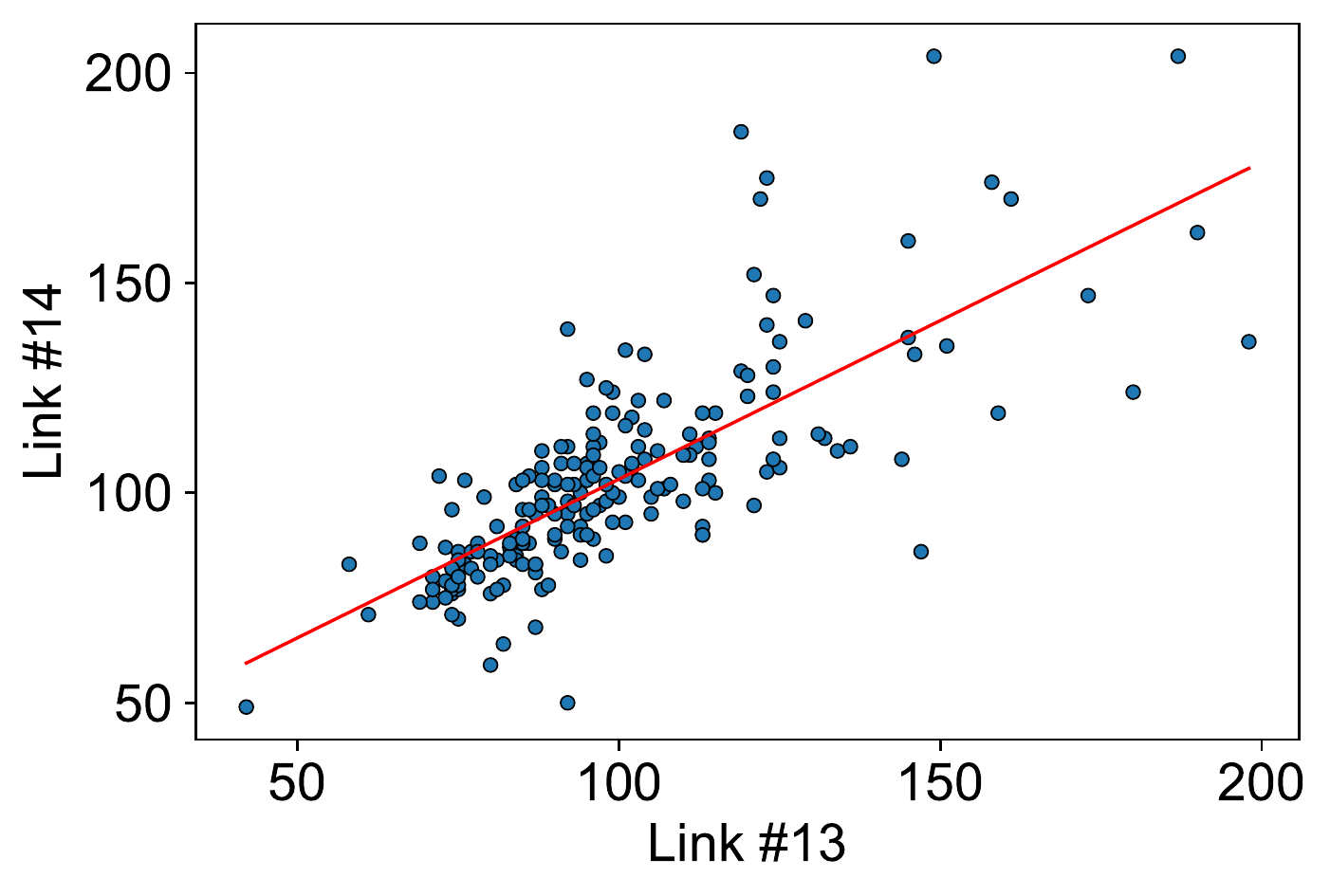}
}
\subfigure[Travel time on links \#13 and \#15.]{
    \centering
    \includegraphics[scale=0.33]{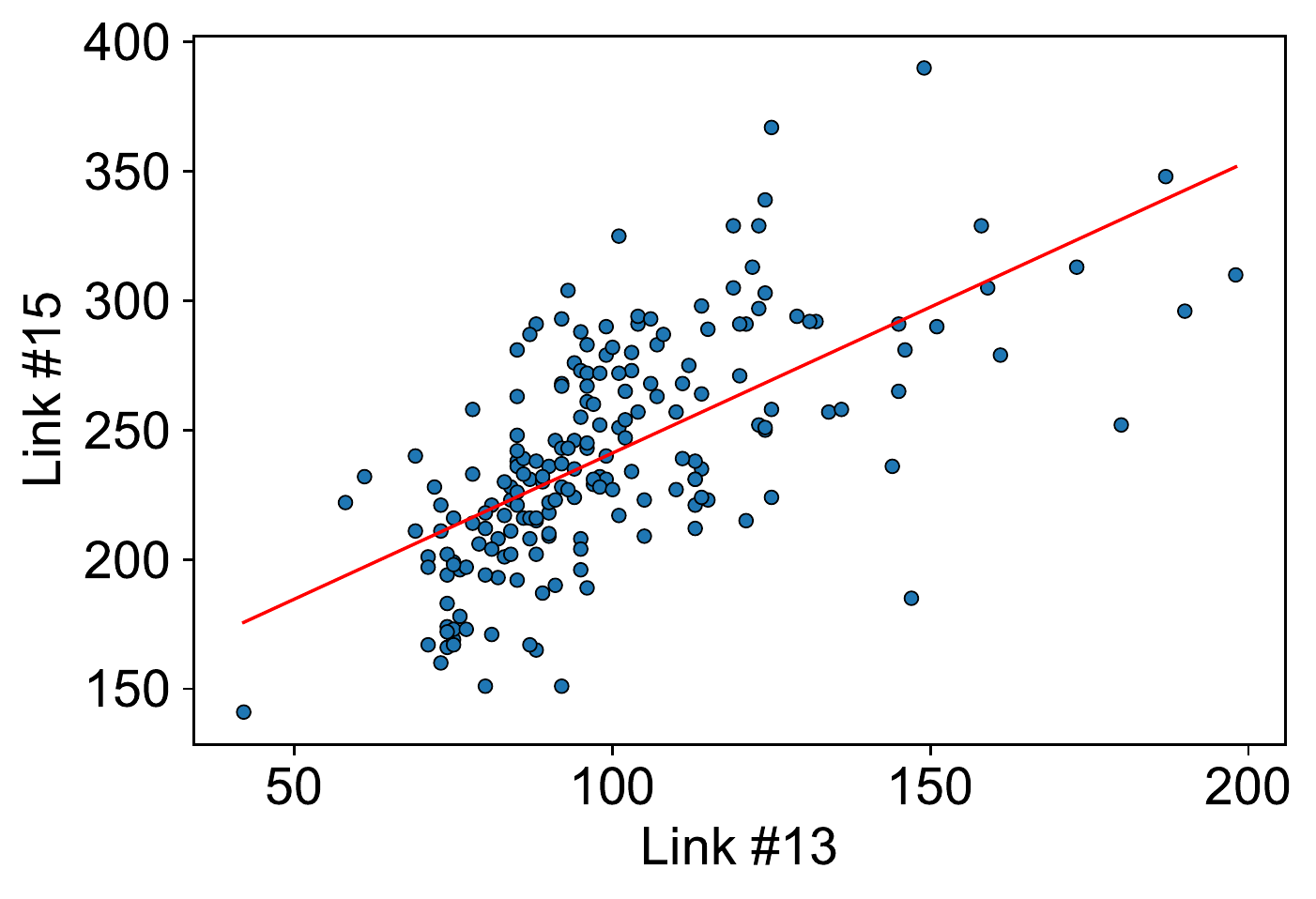}
}
\subfigure[Travel time on links \#13 and \#16.]{
    \centering
    \includegraphics[scale=0.33]{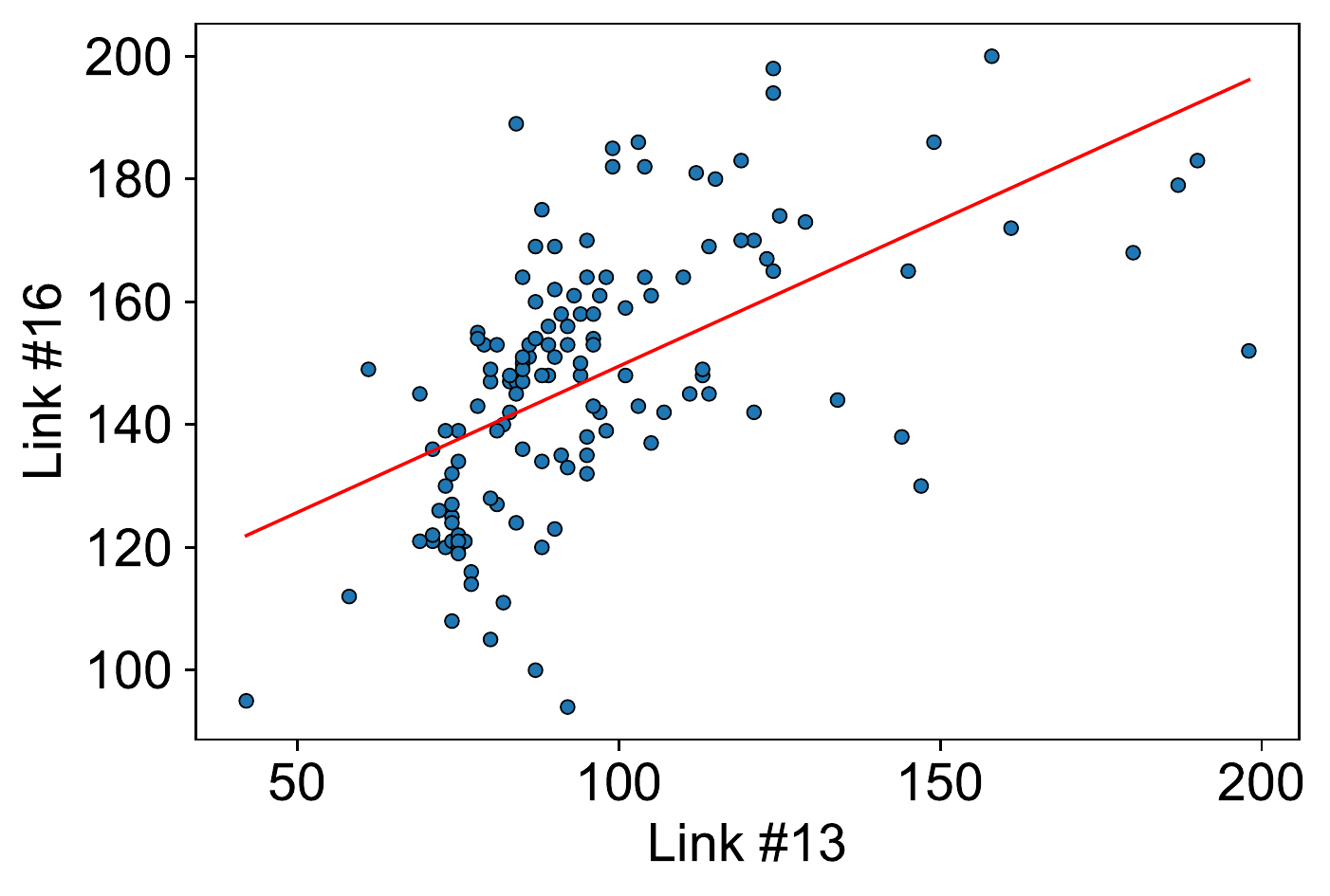}
}
\subfigure[Travel time on links \#14 and \#15.]{
    \centering
    \includegraphics[scale=0.33]{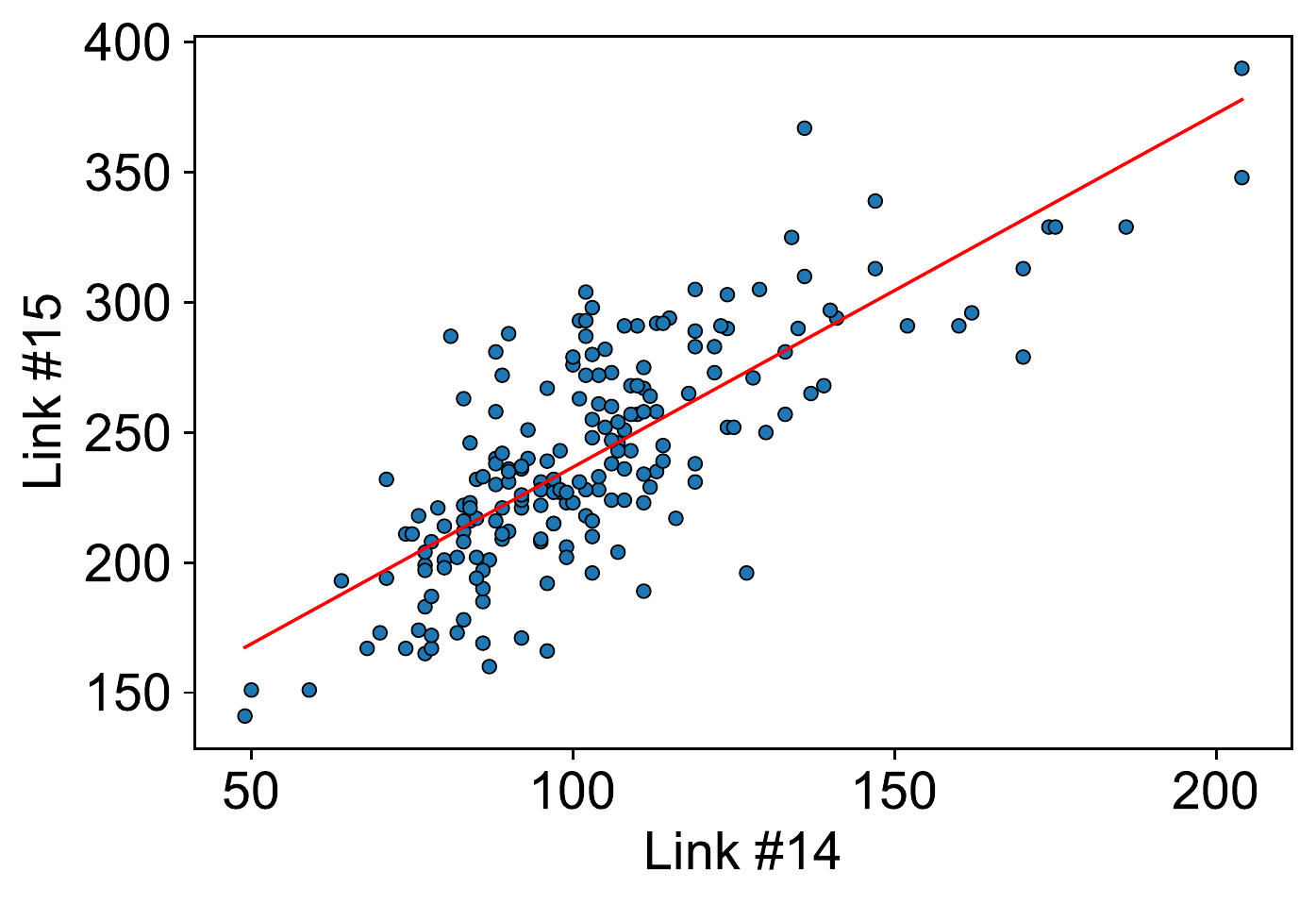}
}
\subfigure[Travel time on links \#14 and \#16.]{
    \centering
    \includegraphics[scale=0.33]{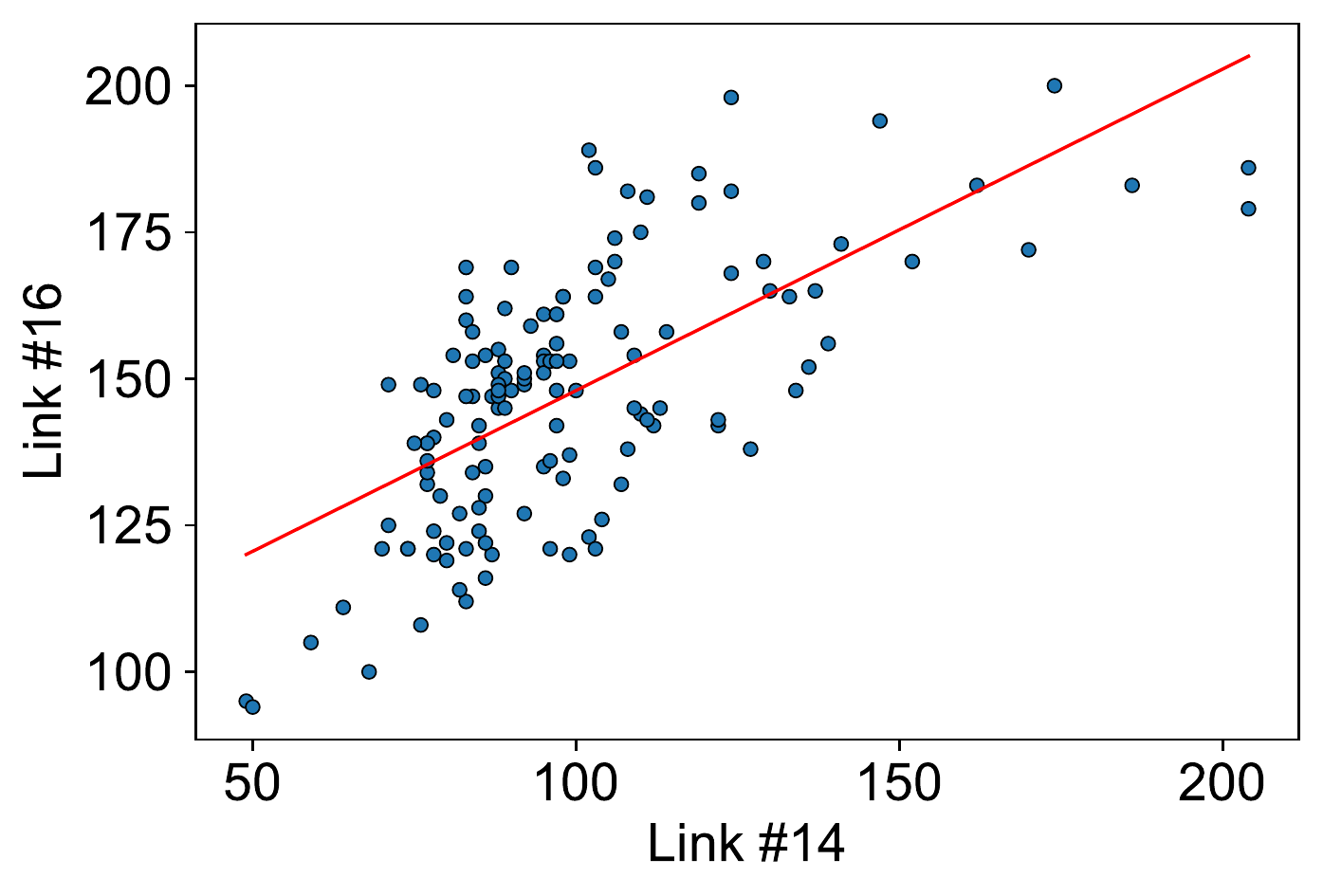}
}
\subfigure[Travel time on links \#15 and \#16.]{
    \centering
    \includegraphics[scale=0.33]{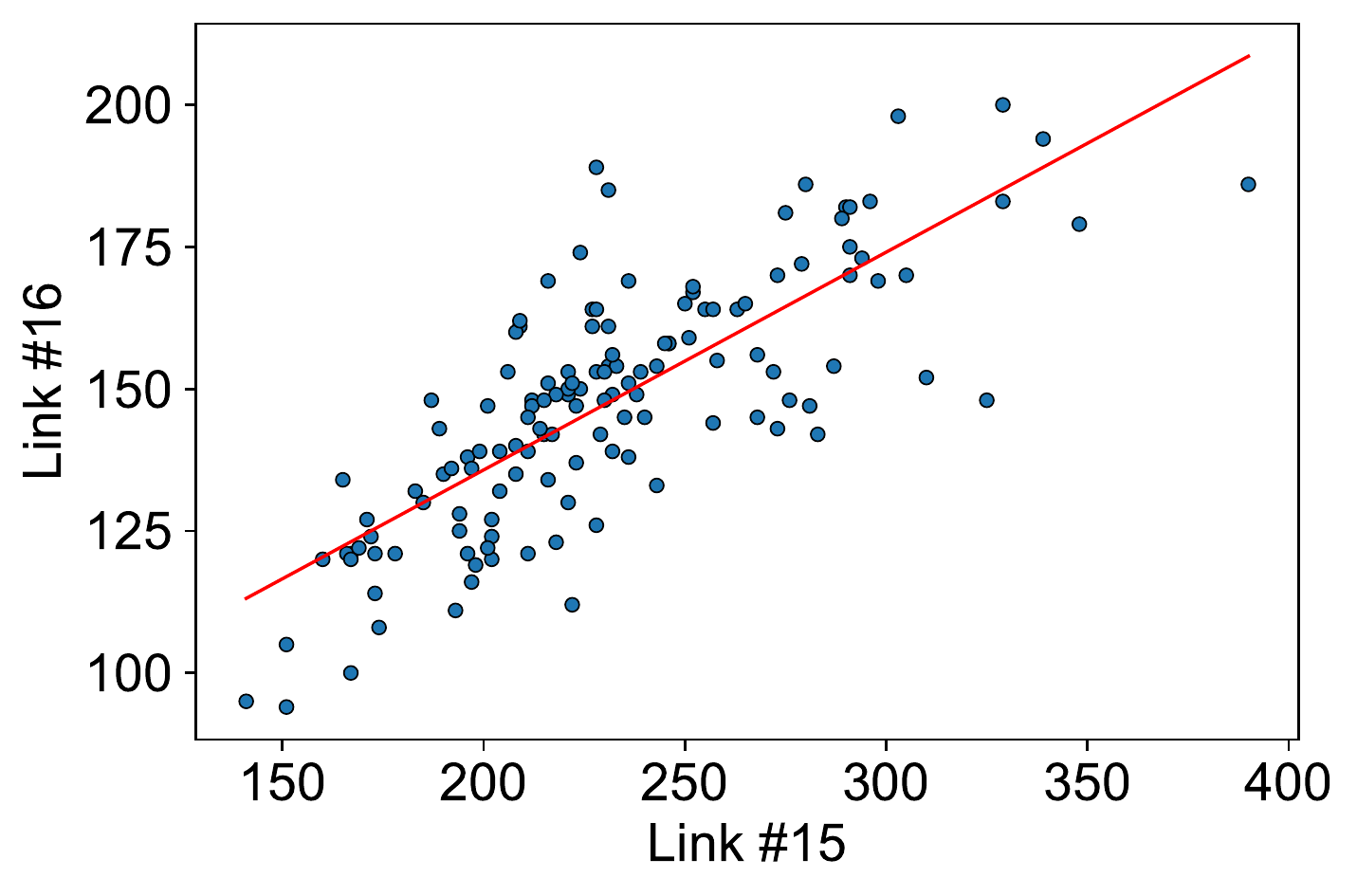}
}
\caption{The travel time scatter plots of some strong correlated link pairs during afternoon peak.}
\label{Scatter_Afternoon}
\end{figure*}

\begin{figure*}[!ht]
\centering

\subfigure[Travel time on links \#11 and \#12.]{
    \centering
    \includegraphics[scale=0.33]{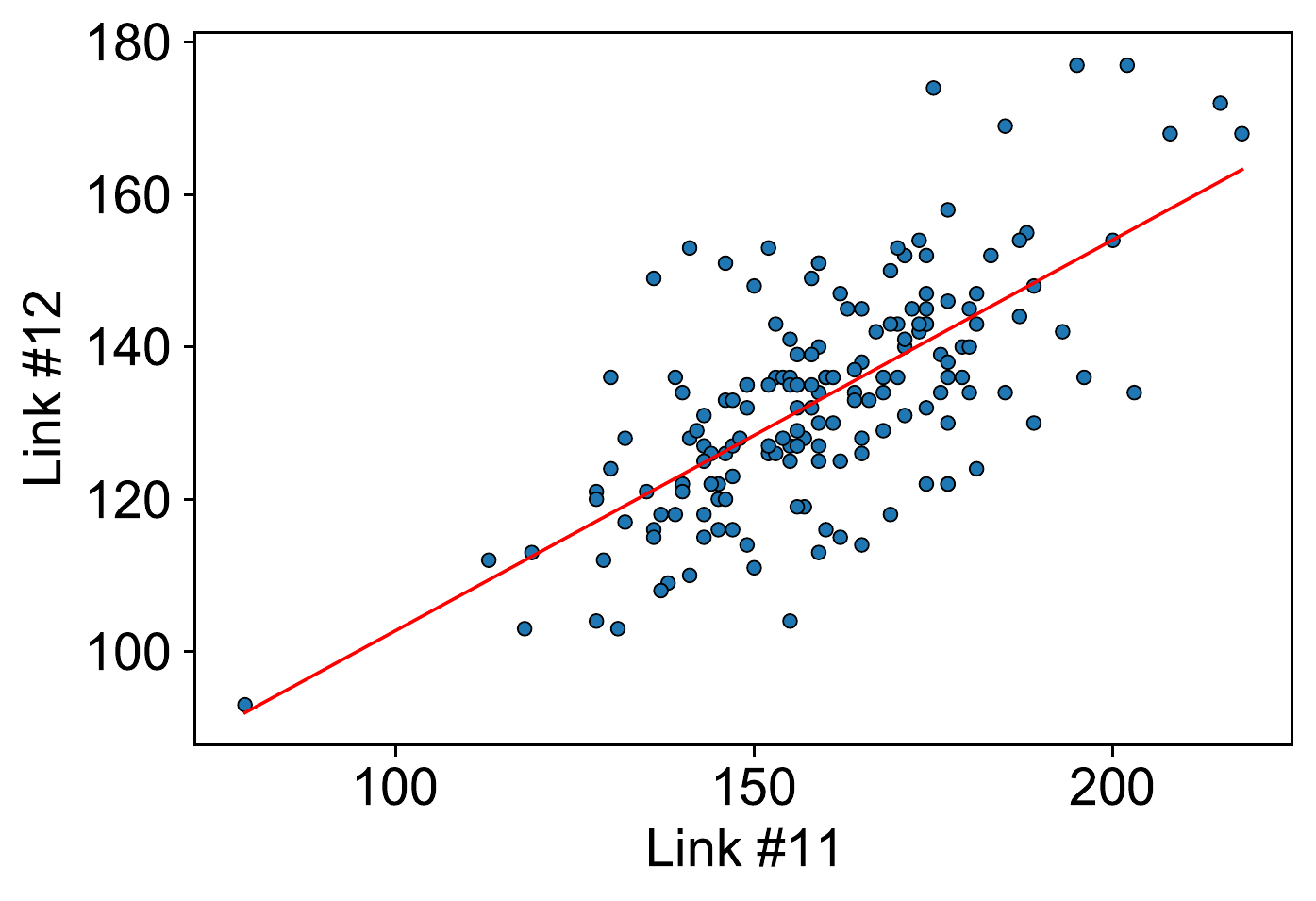}
}
\subfigure[Travel time on links \#11 and \#13.]{
    \centering
    \includegraphics[scale=0.33]{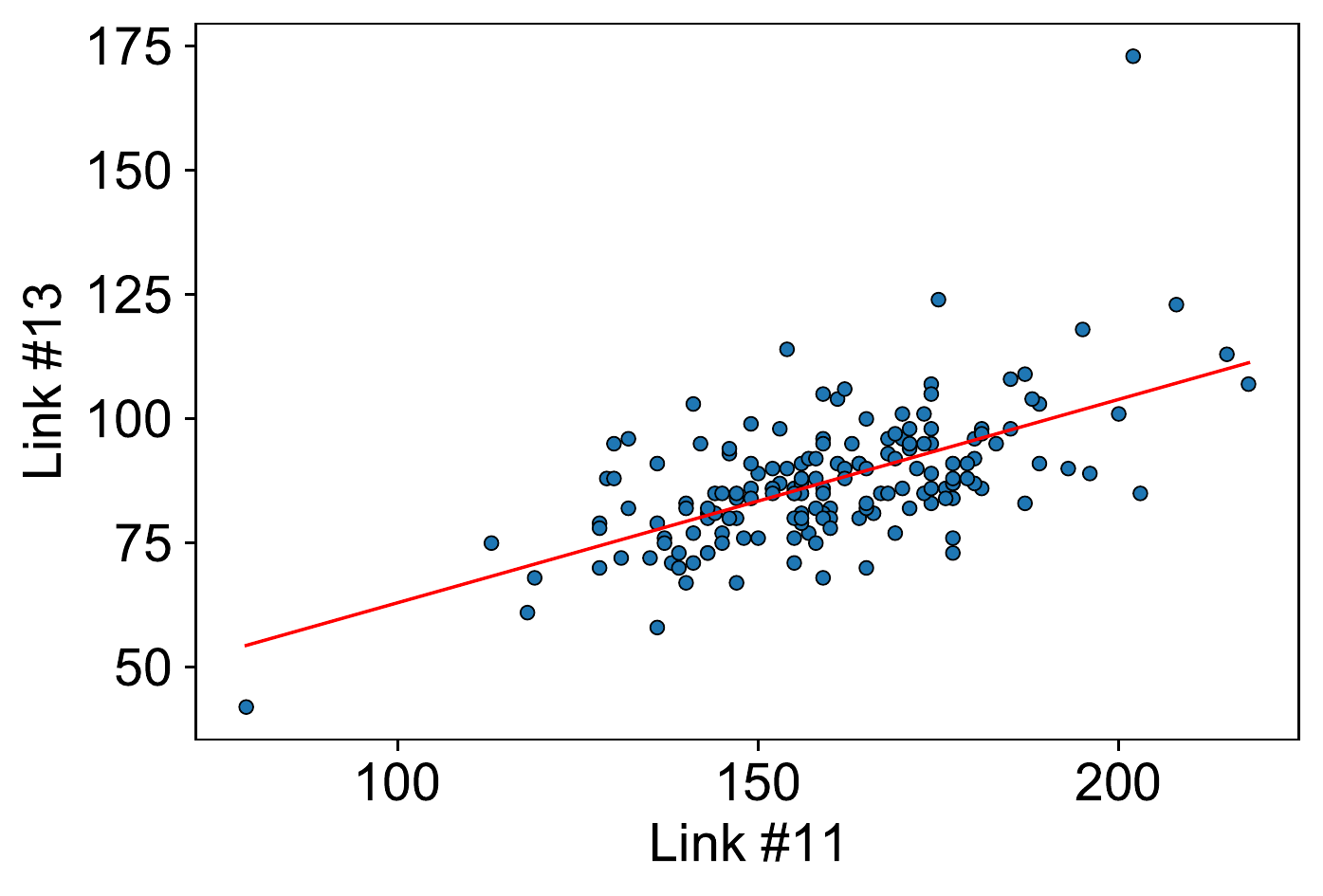}
}
\subfigure[Travel time on links \#12 and \#13.]{
    \centering
    \includegraphics[scale=0.33]{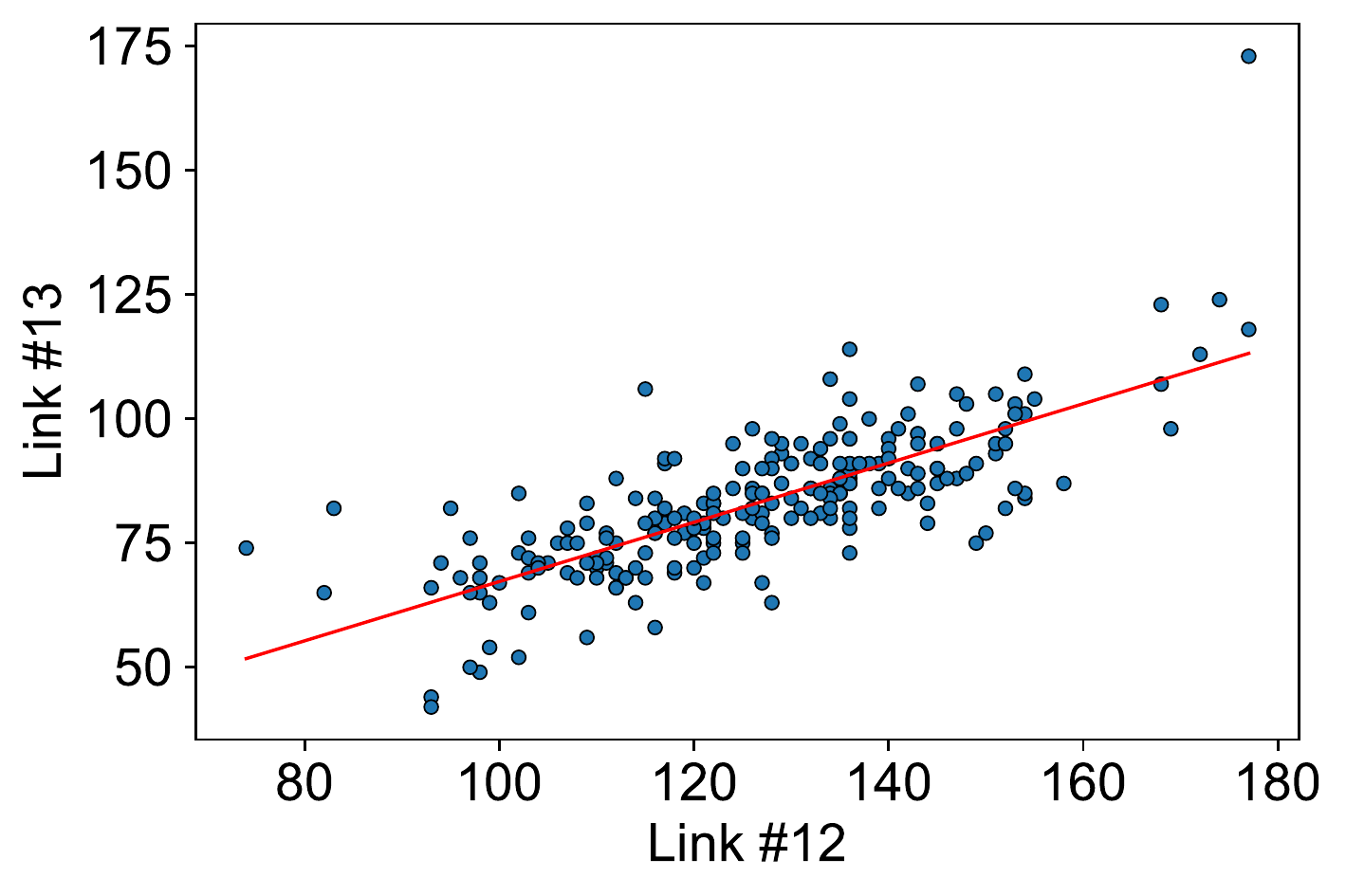}
}
\subfigure[Travel time on links \#12 and \#14.]{
    \centering
    \includegraphics[scale=0.33]{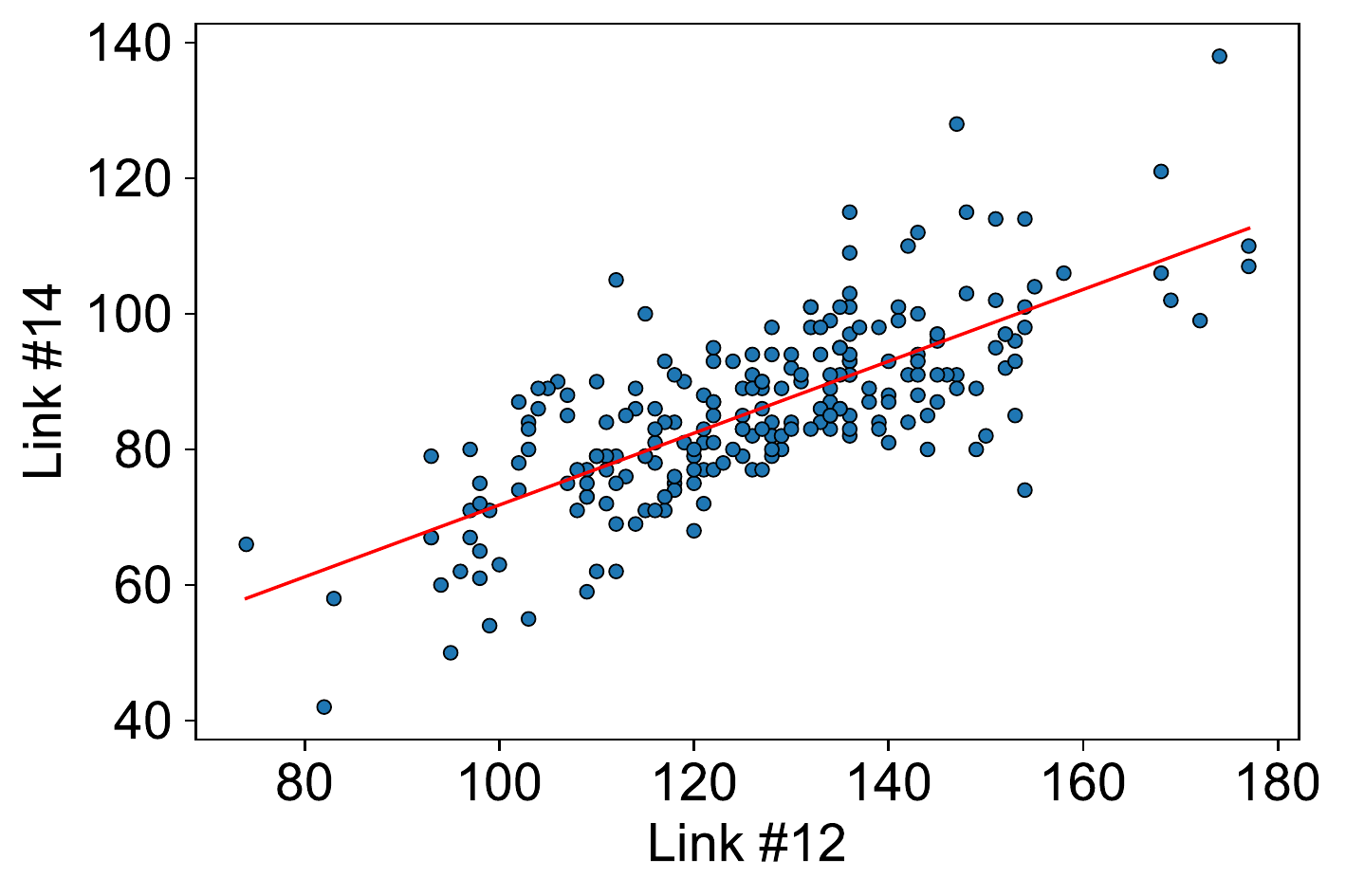}
}
\subfigure[Travel time on links \#12 and \#15.]{
    \centering
    \includegraphics[scale=0.33]{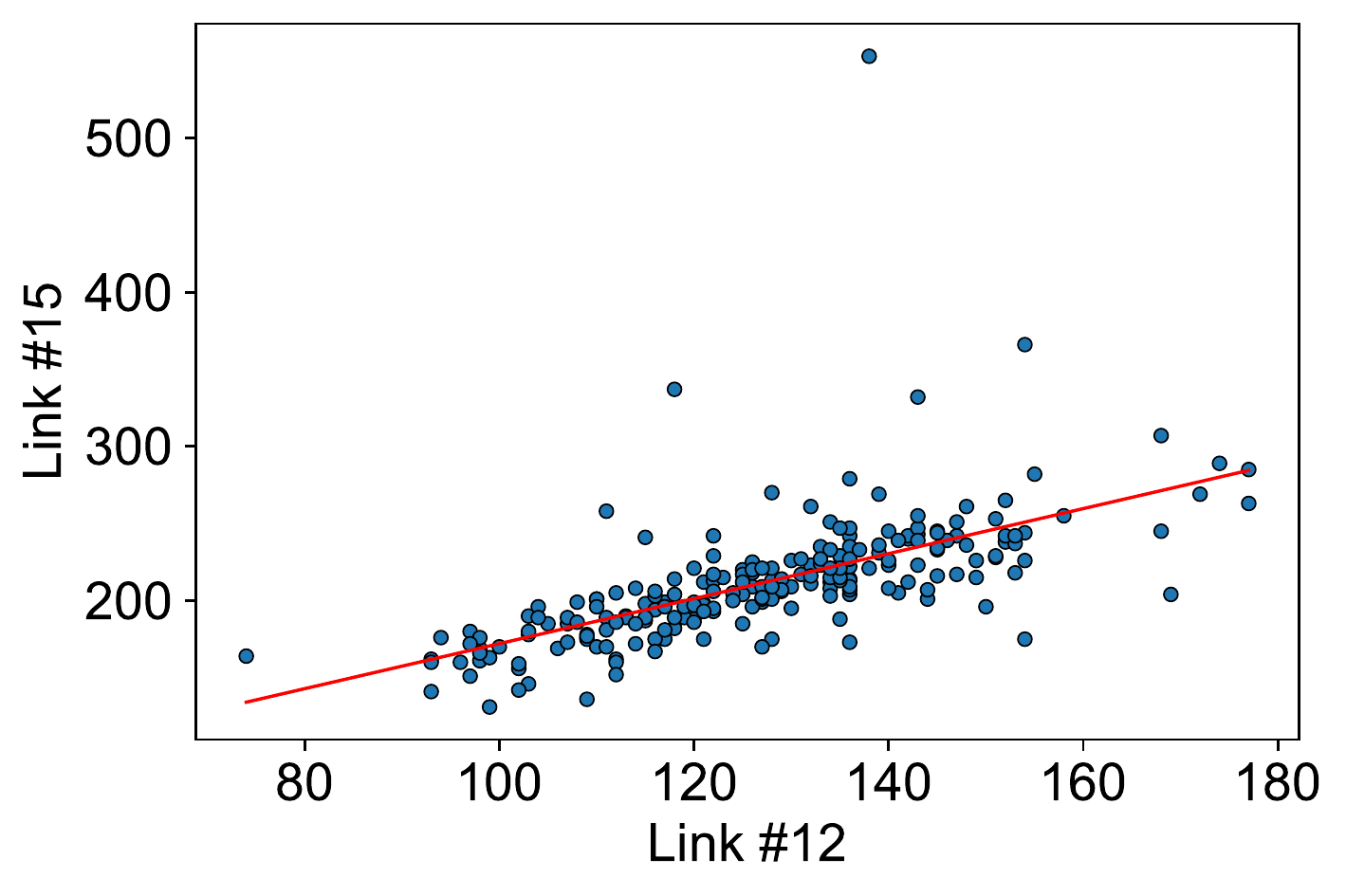}
}
\subfigure[Travel time on links \#13 and \#14.]{
    \centering
    \includegraphics[scale=0.33]{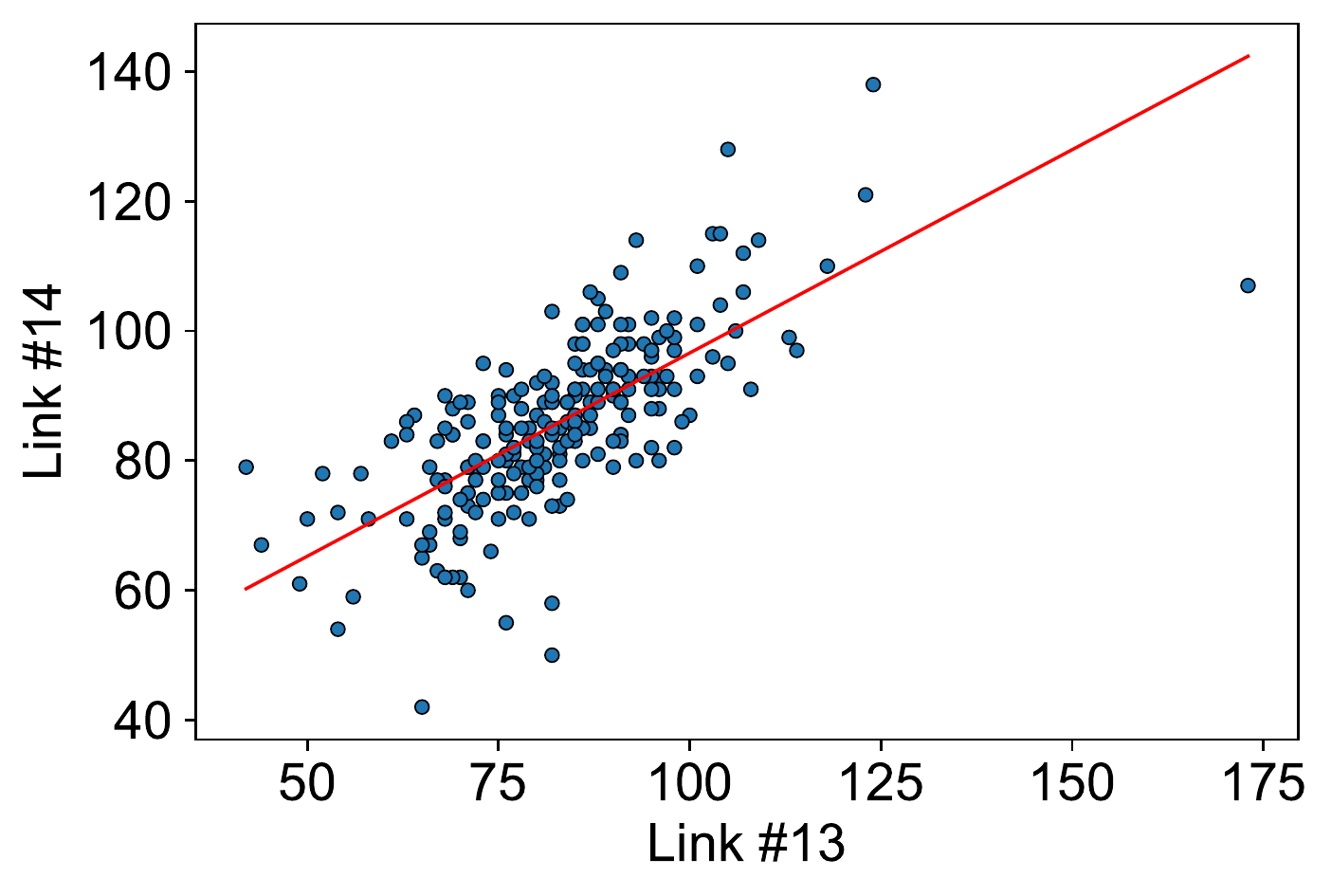}
}
\subfigure[Travel time on links \#13 and \#15.]{
    \centering
    \includegraphics[scale=0.33]{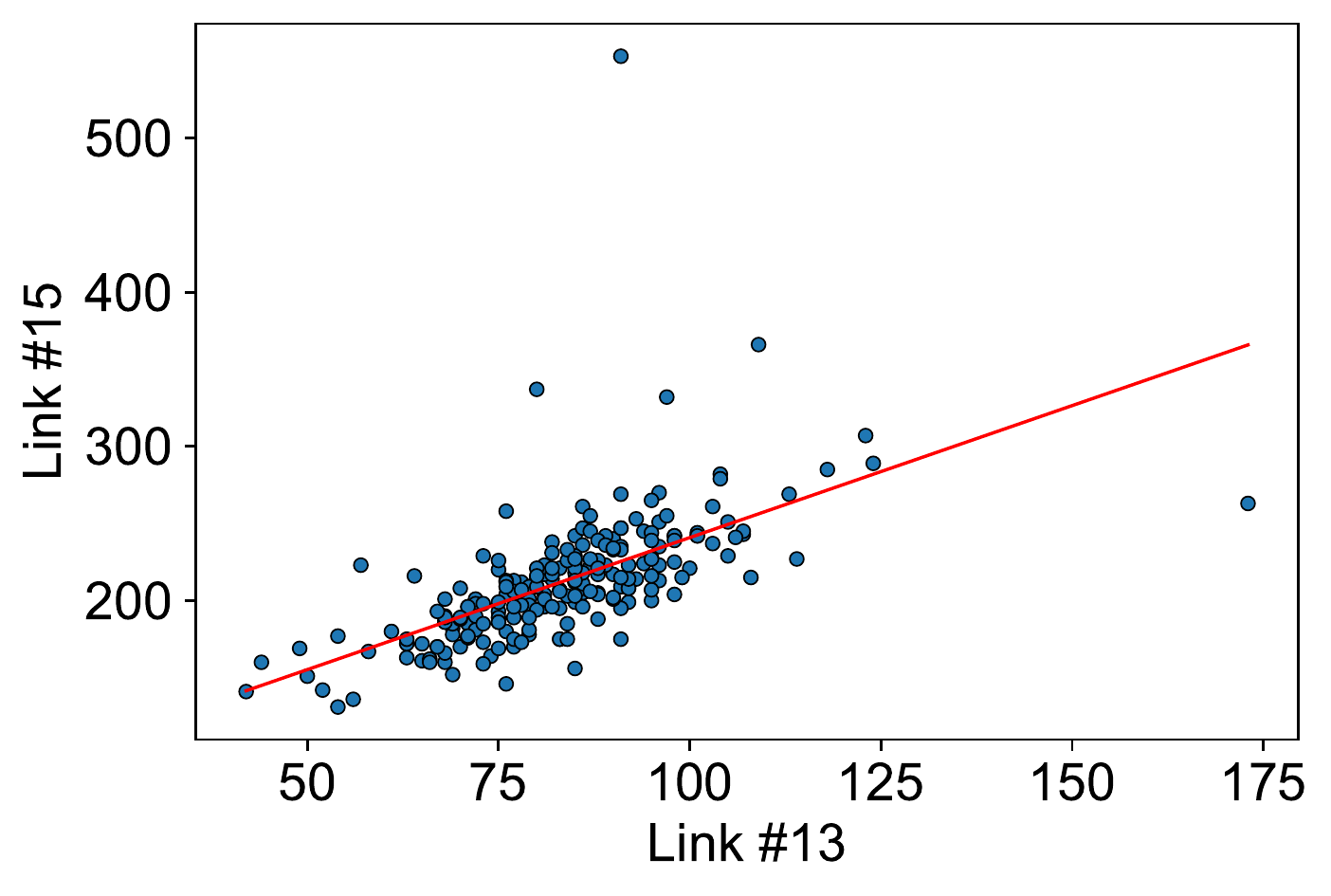}
}
\subfigure[Travel time on links \#14 and \#15.]{
    \centering
    \includegraphics[scale=0.33]{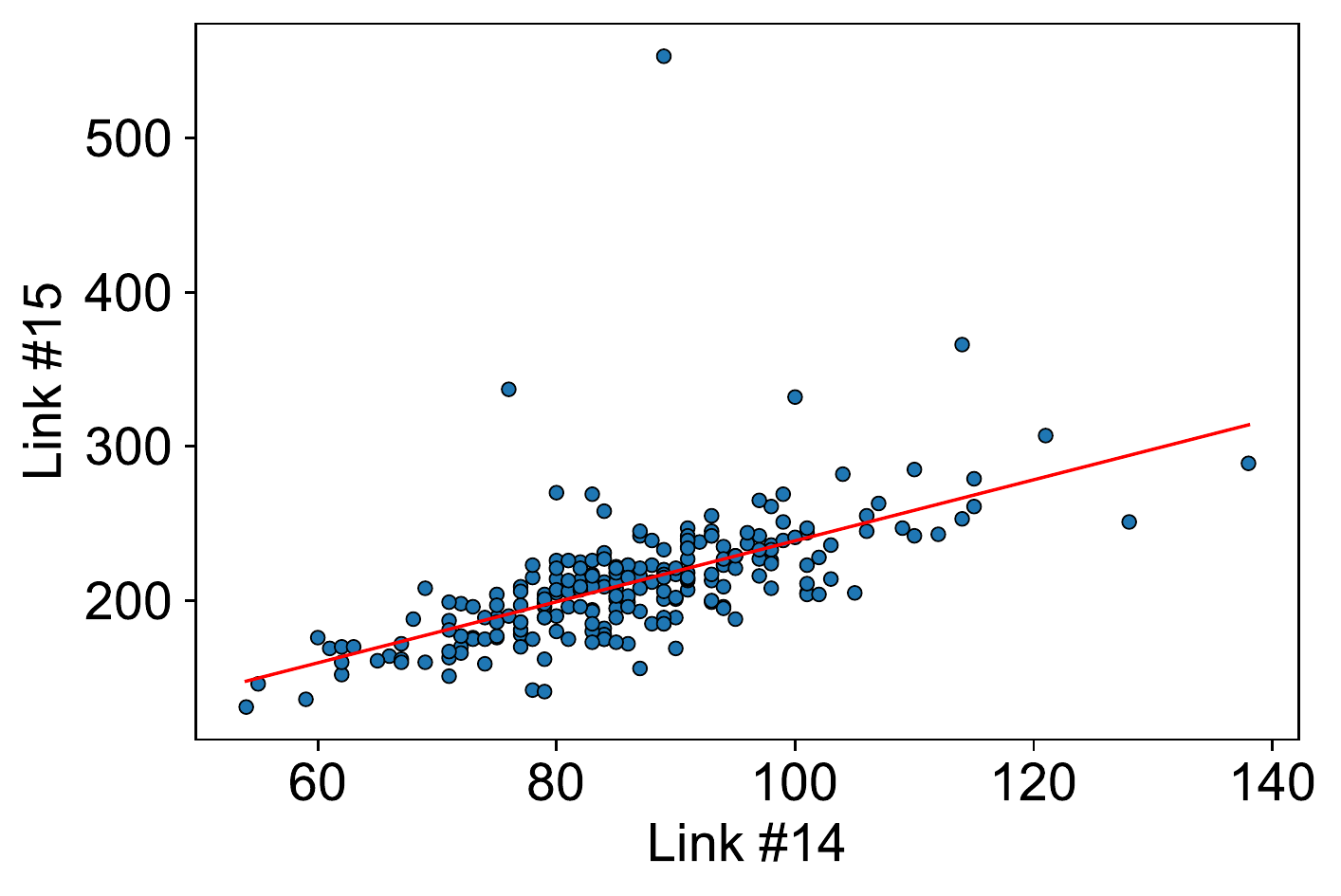}
}
\subfigure[Travel time on links \#14 and \#17.]{
    \centering
    \includegraphics[scale=0.33]{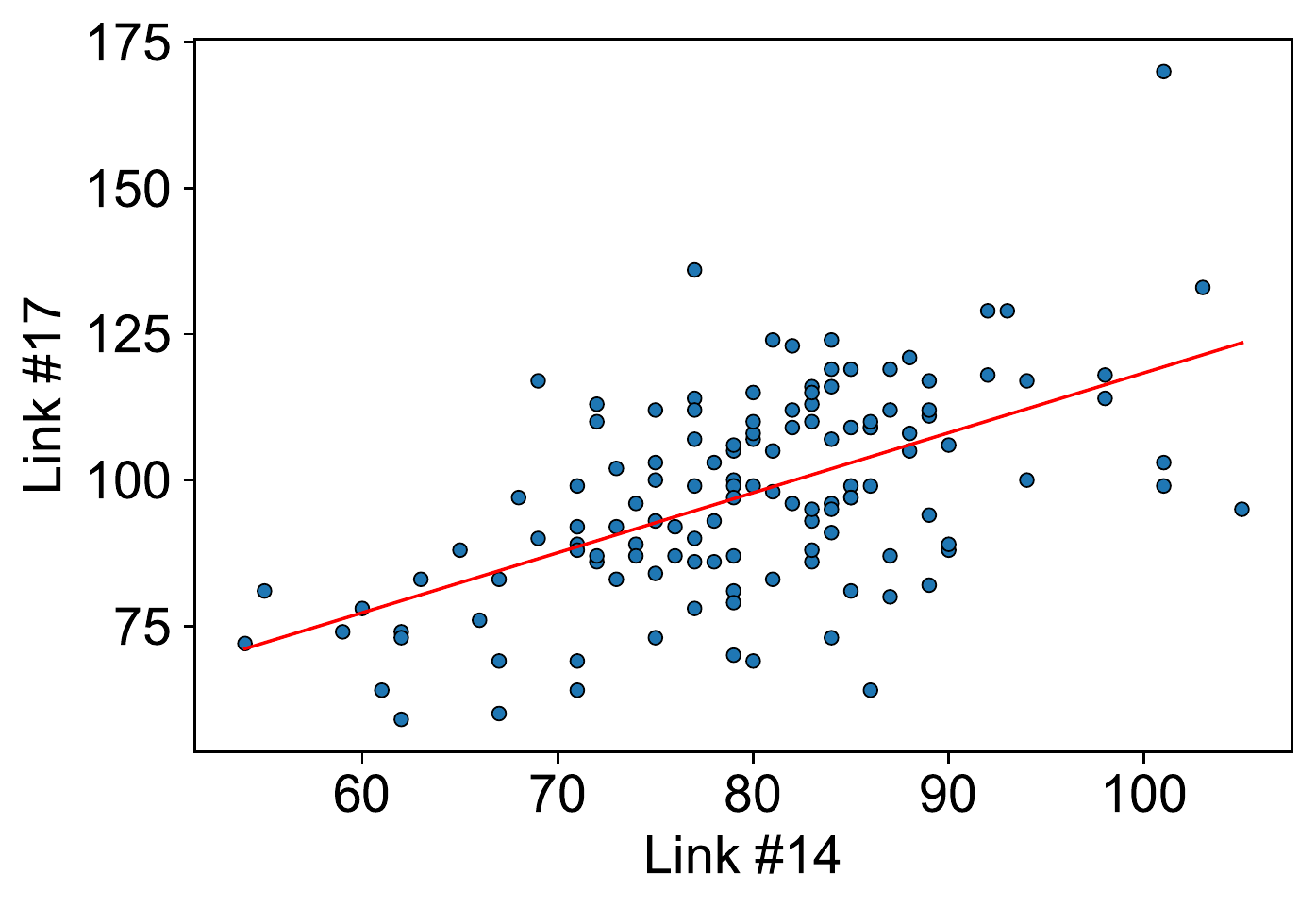}
}
\caption{The travel time scatter plots of some strong correlated link pairs during night period.}
\label{Scatter_Night}
\end{figure*}

\end{document}